\documentclass[11pt, oneside]{article}   	
\usepackage{slashed}
\usepackage{jheppub}
\usepackage{amsmath}
\usepackage{amssymb}
\usepackage{amsfonts}
\usepackage{amsthm}
\usepackage{amsbsy}
\usepackage{array}
\usepackage{mathtools}
\usepackage[usenames,dvipsnames]{xcolor}
\usepackage{subcaption}
\usepackage{dsdshorthand}
\usepackage{graphicx}
\usepackage[vcentermath]{youngtab}
\usepackage{hyperref}
\usepackage{cleveref}
\usepackage{cancel}
\usepackage[normalem]{ulem}

\usepackage{tikz}
\usetikzlibrary{snakes}
\usetikzlibrary{decorations}
\usetikzlibrary{trees}
\usetikzlibrary{decorations.pathmorphing}
\usetikzlibrary{decorations.markings}
\usetikzlibrary{external}
\usetikzlibrary{intersections}
\usetikzlibrary{shapes,arrows}
\usetikzlibrary{arrows.meta}
\usetikzlibrary{calc}
\usetikzlibrary{shapes.misc}
\usetikzlibrary{decorations.text}
\usetikzlibrary{backgrounds}
\usetikzlibrary{fadings}
\usetikzlibrary{tikzmark,calc,arrows,shapes,decorations.pathreplacing}
\usetikzlibrary{patterns}
\tikzset{
        cross/.style={cross out, draw=black, minimum size=2*(#1-\pgflinewidth), inner sep=0pt, outer sep=0pt},
	branchCut/.style={postaction={decorate},
		snake=zigzag,
		decoration = {snake=zigzag,segment length = 2mm, amplitude = 2mm}	
    }}

\newcommand{\bea}{\setlength\arraycolsep{2pt} \begin{eqnarray}}
\newcommand{\eea}{\end{eqnarray}}
\def\ft#1#2{{\textstyle{\frac{\scriptstyle #1}{\scriptstyle #2} } }}
\def\fft#1#2{{\frac{#1}{#2}}}

\newcommand{\baa}{\begin{align}}
\newcommand{\eaa}{\end{align}}

\makeatletter
\def\@fpheader{\ }
\makeatother

\title{Regge trajectories, detectors, and distributions in the critical ${\rm O}(N)$ model
}
\author{Yue-Zhou Li$^\Delta$, David Simmons-Duffin$^J$}
\affiliation{
${}^\Delta$ Department of Physics, Princeton University, Princeton, NJ 08544, USA \\
${}^J$ Walter Burke Institute for Theoretical Physics, Caltech, Pasadena, California 91125, USA
}
\emailAdd{liyuezhou@princeton.edu}
\emailAdd{dsd@caltech.edu}
\date{}
\abstract{We explore light-ray operators in the critical O$(N)$ model in the large-$N$ limit, focusing on leading-twist and leading ``horizontal" trajectories. We distinguish between light-ray operators in two conformal frames: detector operators, which characterize event shapes of final states, and distribution operators, which probe initial-state distributions. In particular, we identify parton distribution functions (PDFs) and collinear functions as matrix elements of appropriate distribution operators. We renormalize some simple detector operators at leading order in $1/N$, allowing us to extract the Regge intercept and the anomalous spin of the leading horizontal trajectory. We furthermore renormalize distribution versions of these operators, obtaining the leading-twist splitting function and a BFKL-type kernel, which match results from the detector frame. Finally, we show how these results can be read off from OPE data encoded in the Bethe-Salpeter resummation of conformal four-point functions.}

\begin{document}

\maketitle
\pagenumbering{roman}
\setcounter{page}{2}
\newpage
\pagenumbering{arabic}
\setcounter{page}{1}

\section{Introduction}

Local operators play a fundamental role in quantum field theory (QFT). Their correlation functions encode the causal structure of spacetime, the spectrum of excitations and their interactions, and serve as probes of the microscopic structure of quantum states. In collider physics, local operators can create states that are accessible in experiments. In conformal field theory (CFT), quantum numbers of local operators determine critical exponents that govern universal scaling behavior at long distances.

In perturbative QFTs, computations involving bare local operators quickly encounter ultraviolet (UV) divergences, requiring renormalization to remove them. It is the renormalized operators that construct physical observables, as described above. For example, the scaling behavior of renormalized operators at a fixed point determines the critical exponents of a CFT.

However, non-local operators can capture nontrivial physics that is not manifest in the spectrum of local operators alone. This insight dates back to studies in perturbative quantum chromodynamics (QCD), which found that hadronic wavefunctions, distribution and density functions, and fragmentation functions can be defined in terms of non-local operators along a null line \cite{Lipatov:1974qm,Altarelli:1977zs,Lepage:1980fj,Collins:1981uw,Chernyak:1983ej,Mankiewicz:1997uy,Geyer:1999uq,Geyer:1994zu,Muller:1994ses}. These operators are essential for describing cross sections with large scale separations, such as those encountered in Regge physics \cite{Balitsky:1987bk,Balitsky:2001gj}.\footnote{There are other types of non-local operators that model impurities in quantum systems, known as defects. These operators are important for understanding the symmetries and phases of quantum systems, which will not be discussed in this paper. See, e.g.~\cite{Andrei:2018die} for a review.}

Many collider measurements are associated with non-local operators. For example, the energy flux operator measured in collider experiments can be written as a null integral of the stress tensor \cite{Basham:1978zq,Basham:1978bw,Basham:1979gh,Hofman:2008ar}, known as the averaged null energy condition (ANEC) operator. The ANEC operator and its correlations are important in phenomenological studies, for example providing windows into infrared-safe precision physics \cite{Sveshnikov:1995vi,Tkachov:1995kk,Korchemsky:1999kt,Larkoski:2013eya,Belitsky:2013xxa,Lee:2022uwt,Komiske:2022enw,CMS:2024mlf}. They also play a crucial role in quantum information \cite{Ford:1994bj,Hartman:2016lgu,Faulkner:2016mzt}, holography \cite{Hofman:2008ar,Camanho:2014apa,Buchel:2009sk,Larkoski:2013eya,Kologlu:2019bco,Chen:2024iuv}, Lorentzian CFT \cite{Hartman:2015lfa,Hartman:2016lgu,Hartman:2016dxc,Afkhami-Jeddi:2017rmx,Kravchuk:2018htv} (and SCFT \cite{Belitsky:2013ofa,Belitsky:2013bja,Henn:2019gkr}), and in constraints on RG flows such as the $a$-theorem \cite{Hartman:2023qdn,Hartman:2023ccw,Hartman:2024xkw}. See \cite{Moult:2025nhu} for a beautiful review.

In parallel, recent progress in Lorentzian CFT has emphasized the significance of additional non-local operators associated to a null line, known as light-ray operators. Light-ray operators make analyticity in spin \cite{Caron-Huot:2017vep,Simmons-Duffin:2017nub} manifest and provide useful tools for understanding conformal Regge theory \cite{Costa:2012cb,Caron-Huot:2020nem,Kravchuk:2018htv,Homrich:2022cfq,Homrich:2024nwc}. 
This development in CFT has interesting applications to collider physics. For example, the light-ray operator product expansion (OPE) \cite{Hofman:2008ar,Kologlu:2019mfz,Chen:2019bpb,Chang:2020qpj,Korchemsky:2021htm} can help organize and understand energy correlators \cite{Chang:2020qpj,Chang:2022ryc,Dixon:2019uzg,Chen:2020vvp,Chen:2020adz,Chen:2021gdk,Komiske:2022enw,Cuomo:2025pjp}.

It is thus interesting to find connections between the non-local operators traditionally used to describe measurements and cross sections in generic QFTs with light-ray operators in CFTs. We can first explore these connections in perturbative theories. We may then hope to transfer lessons from perturbative studies to elucidate nonperturbative structures in QFT that go beyond local operators.

This approach was taken in \cite{Caron-Huot:2022eqs}, where the authors studied non-local detector operators in weakly-coupled $\phi^4$-theory. These detector operators become CFT light-ray operators at the Wilson-Fisher fixed point. Bare detectors exhibit infrared (IR) divergences, analogous to the UV divergences of bare local operators \cite{Caron-Huot:2022eqs}. These IR divergences can be renormalized, leading to a set of IR-finite detector operators in $\phi^4$-theory. Among these detectors are novel ``horizontal trajectories" which are analogous to the BFKL trajectory that appears in perturbative gauge theories. There have also been recent studies of higher-twist light-ray operators in perturbative theories \cite{Homrich:2022cfq,Henriksson:2023cnh}.


In this paper, we adapt techniques from \cite{Caron-Huot:2022eqs} to study detector operators in the critical O$(N)$ model in the large-$N$ limit. This suggests that it should be possible to characterize the space of detectors even in strongly coupled theories, though we retain perturbative control via the $1/N$ expansion. We also perform an RG analysis of detector operators to extract the Regge intercept and study a leading horizontal Regge trajectory.

We note that detectors, as probes of collider observables, are placed at future null infinity $\mathcal{J}^+$. In this work, we also study another class of non-local operators, which we refer to as ``distribution" operators, which live on a null plane inside Minkowski space. In CFTs, detector operators and distribution operators are related by a change of conformal frame --- they are both examples of light-ray operators, inserted at different locations. However, in non-conformal theories, these operators have different physical interpretations and can exhibit different physics. We explain how distribution operators are related to several non-local objects appearing in the QCD context, 
%
%
 such as those defining parton distribution functions (PDFs) \cite{Collins:1981uw,Collins:1989gx,Sterman:1995fz} and collinear functions \cite{Rothstein:2016bsq,Neill:2023jcd,Gao:2024qsg}. All of these objects thus have natural analogies in perturbative CFT. To illustrate this idea, we study PDFs and collinear functions in the $O(N)$ model, defined as matrix elements of the appropriate distribution operators. We explain how to renormalize them and study their RG evolution, finding results that agree with our study of detector operators (because the theory is conformal).
 

The rest of this paper is organized as follows: In Section \ref{sec: review O(N)}, we review the critical O$(N)$ model in the large-$N$ limit and analyze its Regge trajectories and Regge intercepts based on existing data. In Section \ref{sec: lightray operator}, we revisit light-ray operators in CFTs, discuss the physical distinction between different frames, and accordingly define detector operators and distribution operators. We further establish an explicit relation linking distribution operators to PDFs and collinear functions. In Section \ref{sec: detector operator}, we renormalize some detector operators in the $O(N)$ model at leading nontrivial order in $1/N$. Renormalizing the leading-twist trajectory requires understanding its mixing with its shadow, allowing us to extract the Regge intercept of the theory. The renormalization of a simple class of horizontal detectors gives rise to the anomalous spin of their trajectory. In Section \ref{sec: RG distribution operator}, we study renormalization of distribution operators, following closely the treatment of PDFs and collinear functions. The former yields the leading-twist splitting function, while the latter is governed by rapidity renormalization and leads to the same anomalous spin as that of horizontal detector operators. In Section \ref{sec: BS equation}, we show that the leading-twist anomalous dimensions and horizontal anomalous spin are both encoded in the Bethe-Salpeter resummation of conformal correlators. We summarize in Section \ref{sec: summary}.

In Appendix \ref{app: wavefunction RG}, we review wavefunction renormalization in the critical O$(N)$ model. In Appendix \ref{app: Lorentzian inversion}, we review the Lorentzian inversion formula and compute the OPE data used in the main text. In Appendix \ref{app: more horizontal}, we provide further details on renormalizing the horizontal detectors. In Appendix \ref{app: rapidity}, we conceptually revisit rapidity renormalization.


\section{Critical ${\rm O}(N)$ vector model and its Regge trajectories}
\label{sec: review O(N)}

\subsection{The critical ${\rm O}(N)$ vector model}

The critical ${\rm O}(N)$ vector model is the IR fixed point of a theory of $N$ scalars $\f^i$ transforming in the vector representation of $O(N)$, with the action
\be
S=-\fft{1}{2}\int d^dx\big((\partial_\mu \phi^i)^2+m^2 \phi^i \phi^i+\fft{\lambda}{2}(\phi^i \phi^i)^2\big)\,.
\ee
This theory is weakly-coupled in the large-$N$ limit. To develop perturbation theory, we can apply the Hubbard-Stratonovich trick by introducing an auxiliary field $\sigma$, and replacing the $\phi^4$ interaction with a quadratic term $\s^2$ and a cubic interaction $\sigma \phi^i \phi^i$. For $d<4$, we can neglect the $\sigma^2$ term in the IR, and we obtain an effective action for the critical theory  (see, e.g., \cite{Giombi:2016ejx,Henriksson:2022rnm} for a review)
\be
S_{\rm cri}=-\fft{1}{2}\int d^d x\big((\partial_\mu \phi^i)^2+\fft{1}{\sqrt{N}}\sigma \phi^i\phi^i\big)\,.\label{eq: critical action}
\ee
The propagator for $\sigma$ is determined by its effective action after integrating out $\phi^i$:
\be
G_\sigma(p^2) = C_\sigma (p^2)^{\fft{d-4}{2}}\,,\quad \langle\sigma(x)\sigma(y)\rangle= N_\sigma \fft{1}{|x-y|^4}\,,
\ee
where
\be
C_\sigma=2^{d+1} (4\pi)^{\fft{d-3}{2}}\Gamma(\tfrac{d-1}{2})\sin(\tfrac{\pi d}{2}) \,,\quad N_\sigma=\fft{16}{(4\pi)^{\fft{d}{2}}\Gamma(\fft{d-4}{2})}C_\sigma\,.
\ee
The auxiliary primary $\sigma$ essentially replaces the $\phi^2$ operator at the critical point. Thus, in the UV $\phi^2$ has dimension $d-2$, while in the IR it flows to $\sigma$ which has dimension $2$.

We have motivated the effective action \eqref{eq: critical action} by thinking about $3 \leq d < 4$, but it turns out that this theory can be analytically continued to $d < 6$, where it serves as the UV critical point of the standard O$(N)$ model as well as the IR critical point of the ${\rm O}(N)$ cubic model \cite{Parisi:1975im,Fei:2014yja,Fei:2014xta,Gracey:2015tta}. We will thus study the critical action \eqref{eq: critical action} in this paper for general dimensions $3 \leq d < 6$ (excluding $d=4$ since in that case there is no IR fixed point). 

Although we work in general $d$, in perturbation theory one encounters divergences associated with the $\sigma$ propagator that are not regulated by dimensional regularization. A standard scheme to deal with these divergences is to deform $\Delta_\sigma\to 2-\e$, and subtract poles in $\e$ before taking $\e\to 0$. We emphasize that $\e$ is not related to $d$. We review details of this regularization scheme in appendix~\ref{app: wavefunction RG}.

\subsection{Regge trajectories and anomalous dimensions}

Let us list some families of local operators that will be important in this work. The lowest twist families are ``double-twist" operators built from two scalars $\f^i$. These can be grouped into $O(N)$ singlets, antisymmetric tensors, and symmetric tensors as follows:
\be
& [(\phi^i \phi^j)_s]_{J}\sim \phi^i \partial_{\mu_1}\cdots \partial_{\mu_J}\phi^i+\cdots\,,&& \tau=d-2 + O(1/N)\,,
\nn\\
& [(\phi^i \phi^j)_{\rm sym}]_{J}\sim \phi^{(i} \partial_{\mu_1}\cdots \partial_{\mu_J}\phi^{j)}+\cdots\,, && \tau=d-2 + O(1/N)\,,
\nn\\
& [(\phi^i \phi^j)_{\rm asym}]_{J}\sim \phi^{[i} \partial_{\mu_1}\cdots \partial_{\mu_J}\phi^{j]}+\cdots\,, && \tau=d-2 + O(1/N)\,.
\ee 
Here ``$\dots$" represents combinations of derivatives needed to ensure that the operator is primary.  By analogy with QCD, we will sometimes refer to these leading twist operators as DGLAP-type \cite{Gribov:1972ri,Altarelli:1977zs,Dokshitzer:1977sg}. In the limit $N\to \infty$, these operators become higher-spin conserved currents with scaling dimensions $\Delta=J+d-2$. At finite $N$, the higher-spin symmetries are slightly broken and these operators acquire anomalous dimensions, except for the spin-2 singlet current, which is the stress tensor $[(\f^i \f^i)_s]_2=T_{\mu\nu}$. The leading anomalous dimensions are \cite{Lang:1992zw,Giombi:2016hkj}
\be
& \gamma_{{\rm sym},J}=\gamma_{{\rm asym},J}=\fft{1}{N}\frac{16 (J-1) (d+J-2) \sin \left(\frac{\pi  d}{2}\right) \Gamma (d-2)}{\pi  (d+2 J-4) (d+2 J-2) \Gamma \left(\frac{d}{2}-2\right) \Gamma \left(\frac{d}{2}+1\right)}\,,\nn\\
& \gamma_{s,J}=\gamma_{{\rm (a)sym},J}-\fft{1}{N}\frac{8 \sin \left(\frac{\pi  d}{2}\right) \Gamma (d-2) \Gamma (d+1) \Gamma (J+1)}{\pi  (d-1) (d+2 J-4) (d+2 J-2) \Gamma \left(\frac{d}{2}-2\right) \Gamma \left(\frac{d}{2}+1\right) \Gamma (d+J-3)}\,.\label{eq: leading twist ano}
\ee

Some example higher-twist families are
\be
\label{eq:highertwistexamples}
& [\sigma \phi^i]_J \sim \sigma \partial_{\mu_1}\cdots \partial_{\mu_J}\phi^i\,, && \tau=\tfrac{d+2}{2} + O(1/N)\,,
\nn\\
& [\sigma (\phi^i \phi^j)_\rho]_{J} \sim \sigma (\phi^i\partial_{\mu_1}\cdots \partial_{\mu_J}\phi^i)_\rho\,,&& \tau=d + O(1/N)\,,
\nn\\
& [\sigma\sigma]_J \sim \sigma \partial_{\mu_1}\cdots \partial_{\mu_J}\sigma\,,&& \tau=4 + O(1/N)\,.
\ee
Here and below, we use the notation that $\rho$ stands for an $O(N)$ representation that can be either ``$s$", ``$\mathrm{sym}$", or ``$\mathrm{asym}$".
The equation of motion implies that $\Box \phi \sim \sigma \phi$. Thus, $[\sigma(\phi^i\phi^j)_\rho]_J$ plays the role of a subleading double-twist family for $\phi$. Each additional insertion of $\sigma$ or $\phi$ generates additional subleading trajectories. Thus, even in the infinite-$N$ limit, the theory contains a rich set of families that become dense in $\tau$ at large $\tau$ (for generic $d$).

\subsection{A first look at the Chew-Frautschi plot}

A Regge trajectory is a family of light-ray operators depending continuously on spin $J$. Nonvanishing light-ray operators with nonnegative integer $J$ become light-transforms of local operators $\mathbf{L}[\cO]$ (or their shadow transforms). We can visualize  Regge trajectories using a Chew-Frautschi plot. The coordinates of the Chew-Frautschi plot are conventionally chosen to be $\Delta-\frac d 2$ and $J$, where $(\Delta,J)=(1-J_L,1-\Delta_L)$, where $(\Delta_L,J_L)$ are the quantum numbers of the corresponding light-ray operator. In these conventions, the light-transform of a local operator $\mathbf{L}[\cO_{\Delta,J}]$ appears at the point $(\Delta-\frac d 2,J)$. 

\begin{figure}[t]
\centering
\begin{subfigure}[t]{0.45\textwidth}
    \centering
    \includegraphics[width=\textwidth]{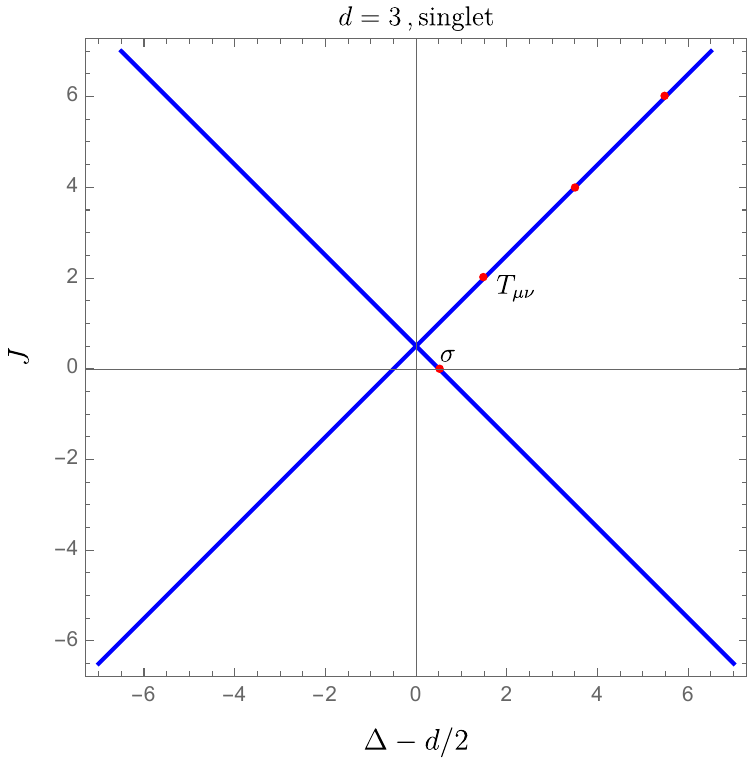} 
    \caption{}
    \label{fig:subfig1}
\end{subfigure}
\hfill
\begin{subfigure}[t]{0.45\textwidth}
    \centering
    \includegraphics[width=\textwidth]{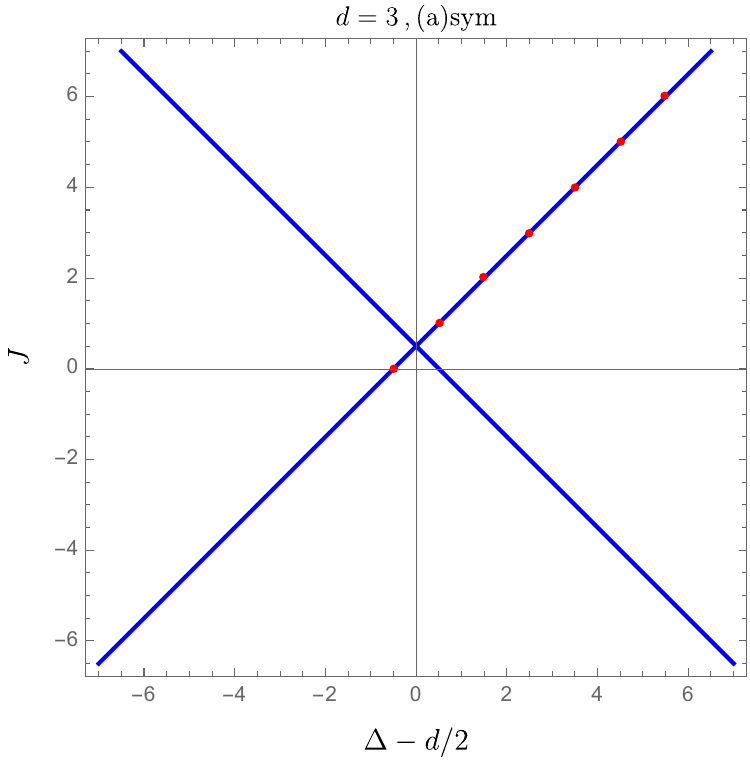} 
    \caption{}
    \label{fig:subfig2}
\end{subfigure}
\caption{The Chew-Frautschi plot of leading-twist Regge trajectories in the critical O$(N)$ model is shown. The red dots represent local operators in the form $[(\phi^i\phi^j)_\rho]_{J}^{\tau=d-2}$. In the singlet representation, there are only even-spin local operators, and we mark the singlet operator $\sigma$ and the stress-energy tensor $T_{\mu\nu}$. On the RHS, we combine the symmetric and anti-symmetric representations, including both even- and odd-spin local operators.}
\label{fig:leading-twist}
\end{figure}

\begin{figure}[t]
\centering
\begin{subfigure}[t]{0.45\textwidth}
\centering
    \includegraphics[width=1\textwidth]{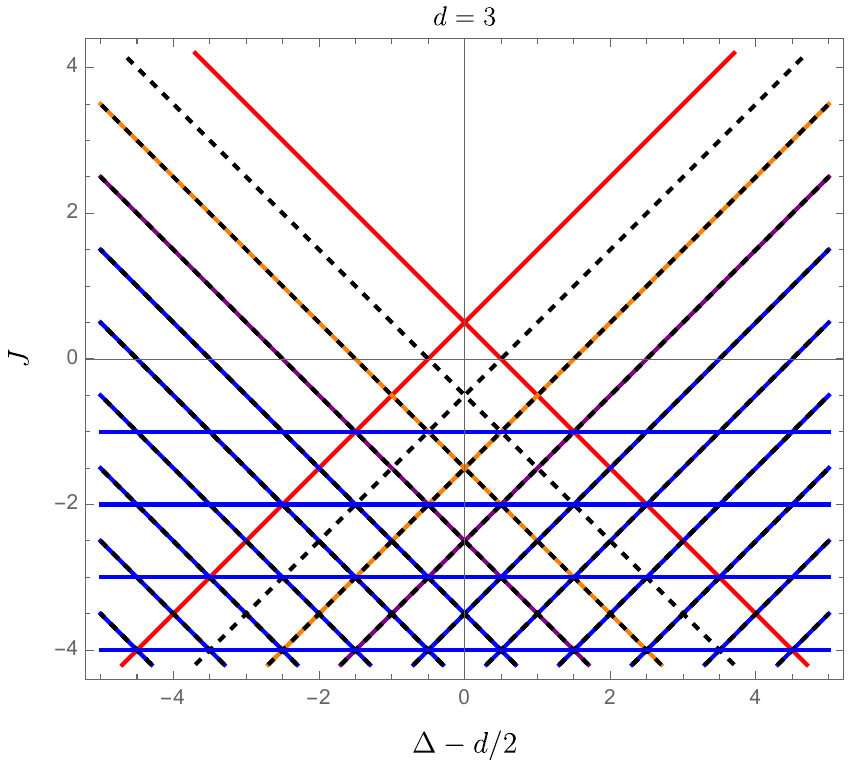} 
    \caption{}
    \end{subfigure}
    \begin{subfigure}[t]{0.45\textwidth}
    \centering
       \includegraphics[width=1\textwidth]{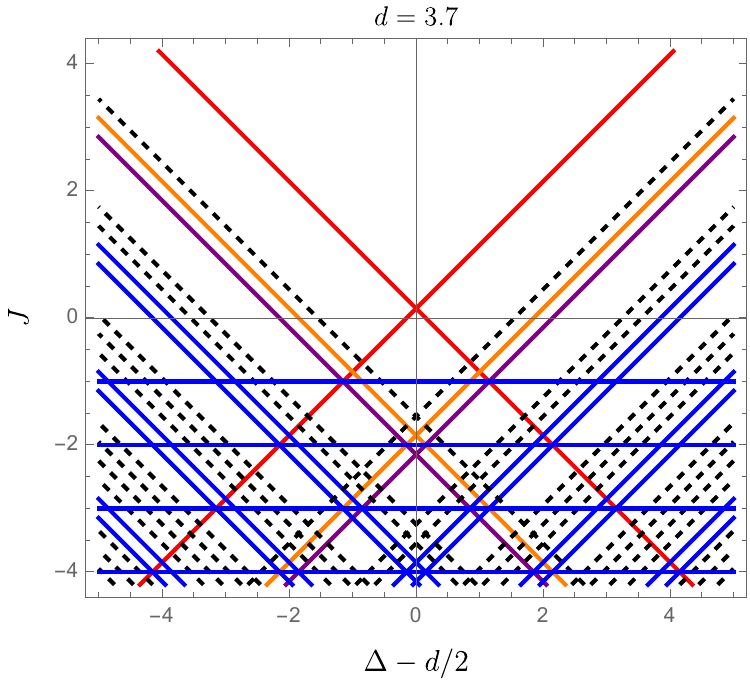} 
       \caption{}
    \end{subfigure}
    \caption{A potentially complete Chew-Frautschi plot in the critical O$(N)$ model in the strict large-$N$ limit for the singlet sector, showing the singlet Regge trajectories generated by $[(\phi^i\phi^j)_s]_J^{\tau=d-2}$, $[\sigma (\phi^i \phi^j)_s]_J^{\tau=d}$, and $[\sigma\sigma]_J^{\tau=4}$, along with their shadows, colored red, orange, and purple, respectively. Their daughter trajectories, as well as possible horizontal trajectories, are also included and marked with blue curves. The multi-twist operators $[(\phi^i\phi^i)^n \sigma^m]_J$ and $[\sigma^n]_J$ are also included and marked with black shaded curves. Different examples in $d=3$ and $d=3.7$ are shown for comparison. 
    }
    \label{fig: free Regge}
\end{figure}

We can obtain some of the Regge trajectories in the strict $N\to \oo$ limit by plotting the points corresponding to local operators and drawing straight lines through them. We must also include their images under the ``spin shadow" transform, which exchanges $\Delta\leftrightarrow d-\Delta$. Doing so for the leading-twist singlet trajectory and its shadow gives figure~\ref{fig: free Regge}. We have marked points corresponding to light transforms of local operators in red. Note that $\sigma$ appears not on the leading trajectory, but on its shadow trajectory. This persists in the presence of interactions, so that the anomalous dimension of $\sigma$ satisfies $\g_{\sigma}=-\g_{s,0}$, as a consequence of the shadow relation $\De_\sigma=d-\De_{s,J=0}$. A similar phenomenon was observed for the Wilson-Fisher theory in \cite{Caron-Huot:2022eqs}. 

The full Chew-Frautschi plot of the $N\to\oo$ theory also includes higher-twist trajectories, such as those containing the light-transforms of the operators (\ref{eq:highertwistexamples}). As in the Wilson-Fisher theory \cite{Caron-Huot:2022eqs}, we also expect horizontal trajectories at negative integer $J$. The operators corresponding to horizontal trajectories are intrinsically nonlocal, and play an important role in controlling the behavior of correlators in the Regge limit. By analogy with QCD, we refer to the leading horizontal trajectory as having Balitsky-Fadin-Kuraev-Lipatov (BFKL) type \cite{Lipatov:1976zz,Kuraev:1976ge,Kuraev:1977fs,Balitsky:1978ic}. In figure~\ref{fig: free Regge}, we show a more complete picture of the Chew-Frautschi plot in the strict large-$N$ limit, including subleading (higher-twist) trajectories and horizontal trajectories.

When we move away from the strict large-$N$ limit, we note that the anomalous dimensions $\g_{s,J}$, etc.\ exhibit poles in $J$. These singularities signal recombination of different Regge trajectories at finite $N$. It is therefore vital to understand and resolve these recombinations to understand the form of the Chew-Frautschi plot at finite $N$ and faithfully describe the underlying physics.


\subsection{Operator mixing, the leading Pomeron and the Regge intercept}

As an example, let us focus on the leading-twist trajectory $[(\phi^i\phi^i)_s]_J$. The leading singularity in its anomalous dimension \eqref{eq: leading twist ano} is located at $J = \frac{4-d}{2}$ (for $d\neq 4$). As explained in \cite{Caron-Huot:2022eqs}, singularities in  anomalous dimensions arise from mixing between light-ray operators that become degenerate at tree level at the locations of the singularities. The location $J = \frac{4-d}{2}$ is precisely where the leading-twist trajectory mixes with its shadow, $d-2+J = d-(d-2+J) \rightarrow J = \frac{4-d}{2}$. This indicates that the leading-twist trajectory and its shadow recombine into a smooth curve near this point. We will describe this recombination at the level of operators later in section~\ref{sec: detector operator mixing}. However, we can already understand what the resolved curve must look like following the simple procedure outlined in \cite{Brower:2006ea,Caron-Huot:2022eqs}. We write the equation for the anomalous dimension in a shadow-symmetric form
\be
\p{\Delta-\fft{d}{2}}^2=\p{d-2+J+\gamma_{s,J}-\fft{d}{2}}^2\,\label{eq: quadratic eq},
\ee
and then expand to leading order in $1/N$. The singularity at $J = \frac{4-d}{2}$ cancels, and we are left with a smooth curve near this location. See also \cite{Manashov:2025kgf} for explorations of this smooth curve in the O(N) model at $d = 4 - \epsilon$ up to four loops, as well as for generalizations to other models\footnote{We are grateful to Sven Moch for pointing this out to us.}. 

From the smooth curve, we can read off the Regge intercept $J^\ast$, which is the largest value of $J$ at which $\Delta = \frac{d}{2}$. The value $J^\ast$ is the spin of the ``pomeron" \cite{Forshaw:1997dc}: a light-ray operator that controls the growth of correlation functions in the Regge limit \cite{Costa:2012cb}. We can also compute analogous Regge intercepts $J_\rho^\ast$ for different $O(N)$ representations $\rho\in\{s,\mathrm{sym},\mathrm{asym}\}$.
%
%
We find
\be
& J^\ast_\rho= c_0 + \fft{c_{1,\rho}}{\sqrt{N}} + \fft{c_{2,\rho}}{N}\,,\quad c_0=\fft{4-d}{2}\,,\quad c_{1,\rho} = \Big(\frac{4^{d-1} \Gamma \left(\frac{d-1}{2}\right)^2}{\pi  \Gamma \left(\frac{d}{2}-2\right)^2}\delta_{\rho,s}+\frac{2^d \sin \left(\frac{\pi  d}{2}\right) \Gamma \left(\frac{d-1}{2}\right)}{\pi ^{3/2} \Gamma \left(\frac{d}{2}-2\right)}\Big)^{\fft{1}{2}}\,,\nn\\
& c_{2,\rho}=\frac{2^{d-4} \Gamma \left(\frac{d-1}{2}\right) \left(\frac{2 \pi  d \left(-d+\pi  (d-4) \cot \left(\frac{\pi  d}{2}\right)+2\right) \Gamma (d-1)}{\Gamma \left(\frac{d}{2}-2\right) \Gamma \left(\frac{d}{2}-1\right)}\delta_{\rho,s}-(d-4) ((d-2) d+4) \sin \left(\frac{\pi  d}{2}\right)\right)}{\pi ^{3/2} \Gamma \left(\frac{d}{2}+1\right)}\,.\label{eq: intercept}
\ee
Specializing to $d=3$, these become
\be
J^\ast_s\big|_{d=3}=\fft{1}{2}+ \fft{2\sqrt{2}}{\pi \sqrt{N}} - \fft{8}{3\pi^2 N}\,,\quad J^\ast_{{\rm (a)sym}}\big|_{d=3}=\fft{1}{2}+\fft{2}{\pi \sqrt{N}} -\fft{14}{3\pi^2 N}\,.
\ee
These results match the analysis in \cite{Liu:2020tpf,Caron-Huot:2020ouj}. Note that the Regge intercept of the singlet representation is larger than that of the symmetric and anti-symmetric representations, implying that the singlet channel dominates in the Regge limit.

This analysis predicts how recombination happens at the level of curves on the Chew-Frautschi plot, but it is worthwhile to understand in detail how recombination works at the level of operators. In particular, an operator analysis will lead to explicit expressions for the pomeron operator which could, for example, be used to study its correlation functions. We undertake this analysis in section \ref{sec: detector operator}, following similar logic to that of \cite{Caron-Huot:2022eqs} in the Wilson-Fisher theory (though many of the details are different).


Besides the pomeron singularity at $J=\frac{4-d}{2}$, additional singularities remain, and we would like to briefly understand their origins. 
The precise pattern of singularities depends on $d$, so let us discuss a few examples on a case by case basis, displaying the corresponding Chew-Frautschi plots as appropriate.

\begin{itemize}
\item $d=3$

After resolving the leading singularity at $J = 1/2$, we obtain the Chew-Frautschi plot shown in Fig. \ref{fig: d3 Regge}. Notably, the singlet trajectory is now free of singularities. However, the symmetric and antisymmetric trajectories exhibit singularities at $J = -1/2$, suggesting mixing between the leading-twist operators and the shadow of subleading-twist operators, and vice versa. To resolve these singularities, the anomalous dimensions of the subleading-twist trajectories $[\sigma(\phi^i \phi^j)_\rho]_J^{\tau=d}$ will be required. To the best of our knowledge, these are currently unknown.

\begin{figure}[t]
\centering
\begin{subfigure}[t]{0.45\textwidth}
    \centering
    \includegraphics[width=\textwidth]{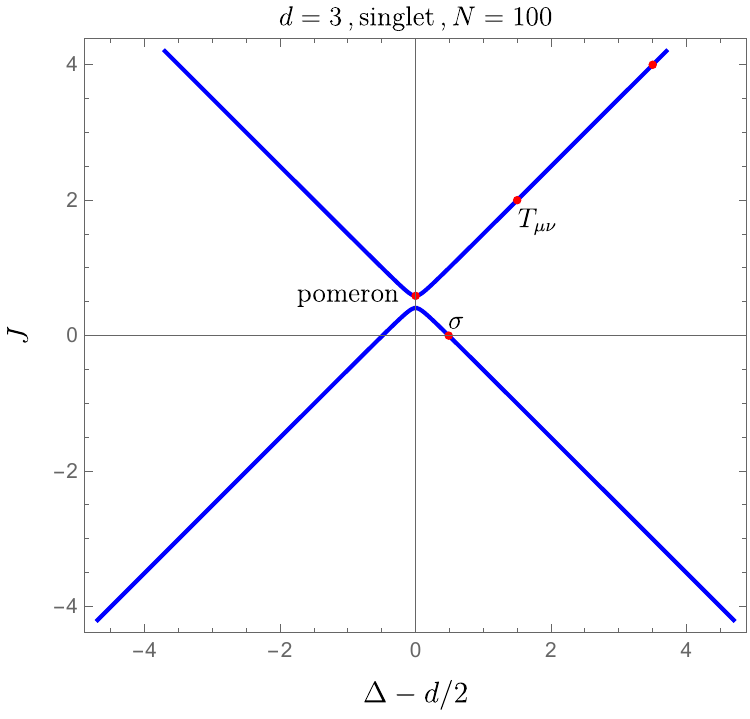} 
    \caption{}
\end{subfigure}
\hfill
\begin{subfigure}[t]{0.45\textwidth}
    \centering
    \includegraphics[width=\textwidth]{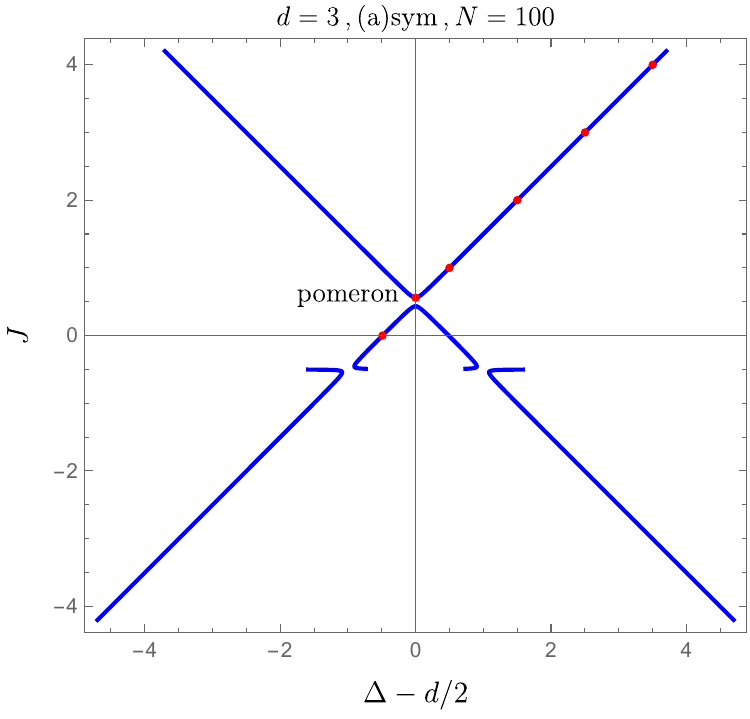} 
    \caption{}
\end{subfigure}
\caption{The Chew-Frautschi plot of leading-twist Regge trajectories with $1/N$ corrections in the critical O$(N)$ model in $d=3$ is shown. Singularities at $J = -1/2$ remain unresolved; their resolution requires information from subleading-twist operators and possibly more. 
Because we have not resolved the mixing of trajectories near the poles in $\gamma_{\rho,J}$, locations in this plot near poles in $J$ are not to be trusted. The fact that poles lead to offshoots in the horizontal direction is an artifact of our choice to use $J$ as a coordinate to make these plots. A proper accounting for the fate of these offshoots requires understanding the other trajectories that they mix with. Similar comments apply to the other plots in this section: nearly horizontal parts of the plot are artifacts.
}
\label{fig: d3 Regge}
\end{figure}

\item $3<d<4$

In generic $3<d<4$, the Chew-Frautschi plot displays even richer structure, as shown in Fig \ref{fig: d7/2 Regge} (with $d=7/2$ as an example).  The first subleading singularity occurs at $J=\frac{2-d}{2}$, corresponding to mixing between the leading twist trajectory $[(\phi^i \phi^j)_\rho]_J$ and the shadow of the subleading-twist trajectory $[\sigma (\phi^i \phi^j)_\rho]_J$.
Furthermore, the singlet anomalous dimension $\g_{s,J}$ exibits additional singularities at all negative integer spins $J=-1,-2,\dots$. These could potentially arise from mixing into the horizontal trajectories. By contrast, the symmetric and antisymmetric representations don't exhibit such singularities at leading order in $1/N$. This is similar to perturbative QCD, where the BFKL trajectory plays a role at leading order only for the color singlet \cite{Forshaw:1997dc}.

\begin{figure}[t]
\centering
\begin{subfigure}[t]{0.45\textwidth}
    \centering
    \includegraphics[width=\textwidth]{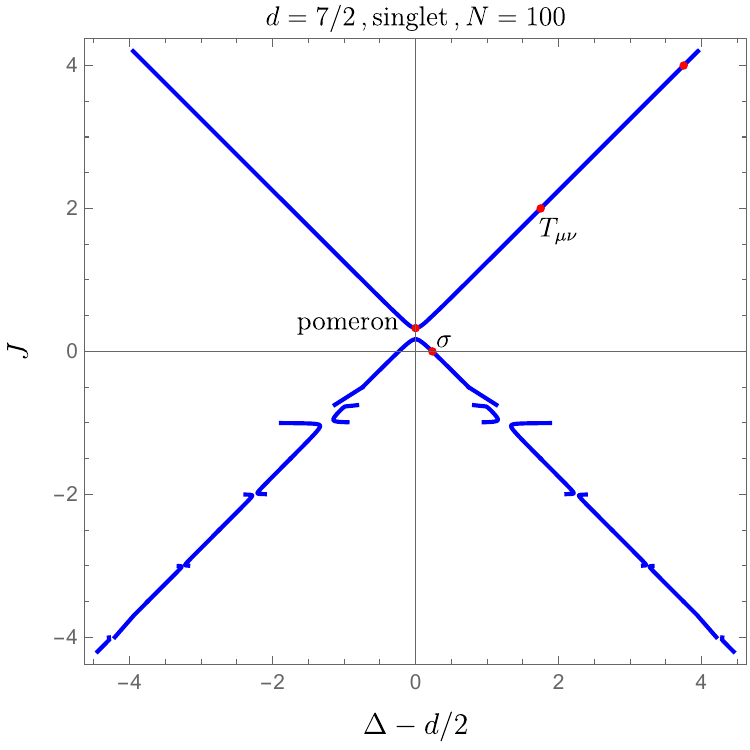} 
    \caption{}
\end{subfigure}
\hfill
\begin{subfigure}[t]{0.45\textwidth}
    \centering
    \includegraphics[width=\textwidth]{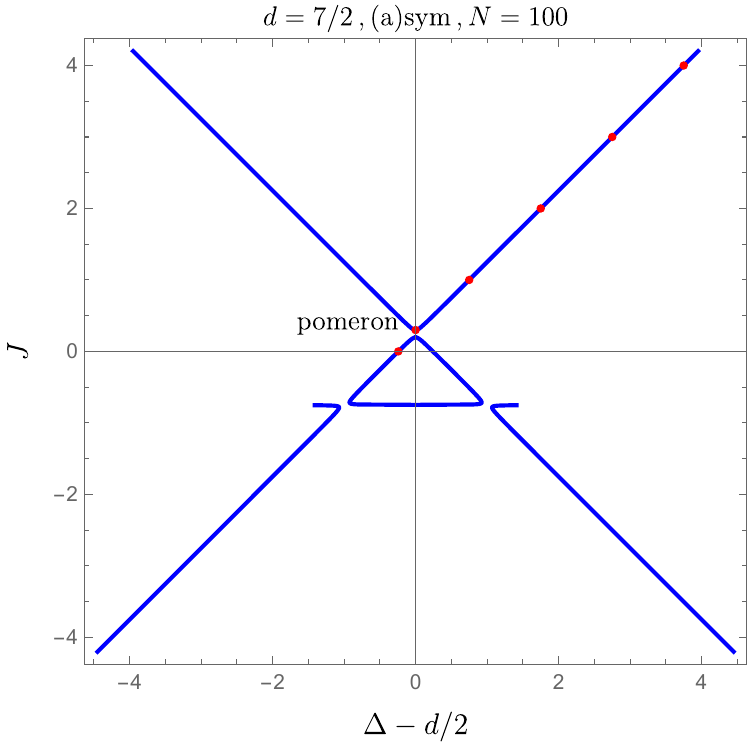} 
    \caption{}
\end{subfigure}
\caption{The Chew-Frautschi plot of leading-twist Regge trajectories with $1/N$ corrections in the critical O$(N)$ model in $d=7/2$. Singularities at $J = 1/2(2-d)$ and negative integers $J=-1,-2,-3,\cdots$ remain unresolved; they indicate the mixing into the subleading Regge trajectories, horizontal Regge trajectories and probably more. For this reason, we emphasize that the trends below $J = 0$ should not be trusted, except to the extent that they indicate singularities and mixings. 
}
\label{fig: d7/2 Regge}
\end{figure}

\item $d=5$

In $d=5$, after resolving the leading singularity at $J=-1/2$, there are two remaining singularities, located at $J=-1$ and $J=-3/2$. The singularity at $J=-1$ only appears in the singlet sector. See Fig \ref{fig: d5 Regge} below.

\begin{figure}[t]
\centering
\begin{subfigure}[t]{0.47\textwidth}
    \centering
\includegraphics[width=\textwidth]{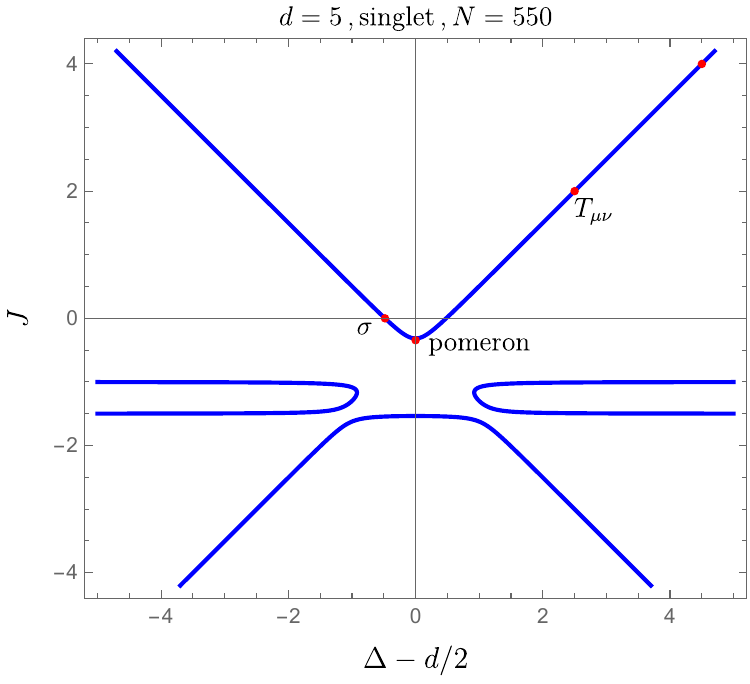} 
    \caption{}
\end{subfigure}
\hfill
\begin{subfigure}[t]{0.45\textwidth}
    \centering
    \includegraphics[width=\textwidth]{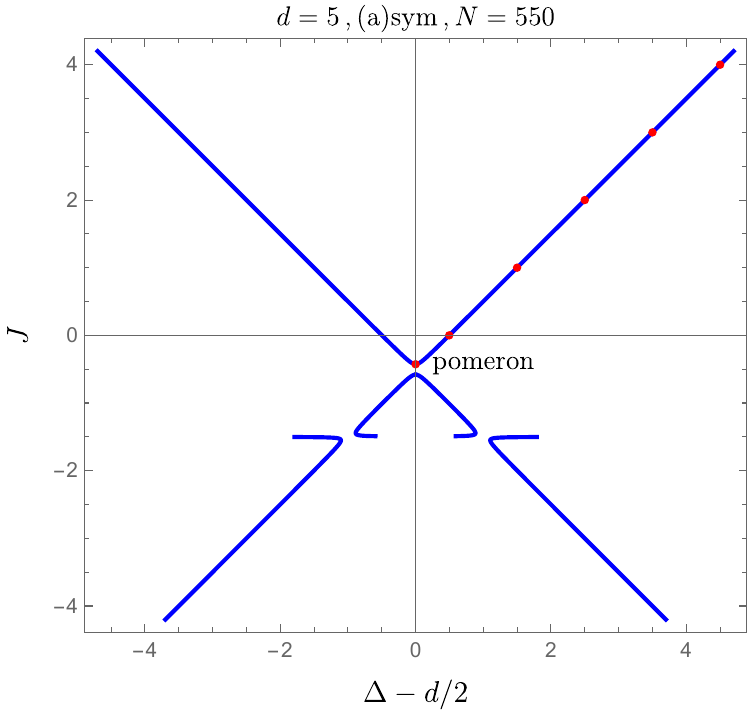} 
    \caption{}
\end{subfigure}
\caption{The Chew-Frautschi plot of leading-twist Regge trajectories with $1/N$ corrections in the critical O$(N)$ model in $d=5$. Singularities at $J = -3/2$ and $J=-1$ are not resolved, as there are mixings into the unknown subleading Regge trajectories and horizontal Regge trajectories, and probably more.}
\label{fig: d5 Regge}
\end{figure}

\item $5<d<6$

The case $5<d<6$ is similar to $3<d<4$, except that mixing into the $J=-1$ horizontal trajectory occurs at higher spin than mixing into the subleading trajectory.

\end{itemize}

We emphasize that we are currently unable to resolve most of the subleading singularities, as we have limited knowledge of operator spectra as a function of spin, despite the availability of a large amount of local operator data \cite{Derkachov:1997ch,Vasiliev:1993ux,Lang:1993ct,Derkachov:1997gc,Lang:1992zw,Lang:1990ni,Lang:1990re,Manashov:2017xtt,Alday:2019clp} (see \cite{Henriksson:2022rnm} for a more comprehensive collection of data). Moreover, constructing a more complete Chew-Frautschi plot also requires knowledge of BFKL-type and even subleading horizontal trajectories, which, to the best of our knowledge, are poorly understood in the critical O$(N)$ model. In later sections, we will take the first steps toward understanding the $1/N$ corrections to the BFKL-type trajectory at $J = -1$. However, we will not resolve the $J = -1$ singularities due to additional mixing with other trajectories beyond our current knowledge, such as $[\sigma\sigma]_{J=-1}$. Determining all this missing data is an important challenge for the future.


\section{Light-ray operators, detectors, and distribution functions}
\label{sec: lightray operator}

In this section, we review some basic ingredients needed to understand light-ray operators in the $O(N)$ model. Along the way, we also establish some connections between CFT terminology and terminology commonly used in perturbative QCD. These connections suggest some potential generalizations of our calculations to other theories like QCD that will be interesting to explore in the future \cite{Chang:2025zib}.
%
For further connections to perturbative QCD, we refer readers to the excellent discussion in \cite{Braun:2003rp,Chen:2023zzh}.

\subsection{Light-ray operators in CFTs}


The simplest kind of light-ray operator in a CFT is the light-transform of a local operator. Consider a local operator $\mathcal{O}$ with dimension $\Delta$ and spin $J$. Its light-transform is defined by
%
\be
\bold{L}[\mathcal{O}](x,z)=\int_{-\infty}^{\infty}(-\alpha)^{-\Delta-J}\mathcal{O}\p{x-\fft{z}{\alpha},z}\,,
\ee
where $z$ is a future pointing null vector $z=(1,\vec{n})$, and $O(x,z)$ is index-free notation for $\cO$ \cite{Costa:2011mg}. More general light-ray operators $\mathbb{O}_{\Delta,J}(x,z)$ come in analytic families that coincide with light transforms of local operators at integer $J$:\footnote{A light-ray operator with integer spin $J>J^\ast$ (where $J^\ast$ is the Regge intercept) must be either the light-transform of a local operator, the shadow of the light-transform of a local operator, or zero \cite{Henriksson:2023cnh}. (The zero case is sometimes called ``missing/magic zeros".)} 
\be
\mathbb{O}_{\Delta,J}(x,z)\propto \bold{L}[\mathcal{O}_{\Delta,J}(x,z)]\,,\quad \text{for integer $J$}.
\ee
The operator $\mathbb{O}_{\Delta,J}(x,z)$ transforms as a primary at $x$ with quantum numbers $(\Delta_L,J_L)=(1-J,1-\Delta)$.

Note that analytic continuation in spin must be performed separately for even and odd spins, as indicated by the Lorentzian inversion formula \cite{Caron-Huot:2017vep} (which we review in appendix~\ref{app: Lorentzian inversion}). The analytic continuation of even(odd)-spin operators have even(odd) signature.
%

In a CFT, light-ray operators at different locations $x$ are essentially equivalent, as we can  move them around using conformal transformations. However, the physical interpretation of different locations is different. More importantly, in non-conformal theories, the actual {\it physics\/} of different locations is different. In this paper, we emphasize and distinguish two configurations: the ``detector frame" where $x$ is placed at spatial infinity, and the ``distribution frame" where $x$ is placed at null infinity. Let us introduce them.
\begin{itemize}
\item[1.] {\bf The detector frame}

In the detector frame, we place $x$ at spatial infinity $\oo=i^0$, thereby defining $\mathbb{O}_{\Delta,J}(\infty,z)$. In the embedding formalism \cite{Costa:2011mg}, this frame can be reached by setting
\be X=(0,1,0)\,,\quad Z=(0,0,z)\,. \ee 
This choice places the light-ray operator along $\mathcal{J}^+$ at a particular angle $\vec n=\vec z/z^0$ on the celestial sphere.  In the case of the light transform of a local operator, this means placing the local operator at null infinity $\mathcal{J}^+$ and integrating from $i^0$ to $i^+$ (see Fig \ref{fig: detector frame}).

We review the symmetries of operators in the detector frame later in section~\ref{sec: detector operator}. Light-ray operators in the detector frame play a role in event shape observables in collider physics. They provide measurements at asymptotic infinity of radiation from local excitations inside Minkowski space. A prominent example is the light transform of the stress tensor $\mathcal{E}(z)=2\mathbb{L}[T](\infty,z)$, referred to as the ANEC operator, which measures energy flux at a specific angle on the celestial sphere.

\begin{figure}
    \centering
    \begin{subfigure}[b]{0.45\textwidth}
        \centering
        \hspace{0mm}\def\svgwidth{80mm}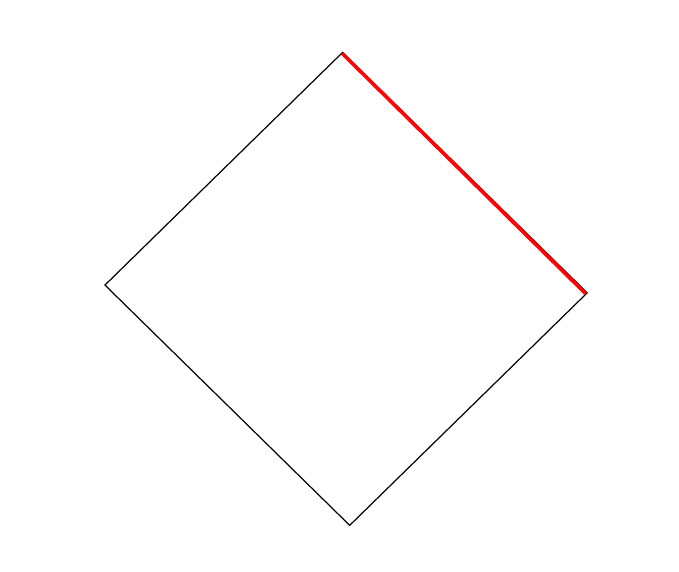
        \caption{}
        \label{fig:sub1}
    \end{subfigure}
    \hfill
    \begin{subfigure}[b]{0.45\textwidth}
        \centering
        \hspace{0mm}\def\svgwidth{80mm}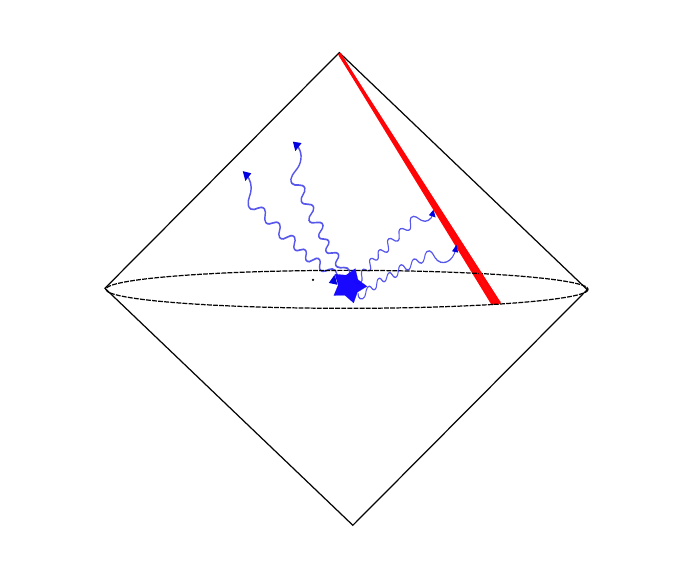
        \caption{}
        \label{fig:sub2}
    \end{subfigure}
    \caption{Light-ray operators in the detector frame, where they measure the event shape observables.
    }
    \label{fig: detector frame}
\end{figure}

\item[2.] {\bf The distribution frame}

In the distribution frame, we place $x$ at past null infinity $\mathcal{J}^-$, thus defining $\mathbb{O}_{\Delta,J}(-\infty z+y^\perp,z)$. The light transform contour then extends from $\mathcal{J}^-$ to $\mathcal{J}^+$. In the embedding space, this frame can be reached by setting, for example,
\be
X=-(0,0,z)\,,\quad Z=(1,x^2,x)\,, \quad z=(0,1,0)\,,\quad x=(0,0,y^\perp)\,. 
\ee
The light-transform of a local operator in this frame takes the form 
 \be 
 \bold{L}[\mathcal{O}](-\infty z+y^\perp,z)=\int dy^- \mathcal{O}_{-\cdots -}(y^+=0,y^-,y^\perp)\,,
 \ee
see Fig.\ref{fig: distribution frame}. This frame is relevant for describing various distribution functions or hadronic wavefunctions (for a particle that moves fast along the $+$ direction) \cite{Geyer:1994zu,Hatta:2008st}. For example, the light transform of the stress-tensor in this frame physically describes the distribution of energy over the transverse plane. As we will explicitly show later, matrix elements of light-ray operators in this frame are related to parton distribution functions (PDFs).  Thus, we refer this frame as ``distribution frame''. 

\begin{figure}
    \centering
    \begin{subfigure}[b]{0.45\textwidth}
        \centering
        \hspace{0mm}\def\svgwidth{80mm}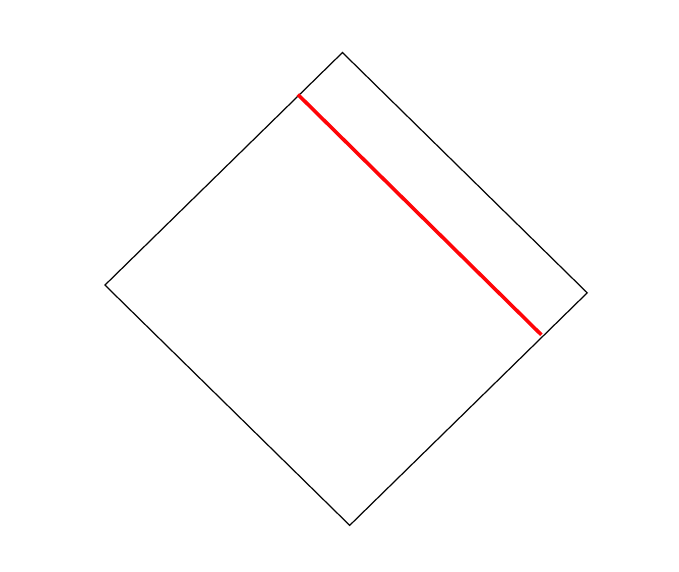
        \caption{}
        \label{fig:sub1}
    \end{subfigure}
    \hfill
    \begin{subfigure}[b]{0.45\textwidth}
        \centering
        \hspace{0mm}\def\svgwidth{80mm}
\begingroup%
  \makeatletter%
  \providecommand\color[2][]{%
    \errmessage{(Inkscape) Color is used for the text in Inkscape, but the package 'color.sty' is not loaded}%
    \renewcommand\color[2][]{}%
  }%
  \providecommand\transparent[1]{%
    \errmessage{(Inkscape) Transparency is used (non-zero) for the text in Inkscape, but the package 'transparent.sty' is not loaded}%
    \renewcommand\transparent[1]{}%
  }%
  \providecommand\rotatebox[2]{#2}%
  \newcommand*\fsize{\dimexpr\f@size pt\relax}%
  \newcommand*\lineheight[1]{\fontsize{\fsize}{#1\fsize}\selectfont}%
  \ifx\svgwidth\undefined%
    \setlength{\unitlength}{215.2431458bp}%
    \ifx\svgscale\undefined%
      \relax%
    \else%
      \setlength{\unitlength}{\unitlength * \real{\svgscale}}%
    \fi%
  \else%
    \setlength{\unitlength}{\svgwidth}%
  \fi%
  \global\let\svgwidth\undefined%
  \global\let\svgscale\undefined%
  \makeatother%
  \begin{picture}(1,0.7225317)%
    \lineheight{1}%
    \setlength\tabcolsep{0pt}%
    \put(0,0){\includegraphics[width=\unitlength,page=1]{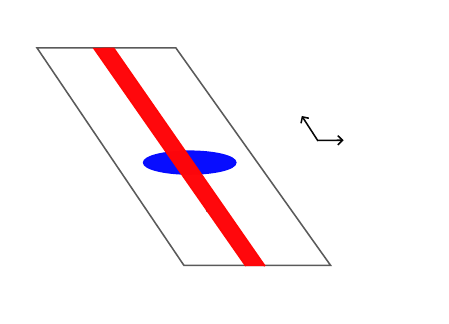}}%
    \put(0.62503216,0.40888043){\color[rgb]{0,0,0}\makebox(0,0)[lt]{\lineheight{1.25}\smash{\begin{tabular}[t]{l}$y^-$\end{tabular}}}}%
    \put(0.6897839,0.35109671){\color[rgb]{0,0,0}\makebox(0,0)[lt]{\lineheight{1.25}\smash{\begin{tabular}[t]{l}$y^\perp$\end{tabular}}}}%
  \end{picture}%
\endgroup%

        \caption{}
        \label{fig:sub2}
    \end{subfigure}
    \caption{Light-ray operators in the distribution frame, where they measure the distribution of initial states.}
    \label{fig: distribution frame}
\end{figure}

\end{itemize}

A third possibility that will not play a major role in this paper is to place the start of a light-ray operator in the middle of Minkowski space. In this case, the light-ray must extend into the next Poincar\'e patch on the Lorentzian cylinder \cite{Kravchuk:2018htv}.\footnote{One can also consider light ray operators that begin at a point in Minkowski space and wrap around into the second sheet in the Schwinger-Keldysh formalism \cite{Caron-Huot:2022lff}.}

We emphasize that in CFTs, the detector and distribution frames are related by a simple conformal transformation \cite{Cornalba:2006xk,Hofman:2008ar,Cornalba:2008qf,Hatta:2008st}
\be
y^+=-\fft{1}{\mu^2 x^+}\,,\quad y^-=x^- - \fft{(x^\perp)^2}{x^+}\,,\quad y^\perp = \fft{x^\perp}{\mu x^+}\,,\label{eq: conformal map}
\ee
where $\mu$ is a reference scale. Via this transformation, the celestial sphere in the detector frame becomes the transverse plane in the middle of Minkowski space in the distribution frame. The two frames therefore encode completely equivalent information in CFTs, which is the main focus of this paper. However, they exhibit different behavior in QFTs \cite{Caron-Huot:2015bja}. In particular, away from a fixed point, the quantum numbers of light-ray operators in the two frames will not be related in a simple way.

Light-ray operators are expected to obey an OPE, schematically expressed as
\be
\mathbb{O}_{\Delta_1,J_1}(x,z_1)\mathbb{O}_{\Delta_2,J_2}(x,z_2)\sim |z_{12}|^{\Delta-\Delta_1-\Delta_2+1}\mathbb{O}_{\Delta,J=J_1+J_2-1}(x,z_2)+\dots\,,
\ee
where ``$\dots$" refers to descendants on the celestial sphere, and higher transverse spin terms \cite{Kologlu:2019mfz}. The spin selection rule $J = J_1 + J_2 - 1$ arises from the fact that the initial points of the light-ray operators are coincident at $x$. (This selection rule is the same as in ``Reggeon cut" analysis in high-energy scattering processes \cite{Gribov:2003nw}.) In the detector frame, the OPE limit corresponds to the collinear limit for event shapes, where the angular separation between detectors goes to zero $\theta_{12} \to 0$. In the distribution frame, the OPE limit corresponds to two operators getting close along the transverse direction $y_{12}^\perp \to 0$. The validity of the light-ray OPE has been systematically explored for light-transforms of local operators in \cite{Kologlu:2019mfz,Chang:2020qpj}, and has applications for energy correlators in collider physics \cite{Chang:2020qpj,Chang:2022ryc,Chen:2020vvp,Chen:2020adz,Lee:2022uwt,Komiske:2022enw}.

The light-ray OPE is expected to make sense only for $J > J^\ast$, where $J^\ast$ is the Regge intercept \cite{Kologlu:2019bco}. For $J < J^\ast$, the limit where the initial points of the light-rays approach each other is singular, and one must perform a careful expansion in that limit. As a consequence, new kinds of light-ray operators can appear that cannot be interpreted in terms of an expansion in the collinear limit alone. These include (approximately) horizontal trajectories, which we will discuss in the next subsection.



\subsection{Detectors, distributions and their renormalization}

In this subsection, we review detector operators and their renormalization, following the discussion in \cite{Caron-Huot:2022eqs}.
%
%
As we mentioned previously, although conformal symmetry relates different frames for light-ray operators, different configurations have distinct physical interpretations. Furthermore, full conformal symmetry is not respected in intermediate steps in perturbation theory, and thus the ingredients involved in defining and renormalizing light-ray operators in different frames will be different.
Nonetheless, at the fixed point, appropriately renormalized detectors should become the light-ray operators of the CFT, regardless of the frame we use.


\subsubsection{Operators in the detector frame}
 

In free theory, it is easy to construct light-ray operators in the detector frame, i.e.\ ``detectors." An important example is a detector that measures the flux of a power of energy $E^{J-1}$ 
\be
\label{eq:ejoperator}
\mathcal{E}_J(z)
\propto \int_0^\infty d\beta \beta^{J+d-4}a^\dagger(\beta z)a(\beta z)\,,
\ee
where $z$ is a null vector and $J\in \C$ is a complex parameter.
For integer $J$, this operator is equivalent (at tree-level) to the light-transform of leading-twist spin-$J$ local operators. Allowing $J$ to be continuous, the operator $\cE_J$ traces out a leading-twist Regge trajectory.

This operator can also be written in terms of fields at null infinity. We start by defining a scalar at null infinity by 
\be
\phi(\alpha,z)=\lim_{L\rightarrow\infty} L^{\Delta_\phi}\phi(x+L z)\,,\quad \alpha=-2x\cdot z\,,
\ee
where $\Delta_\phi=(d-2)/2$ is the tree-level dimension of $\phi$. The leading-twist detectors can be constructed from a double integral of $\phi$'s along null infinity against an appropriate kernel:
\be
\mathcal{D}_{J_L}^{\rm detector}(z)=\fft{1}{C_{J_L}}\int d\alpha_1 d\alpha_2 |\alpha_{12}|^{2(\Delta_\phi-1)+J_L} : \phi(\alpha_1,z)\phi(\alpha_2,z):\,.\label{eq: detector operator leading-twist}
\ee
The normalization is
\be
C_{J_L}=2^{J_L+d-1}\pi \sin\Big(\fft{\pi}{2}(J_L+2\Delta_\phi)\Big)\Gamma(2\Delta_\phi+J_L-1)\,,
\ee
such that for integer spin $J$, with $J_L = 1 - \Delta = 1 - 2\Delta_\phi - J$, this detector reduces to the light transform of local leading-twist operators. For general $J$, $\mathcal{D}_{J_L}^{\rm detector}(z)$ is proportional to $\cE_J(z)$ with the relation $J_L=3-d-J$.


Let us comment on the symmetry properties of this detector and its cousins. Note that $\mathcal{D}_{J_L}^{\rm detector}(z)$ transforms like a primary operator at spatial infinity, which is equivalent to the statement that it is invariant under translations. Indeed, translations act on $\phi$ at null infinity by shifting the retarded time parameter $\alpha$:
\be
[P^\mu,\phi(\alpha,z)]=-2z^\mu\partial_\alpha \phi(\alpha,z).
\ee
Translation-invariance of $\mathcal{D}_{J_L}^{\rm detector}(z)$ is thus ensured in (\ref{eq: detector operator leading-twist}) by the fact that the kernel $|\alpha_{12}|^{2(\Delta_\phi-1)+J_L}$ is translationally invariant in the $\alpha_i$'s. Translation-invariance is also manifest in the expression (\ref{eq:ejoperator}) in terms of creation and annihilation operators.

Because $\mathcal{D}_{J_L}^{\rm detector}(z)$ is a homogeneous function of $z$ with homogeneity $J_L$, it transforms in an irrep of the Lorentz group with spin $J_L$. Finally, its scaling dimension is given by
\be
[D,\mathcal{D}_{J_L}^{\rm detector}(z)]=-\Delta_L \mathcal{D}_{J_L}^{\rm detector}(z)\,.
\ee
In the free theory (or the strict large-$N$ limit of the $O(N)$ model), we have $\Delta_L = J_L + d - 2 = 1 - J$. 


Because perturbation theory respects Poincare symmetry, the condition of primariness $[P^\mu,\cD^\mathrm{detector}_{J_L}(z)]=0$ and the Lorentz spin $J_L$ do not receive corrections at any order in perturbation theory. By contrast, the scaling dimension $\Delta_L$ gets perturbative corrections: $\Delta_L=J_L+d-2+\g_L(J_L)$. In other words, in the detector frame, quantum corrections move points on the Chew-Frautschi plot up or down, with fixed $J_L$.

By contrast, perturbation theory for local operators fixes $J$ and generates corrections to $\Delta$: $\Delta=\Delta_0+\g(J)$. This corresponds to moving points left or right on the Chew-Frautschi plot. In the QCD literature, $\g(J)$ and $\g_L(1-\Delta)$ are referred to as spacelike and timelike anomalous dimensions, respectively. In CFT, these two anomalous dimensions are related by the reciprocity \cite{Basso:2006nk,Caron-Huot:2022eqs} (which was earlier found by perturbative QCD studies \cite{mueller1983multiplicity})\footnote{See, e.g., \cite{Mitov:2006ic,Almasy:2011eq,Moch:2017uml,He:2025hin,Kniehl:2025ttz} for applications of the reciprocity relation in QCD at higher loops.}
\be
\gamma_L(J_L)=\gamma\Big(\Delta-2\Delta_\phi-\gamma_L(J_L)\Big)\,,\label{eq: reciprocity relation}
\ee
which simply expresses the statement that the same light-ray operators can be accessed using either conformal frame.


The operators (\ref{eq: detector operator leading-twist}) generate a 45$^\circ$ trajectory on the Chew-Frautschi plot at tree level. Another class of operators that will be important in this work are horizontal trajectories. They can be constructed, for example, from a product of leading-twist detector operators by integrating against a kernel to ensure that the resulting expression has definite Lorentz spin $J_L$:
\be
\mathcal{H}_{J_{L1},J_{L2},J_L}^{\rm detector}(z)=\int D^{d-2}z_1 D^{d-2}z_2 \langle \varphi_{-J_L}(z)\varphi_{d-2+J_{L1}}(z_1)\varphi_{d-2+J_{L2}}(z_2)\rangle \mathcal{D}_{J_{L1}}^{\rm detector}(z_1)\mathcal{D}_{J_{L2}}^{\rm detector}(z_2)\,,\label{eq: detector H}
\ee
where the Lorentz-invariant measure is
\be
\int D^{d-2}z=\int \fft{2d^dz}{{\rm vol}\, \mathbb{R}^+} \delta(z^2)\theta(z^0)\,.
\ee
The kernel is a three-point function of fictitious primary operators on the $(d-2)$-dimensional celestial sphere, with $z_1,z_2,z$ viewed as embedding coordinates:
\be
 \langle \varphi_{\delta_1}(z)\varphi_{\delta_2}(z_1)\varphi_{\delta_3}(z_2)\rangle= \fft{1}{(-2z\cdot z_1)^{\fft{\delta_1+\delta_2-\delta_3}{2}}(-2z\cdot z_2)^{\fft{\delta_1+\delta_3-\delta_2}{2}}(-2z_1\cdot z_2)^{\fft{\delta_2+\delta_3-\delta_1}{2}}}\,.
\ee

There are infinitely many such horizontal trajectories, labeled by the three parameters $(J_{L1}, J_{L2}, J_L)$. However, when we turn on interactions, it turns out that only a 1-dimensional subfamily $J_{L1}=J_{L2}=3-d$ of these operators are multiplicatively renormalized at leading order. This same phenomenon occurs in the Wilson-Fisher theory \cite{Caron-Huot:2022eqs} and in QCD \cite{Chang:2025zib}. We call the resulting operator in this case a BFKL operator
\be
\cH^\mathrm{detector}_{J_L,\mathrm{BFKL}}(z) \propto \mathcal{H}_{3-d,3-d,J_L}^{\rm detector}(z).
\ee
Multiplicative renormalization leads to nontrivial running
\be
\mu \fft{d}{d\mu}\mathcal{H}_{J_L,{\rm BFKL}}^{\rm detector}=\gamma_{\rm BFKL}(J_L)\mathcal{H}_{J_L,{\rm BFKL}}^{\rm detector}\,.\label{eq: RG BFKL detector}
\ee
corresponding to a $J_L$-dependent vertical shift on the Chew-Frautschi plot.


It is important to emphasize that the divergences of detector operators are IR divergences. Intuitively, this is because event shapes are measured by detectors placed at null infinity, which are generally not IR safe, as soft and collinear radiation can be emitted into regions that are not measured (except for energy detectors, which are IR safe).\footnote{Observables such as \eqref{eq: RG BFKL detector}, which have only soft divergences, can be identified with non-global observables, where the RG resums non-global large logarithms under the so-called Banfi-Marchesini-Smye (BMS) evolution equation \cite{Banfi:2002hw}. BMS can be conformally mapped to the distribution frame and identified with BFKL evolution \cite{Hatta:2008st,Caron-Huot:2015bja} using \eqref{eq: conformal map}, up to $\beta$-function contributions. See \cite{Caron-Huot:2015bja} for an excellent discussion.} The IR divergences in the detector frame are sensitive to both soft and collinear radiation, and these two types of divergences are sometimes entangled and enhanced.


\subsubsection{Operators in the distribution frame}

We now come to the perturbative construction of distribution operators. 
We can construct distribution analogs of $\cD^\mathrm{detector}_{J_L}(z)$ by integrating $\phi$'s against the same kernel, along a light-ray through the middle of Minkowski space:
%
\be
\mathcal{D}_{J}^{\rm dis}(-\infty z,x^\perp)=\fft{1}{C_{3-d-J}} \int d\alpha_1 d\alpha_2 |\alpha_{12}|^{-1-J} :\phi(\alpha_1 z + x^\perp)\phi(\alpha_2 z + x^\perp):\,.
\ee
In perturbation theory, primariness with respect to translations along the $z$ direction is ensured to all orders in perturbation theory by choosing a translation-invariant kernel $|\alpha_{12}|^{-1-J}$. For general light-ray operators, primariness with respect to other conformal generators may need to be ensured by correcting their definitions order-by-order in perturbation theory. This is in contrast to the detector frame, where primariness with respect to the full conformal group is exact in perturbation theory. 


The symmetry properties of this type of detector are more involved. We can study transformation properties with respect to SL$(2,\mathbb{R})$, which controls the collinear behavior, or SO$(d-1,1)$, which controls the transverse behavior. However, these subgroups of the conformal group are not disjoint. We refer the readers to \cite{Braun:2003rp} for a comprehensive discussion.

Importantly, the boost symmetry $z\to \lambda z$ is respected in perturbation theory, and this implies that the boost weight $J-1=-\Delta_L$ does not receive perturbative corrections in this frame. Thus, the renormalization of distribution opertators
%
gives rise to a standard spacelike anomalous dimension
\be
\mu \fft{d}{d\mu} \mathcal{D}_J^{\rm dis}(-\infty z,x^\perp)=-\gamma(J)   \mathcal{D}_J^{\rm dis}(-\infty z,x^\perp)\,.\label{eq: ren D dis}
\ee

In the distribution frame, we can also make a horizontal trajectory from a product of two detectors $\mathcal{D}_{J_1}^{\rm dis}(-\infty z,x_1^\perp)\mathcal{D}_{J_2}^{\rm dis}(-\infty z,x_2^\perp)$. Again, it may be convenient to decompose this detector into irreducible representations with respect to the transverse SO$(d-1,1)$. However, this symmetry is not preserved in perturbation theory, and consequently it is not straightforward to use $\mathrm{SO}(d-1,1)$ symmetry to derive selection rules for how operators can mix in the presence of interactions. This is in contrast to the detector frame, where $\mathrm{SO}(d-1,1)$ is the Lorentz group, and it provides simple, exact selection rules on mixing.

Concretely, to project onto $\mathrm{SO}(d-1,1)$ irreps, we can integrate against a conformal 3-point function in the transverse plane:
\be
&\mathcal{H}_{J_L,J_1,J_2}^{\rm dis}(-\infty z,x^\perp)\nn\\
&=\int d^{d-2}x_1^\perp d^{d-2}x_2^\perp \langle \varphi_{-J_L}(x^\perp)\varphi_{1-J_1}(x_1^\perp)\varphi_{1-J_2}(x_2^\perp)\rangle \mathcal{D}_{J_1}^{\rm dis}(-\infty z,x_1^\perp)\mathcal{D}_{J_2}^{\rm dis}(-\infty z,x_2^\perp)\,.\label{eq: horizontal distribution}
\ee
As we will see, the BFKL-type trajectory corresponds to $J_1=J_2=0$, so we may write it more simply in terms of light-transforms of $\phi^2$:
\be
& \mathcal{H}_{J_L,{\rm BFKL}}^{\rm dis}(-\infty z,x^\perp)\nn\\
&\sim \int d^{d-2}x_1^\perp d^{d-2}x_2^\perp \langle \varphi_{-J_L}(x^\perp)\varphi_{1}(x_1^\perp)\varphi_{1}(x_2^\perp)\rangle \bold{L}[\phi^2](-\infty z,x_1^\perp)\bold{L}[\phi^2](-\infty z,x_2^\perp)\,,\label{eq: BFKL dis}
\ee
(Note that in the critical O$(N)$ model, we should replace $\phi^2$ by $\sigma$.)

As we will see, the renormalization of $\mathcal{H}_{J_L,{\rm BFKL}}^{\rm dis}$ is subtle. If we compute perturbative corrections using the definition \eqref{eq: BFKL dis}, we find divergences that cannot be regulated by dim-reg, arising from the presence exactly null lines. To address this, we can slightly tilt $z$ away from being null. This adjustment introduces a ``rapidity scale" $\nu$ that controls the evolution of the light-ray operator under boosts. 
Running of the rapidity scale is known as rapidity evolution \cite{Chiu:2011qc,Chiu:2012ir} (see also appendix \ref{app: rapidity}):
\be
\nu \fft{d}{d\nu}\mathcal{H}_{J_L,{\rm BFKL}}^{\rm dis}=-\gamma_{\rm BFKL}(J_L)\mathcal{H}_{J_L,{\rm BFKL}}^{\rm dis}\,.
\label{eq:rapidityevolution}
\ee
The right-hand side is a correction to the boost eigenvalue of the light-ray operator.
Conformal symmetry then predicts that the same function $\gamma_{\rm BFKL}(J_L)$ appears in (\ref{eq:rapidityevolution}) as in (\ref{eq: RG BFKL detector}), since a boost in the distribution frame is the same as a dilatation in the detector frame.



\subsection{Links to parton distribution functions and collinear functions}

In this section, we briefly describe some relationships between light-ray operators in the distribution frame and objects appearing in QCD factorization theorems. For simplicity, we discuss a toy model of a single scalar field.

Although we focus only on the distribution frame in this section, detector operators also play an interesting role in factorization theorems relevant to phenomenological observables. For example, see \cite{Chen:2023zzh} for a relation between factorization theorems for the EEC \cite{Dixon:2019uzg} and the light-ray OPE \cite{Hofman:2008ar,Kravchuk:2018htv}.


\subsubsection{Relating leading-twist distribution operators to PDFs}

Consider deep inelastic scattering (DIS) between a hadronic state $|p\rangle$ and a light particle with finite Bjorken variable $x$. A factorization theorem relates the two-point function of an operator in the hadronic state to the integral of a hard function $H(q,\xi p)$ (which encodes perturbative partonic scattering) against a PDF $f(\xi)$ (which measures the probability of finding a parton with a certain momentum fraction inside the hadron):
%
\be
F(q,x) = \langle p|J(q)J(0) |p\rangle \sim \int_x^1 d\xi H(q,\xi p)f(\xi).
\ee
The Bjorken variable is defined by $x=q^2/(2p\cdot q)$. 
The operator definition of a PDF in our toy scalar model is
\be
f(\xi)= \fft{\xi p^+}{4\pi}\int dx^- \langle p| \phi(x^+=0,x^-,x^\perp=0)\phi(0)|p\rangle e^{-i \fft{1}{2} \xi p^+ x^-}\,,
\ee
which can be derived, for example, using soft-collinear effective theory (SCET) \cite{Bauer:2000ew,Bauer:2000yr,Bauer:2001ct,Bauer:2001yt,Becher:2014oda}.\footnote{This formula for the scalar is rather straightforward by a multipole expansion of $\phi(x)\phi(0)$ with collinear external states $|p\rangle$: any multipole derivatives along $x^+$ and $x^\perp$ are small by kinematics, while the derivatives along $x^-$ remain order 1 and are thus exponentiated.} The renormalization of the PDF is controlled by the DGLAP evolution equation
\be
\mu \fft{d}{d \mu} f(\xi)= \int_{\xi}^1 \fft{d\xi'}{\xi'} P\Big(\fft{\xi}{\xi'}\Big) f(\xi')\,,\label{eq: DGLAP evol}
\ee
where $P\big(\xi/\xi'\big)$ is the spacelike DGLAP splitting function, and it is well known that its moments give the anomalous dimensions of leading-twist operators
\be
\gamma(J)=-\int_0^1 dx x^{J-1} P(x)\,.\label{eq: anomalous to P}
\ee

Now it is clear how to relate this to leading-twist distribution operators. Consider the forward matrix element of $\mathcal{D}_J^{\rm dis}$ (divided by an infinite volume factor due to translation invariance): 
\be
\fft{\langle p| \mathcal{D}_J^{\rm dis}(-\infty z,x^\perp)|p\rangle}{{\rm Vol}\,\mathbb{R}}  
&=\fft{1}{C_{3-d-J}} \int d\alpha |\alpha|^{-1-J}\langle p| \phi(\alpha z,0)\phi(0)|p\rangle 
\nn\\
&= \fft{(p^+)^{J}}{2\pi} \int_0^1 d\xi f(\xi)\xi^{J-1}.
\ee
In the last line, we rewrote the integral over $\alpha$ in Fourier space, obtaining a convolution of the momentum-space matrix element $f(\xi)/\xi$ with the Fourier-transformed kernel $\xi^{J}$.
We also set $z=(0,1,0)$ so that $-2p\cdot z= p^+$. Performing an inverse Mellin transform, we can also write
\be
f\Big(\fft{k\cdot z}{p\cdot z}\Big)=2\pi \oint \fft{dJ}{2\pi i} \fft{\langle p|\mathcal{D}_J^{\rm dis}(-\infty z,0)|p\rangle}{{\rm Vol}\,\mathbb{R}} (-2k\cdot z)^{-J}\,,\label{eq: from D to f}
\ee
We can summarize this result as:

\begin{itemize}
\item[] \textbf{Forward matrix elements of DGLAP-type distribution operators are moments of PDFs.}
\end{itemize}

We have established this relation at the level of bare operators. In the persence of interactions, both the distribution detector and the PDF are renormalized in a compatible way. Indeed, taking the moment of \eqref{eq: DGLAP evol} and using \eqref{eq: ren D dis}, we derive \eqref{eq: anomalous to P}. This also justifies our choice to call $\mathcal{D}_{J}(-\infty z,x^\perp)$ ``distribution operators."

As we discuss more in section \ref{sec: RG distribution operator}, it is challenging to think about mixing between Regge trajectories in the distribution frame. For example, mixing between the leading trajectory and its shadow is complicated by the fact that conformal symmetry is not preserved order-by-order in perturbation theory, so the shadow transform in the transverse plane must be perturbatively corrected.

\subsubsection{BFKL-type distribution operators and collinear functions}

BFKL physics is important for understanding the small $x$ limit of DIS and Regge scattering in QCD. Consider DIS again, but now with small Bjorken variable (though not too small, in order to avoid the saturation regime \cite{Iancu:2003xm}). Using SCET, \cite{Neill:2023jcd} derived a factorization theorem for this process in terms of a collinear function $C_z$ and a soft function $S$\footnote{Note that collinear, in our convention, means along $p^+$, in contrast to \cite{Neill:2023jcd}, who took $p^-$.}
\be
F(q,x)\sim \int d^{d-2}k^\perp  C_z(k^\perp,p^+) S(q^\perp,k^\perp)\,.
\ee
Similarly, for Regge and forward scattering, \cite{Rothstein:2016bsq,Gao:2024qsg} proved a factorization theorem
\be
\mathcal{M}_{\rm Regge}\sim \int d^{d-2}k_1 d^{d-2}k_2 C_z(k_1^\perp,p_1^+)C_{\bar{z}}(k_2^\perp,p_2^-) S_G(k_1^\perp,k_2^\perp)\,,
\ee
with a different soft function $S_G$.
The soft functions $S,S_G$ will not be our focus in this section.

Both factorization theorems involve the same collinear function, which is defined as
\be
C_z(k^\perp,p^+)\sim\fft{1}{(k^\perp)^{d-2}}\langle p| \Big(\int dx'^- \mathcal{O}(x'^+=0,x'^-,k^\perp) \Big)\Big(\int dx^- \mathcal{O}(x^+=0,x^-,0)\Big)|p\rangle\,.\label{eq: collinear}
\ee
We have not kept track of the overall normalization, as it can be absorbed into the soft functions. The collinear function obeys a rapidity RG equation
\be
\nu\fft{d}{d\nu}C_z(q^\perp,p^+)=\int d^{d-2}k^\perp K_{\rm BFKL}(q^\perp,k^\perp)C_z(k^\perp,p^+)\,,
\ee
where $K_{\rm BFKL}$ is the forward BFKL evolution kernel.

With these definitions, it is clear how to relate the collinear function to the horizontal distribution operator in \eqref{eq: BFKL dis}, as we essentially have
\be
C_z(k^\perp,p^+)\sim\fft{1}{(k^\perp)^{d-2}} \int \fft{d^{d-2}x^\perp}{(2\pi)^{d-2}}\langle p| \bold{L}[\mathcal{O}](-\infty z,x^\perp)\bold{L}[\mathcal{O}](-\infty z,0)|p\rangle e^{ik^\perp\cdot x^\perp}\,.
\ee
It is just a product of two light-transforms, written in momentum space in the transverse plane. By contrast, the BFKL-type operator (\ref{eq: BFKL dis}) is obtained by integrating against a transverse three-point function that projects onto an irreducible representation with respect to the transverse conformal group. However, these choices of kernel (Fouerier space or transverse three-point function) are related in a simple way. Let us take $x^\perp=\oo$ in (\ref{eq: BFKL dis}). The three-point function becomes simply a power $|x_{12}^\perp|^{-2-J_L}$.
With this choice, $\mathcal{H}^\mathrm{dis}_{J_L,\mathrm{BFKL}}(-\oo z,x^\perp=\oo)$ becomes a Mellin-transform of $C_z(k^\perp,p^+)$ with respect to $k^\perp$.
We will see later that the transverse three-point function, acting as a Clebsch-Gordan kernel, diagonalizes the BFKL evolution kernel to yield the BFKL anomalous spin. To summarize,
\begin{itemize}
\item[] \textbf{Forward matrix elements of BFKL-type distribution operators are moments of collinear functions.}
\end{itemize}



\section{Renormalization of detector operators}
\label{sec: detector operator}


In this section, we follow the techniques of \cite{Caron-Huot:2022eqs} to renormalize detector operators in the $O(N)$ model and compute their anomalous dimensions $\gamma_L(J_L)$, for both the leading-twist trajectory (and its shadow) as well as a horizontal trajectory.

\subsection{Leading-twist detectors}

Let us first consider a leading-twist detector in the critical O$(N)$ model. We can generalize \eqref{eq: detector operator leading-twist} to multiple scalars as follows:
\be
\mathcal{D}_{J_L}^{{\rm detector},k,ij}(z)\propto \int d\alpha_1 d\alpha_2 |\alpha_{12}|^{2(\Delta_\phi-1)+J_L} (\mathrm{sign}\,\alpha_{12})^k:\phi^i(\alpha_1,z)\phi^j(\alpha_2,z):\,.
\ee
Detectors $\cD_{J_L,\rho}^\mathrm{detector}$ in specific $O(N)$ represnentations $\rho$ are built from this object.
The singlet and symmetric tensor are given by choosing $k=0$ (so that the kernel is even in $\alpha_{12}$ and the operator has even signature) and projecting onto the appropriate $O(N)$ representation. To describe the antisymmetric tensor, we must choose $k=1$ so that the kernel is odd in $\alpha_{12}$ and the operator has odd signature.

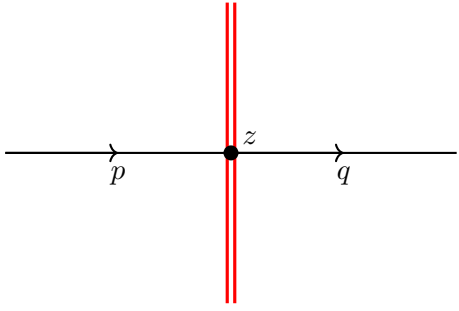
\begin{figure}
    \centering
    \begin{tikzpicture}
    \draw[thick] (-3,0) -- (3,0);
    \draw[red, very thick] (-0.05,2) -- (-0.05,-2);
    \draw[red, very thick] (0.05,2) -- (0.05,-2);
    \draw[->, thick] (-3,0) -- (-1.5,0);
    \draw[->, thick] (0,0) -- (1.5,0);
    \node at (-1.5,-0.3) {$p$};
    \node at (1.5,-0.3) {$q$};
    \fill (0,0) circle (0.1);
    \node at (0.25,0.2) {$z$};
\end{tikzpicture}
\caption{Tree level diagram of the leading-twist detector operators.}
\label{fig: tree detector operator}
\end{figure}

A matrix element
\be
\langle \phi^k(p)|\mathcal{D}_{J_L,\rho}^{{\rm detector}}|\phi^l(q)\rangle\,,
\ee
is computed using the ``in-in"/Schwinger-Keldysh prescription, illustrated in figure~\ref{fig: tree detector operator}.  We represent the cut separating the bra and the ket by a vertical red double-line, with the ket to the left and bra to the right.
Propagators on the right and the left have opposite $i 0$ prescriptions, and propagators that cross the cut are Wightman functions. It is easy to work out the tree-level matrix elements \cite{Caron-Huot:2022eqs}
\be
\label{eq:treeleveldetectormatrixelement}
\langle \phi^k(q) \mathcal{D}_{J_L,\rho}^{\rm detector}(z) \phi^l(p)\rangle
&= (2\pi)^d \delta^d(p-q)V_{J_L}(z,p)T_\rho^{kl},
\ee
where $T_\rho$ is a projector onto the representation $\rho$, explicitly given by
\be
T_s^{kl}&= \delta^{kl}\,, \nn\\
T_{\mathrm{sym},ij}^{kl} &= \frac{1}{2}(\de_i^k \de_j^l + \de_j^k \de_i^l) - \frac 1 {N} \de_{ij} \de^{kl}\,, \nn\\
T_{\mathrm{asym},ij}^{kl} &= \de_i^k \de_j^l - \de_j^k \de_i^l\,, 
\ee
and the vertex $V_{J_L}$ is defined by
\be
V_{J_L}(z,p)&\equiv\int_0^\infty d\beta \beta^{-J_L-1}\delta^d(p-\beta z)\,.
\ee
We have chosen the normalization of $\mathcal{D}_{J_L,\rho}^{\rm detector}(z)$ to make its tree-level matrix elements simple.
We can use the tree-level matrix element (\ref{eq:treeleveldetectormatrixelement}) as a building block in perturbation theory, providing the Feynman rule for the insertion of a bare detector.

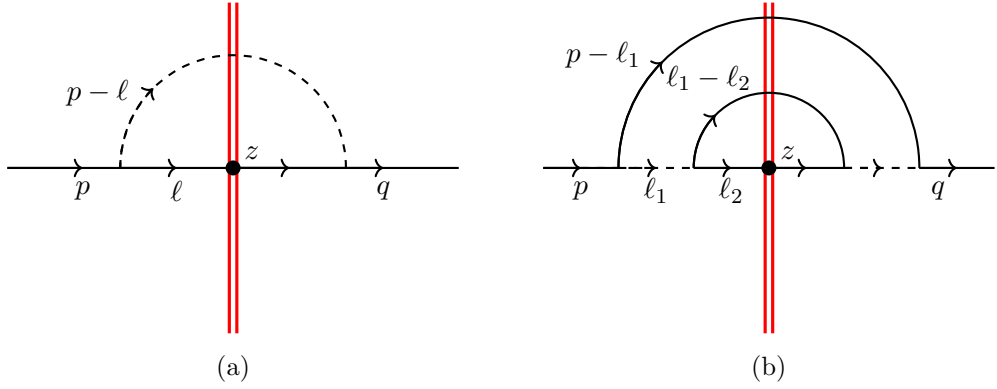
\begin{figure}
\centering
\begin{subfigure}[b]{0.45\textwidth}
\centering
\begin{tikzpicture}
    \draw[thick] (-3,0) -- (3,0);
    \draw[red, very thick] (-0.05,2.2) -- (-0.05,-2.2);
    \draw[red, very thick] (0.05,2.2) -- (0.05,-2.2);
    \draw[dashed, thick] (1.5,0) arc (0:180:1.5);
    \draw[->, thick] (-3,0) -- (-2,0);
     \draw[->, thick] (0,0) -- (2,0);
    \draw[->, thick] (-1.5,0) -- (-0.75,0);
     \draw[->, thick] (0,0) -- (0.75,0);
    \draw[dashed, thick,->] (-1.5,0) arc (0:-45:-1.5);
    \node at (-2,-0.3) {$p$};
     \node at (-0.75,-0.3) {$\ell$};
     \node at (-1.8,1) {$p-\ell$};
     \fill (0,0) circle (0.1);
     \node at (0.25,0.2) {$z$};
      \node at (2,-0.3) {$q$};
    \end{tikzpicture}
\caption{}
\label{fig:oneloopcorrectionsubfigure}
\end{subfigure}
\begin{subfigure}[b]{0.45\textwidth}
\centering
\begin{tikzpicture}
    \draw[thick] (-3,0) -- (-2,0);
     \draw[dashed,thick] (-2,0) -- (-1,0);
     \draw[thick] (-1,0) -- (1,0);
     \draw[dashed,thick] (1,0) -- (2,0);
     \draw[thick] (2,0) -- (3,0);
    \draw[red, very thick] (-0.05,2.2) -- (-0.05,-2.2);
    \draw[red, very thick] (0.05,2.2) -- (0.05,-2.2);
    \draw[thick] (1,0) arc (0:180:1);
    \draw[thick] (2,0) arc (0:180:2);
    \draw[->, thick] (-3,0) -- (-2.5,0);
     \draw[->, thick,dashed] (-2.5,0) -- (-1.5,0);
      \draw[->, thick] (-1,0) -- (-0.5,0);
      \draw[->, thick] (0,0) -- (0.5,0);
      \draw[->, thick,dashed] (1,0) -- (1.5,0);
       \draw[->, thick] (2,0) -- (2.5,0);
       \draw[thick,->] (-1,0) arc (0:-45:-1);
       \draw[thick,->] (-2,0) arc (0:-45:-2);
       \node at (-2.5,-0.3) {$p$};
        \node at (-1.5,-0.3) {$\ell_1$};
         \node at (-0.5,-0.3) {$\ell_2$};
         \node at (-0.8,1.2) {$\ell_1-\ell_2$};
         \node at (-2.2,1.5) {$p-\ell_1$};
         \node at (2.25,-0.3) {$q$};
          \fill (0,0) circle (0.1);
     \node at (0.25,0.2) {$z$};
    \end{tikzpicture}
\caption{}
\label{fig:mysubfigure}
\end{subfigure}
\caption{$1/N$ corrections to the leading-twist detector operators. (a) contributes to singlet, symmetric and anti-symmetric sectors, while (b) only contributes to the singlet.}
\label{fig: detector operators leading}
\end{figure}

Importantly, $O(1/N)$ corrections to detector matrix elements arise from both one- and two-loop diagrams. The two-loop diagrams contain four $\sigma\phi^i\phi^i$ vertices, giving a factor of $1/N^2$. However, one of these two-loop diagrams contains a closed scalar loop,  contributing an additional factor of $N$, resulting in $1/N^2 \times N=1/N$. We find two irreducible diagrams contributing to leading-twist detector matrix elements at $O(1/N)$, see figure~\ref{fig: detector operators leading}.
The two-loop diagram (figure \ref{fig:mysubfigure}) only contributes to the singlet detector matrix element because it contracts $O(N)$ indices between fields in the detector. This is the origin of the additional term in $\gamma_{s,J}$ compared to $\gamma_{{\rm sym},J}=\gamma_{{\rm asym},J}$.


Let us begin with the symmetric and antisymmetric representations, which only get contributions from the diagram in figure~\ref{fig:oneloopcorrectionsubfigure}. Stripping off global symmetry indices for brevity, this diagram evaluates to
\be
\mathcal{F}_{J_L}^1(z,p)=\fft{1}{N} \int \fft{d^d \ell }{(2\pi)^{d}} \fft{1}{p^2+i 0}\fft{1}{p^2-i 0}V_{J_L}(z,\ell) \times \big(-2 \sin (\pi \delta) G_\sigma(-(p-\ell)^2) \theta(p-\ell) \big)\,,
\ee
where we used the regulatization $\delta=1/2(d-2)+\epsilon=(d+2-2\Delta_\sigma)/2$. The last term is the Wightman propagator for $\sigma$, and for convenience we have defined $\theta(p)\equiv \theta(p^0) \theta(-p^2)$, which is supported on future-pointing timelike vectors. This $\theta$-function constrains the integration range in the vertex $V_{J_L}(z,\ell)$ to be $\beta\in (0,p^2/(2p\cdot z)$. Computing the integral, we find 
\be
\mathcal{F}_{J_L}^1(z,p)&=-\fft{\pi ^{-\frac{d}{2}+\epsilon -\frac{1}{2}} \sin \left(\frac{1}{2} \pi  (d+2 \epsilon -2)\right) \Gamma \left(\frac{d}{2}+\epsilon -\frac{1}{2}\right) \Gamma \left(1-J_L\right) 2^{d+4 \epsilon -1}}{N J_L \Gamma \left(\frac{d}{2}+\epsilon -2\right) \Gamma \left(-\frac{d}{2}-\epsilon -J_L+3\right)}\nn\\
&\quad \x  \left(-p^2\right)^{-\frac{d}{2}-J_L-\epsilon } (-2p\cdot z)^{J_L} \theta(-p^2)\,.\label{eq: F1}
\ee

This expression naively appears to be finite, but as emphasized in \cite{Caron-Huot:2022eqs}, it is actually divergent as a distribution. (We can confirm this and make the divergences manifest by Fourier transforming to position space.) The $1/\epsilon$ pole is given by
\be
\label{eq:distributionpole}
\left(-p^2\right)^{-\frac{d}{2}-J_L-\epsilon } (-2p\cdot z)^{J_L}\theta(-p^2)=-{\rm vol}(S^{d-3})\frac{\Gamma \left(\frac{d}{2}+\epsilon -1\right) \Gamma \left(-\frac{d}{2}-\epsilon -J_L+1\right)}{4 \epsilon  \Gamma \left(-J_L\right)} V_{J_L}(z,p)+\dots\,.
\ee
This gives a $1/\epsilon$ divergence in the diagram proportional to the tree-level vertex function:
\be
\mathcal{F}_{J_L}^1(z,p)\Big|_{\epsilon\,{\rm div}}=-\fft{1}{N}\frac{\pi ^{\epsilon -\frac{3}{2}} 2^{d+4 \epsilon -1} (d+2 \epsilon -4) \mu ^{2 \epsilon } \sin \left(\frac{1}{2} \pi  (d+2 \epsilon -2)\right) \Gamma \left(\frac{d}{2}+\epsilon -\frac{1}{2}\right)}{\epsilon  \Gamma \left(\frac{d}{2}-1\right) \left(d+2 \left(J_L+\epsilon -2\right)\right) \left(d+2 \left(J_L+\epsilon -1\right)\right)} V_{J_L}(z,p)\,.
\ee
We must renormalize the detectors to cancel this IR divergence. Including a wavefunction renormalization factor for the external states (see appendix \ref{app: wavefunction RG}), we have
\be
\mathcal{D}_{J_L,{\rm (a)sym},R}^{{\rm detector}}=Z_{J_L,{\rm (a)(sym)}}^{-1}\mathcal{D}_{J_L,{\rm (a)sym}}^{{\rm detector}}\,,\quad Z_{J_L,{\rm a(sym)}}=
Z_\phi^{-1}\Big(1+\fft{\mathcal{F}_{J_L}^1(z,p)\Big|_{\epsilon\,{\rm div}}}{V_{J_L}(z,p)}\Big)\,,
\ee
which gives the anomalous dimension
\be
\gamma_{{\rm (a)sym},L}(J_L)&=-\mu \fft{d}{d\mu}\log Z_{J_L,{\rm (a)sym}}=\fft{1}{N}\frac{2^{d+3} \sin \left(\frac{\pi  d}{2}\right) \Gamma \left(\frac{d-1}{2}\right) \left(J_L-1\right) \left(d+J_L-2\right)}{\pi ^{3/2} (d-2) d  \Gamma \left(\frac{d}{2}-2\right) \left(d+2 J_L-4\right) \left(d+2 J_L-2\right)}\,.\label{eq: gammaL sym}
\ee
Using the reciprocity relation \eqref{eq: reciprocity relation}, we reproduce $\gamma_{{\rm (a)sym},J}$ from \eqref{eq: leading twist ano}.

For the singlet detector, we must include the two-loop diagram in figure \ref{fig:mysubfigure}:
\be
 \mathcal{F}^2(z,p)&=\fft{1}{N} \int \fft{d^d\ell_1 d^d\ell_2}{(4\pi^2)^{d-1}} \fft{1}{p^2-i0}\fft{1}{p^2+i0}G_\sigma(\ell_1^2-i 0)G_{\sigma}(\ell_1^2+i 0) 
 \nn\\
 &\quad 
 \qquad\qquad
\x \delta^+((p-\ell_1)^2)\delta^+((\ell_1-\ell_2)^2) V_{J_L}(z,\ell_2)\nn\\
 & =\ft{2^{2 d+8 \epsilon -3} \pi ^{2 \epsilon -\frac{d}{2}} \Gamma \left(\frac{d}{2}-1\right) \sin ^2\left(\frac{1}{2} \pi  (d+2 \epsilon -2)\right) \csc \left(\frac{1}{2} \pi  (d+4 \epsilon )\right) \Gamma \left(\frac{d}{2}+\epsilon -\frac{1}{2}\right)^2 \csc \left(\pi  \left(d+J_L+2 \epsilon -4\right)\right)}{N\Gamma (2-2 \epsilon ) \Gamma \left(\frac{d}{2}+2 \epsilon -2\right) \Gamma \left(-\frac{d}{2}-2 \epsilon -J_L+3\right) \Gamma \left(d+2 \epsilon +J_L-3\right)}
 \nn\\
&\quad \x \left(-p^2\right)^{-\frac{d}{2}-J_L-2 \epsilon } (-2p\cdot z)^{J_L} \theta(-p^2)\,,\label{eq: F2}
 \ee
where $\delta^+(p^2)=\delta(p^2) \theta(p^0)$. 
Its $1/\epsilon$ pole can be obtained using (\ref{eq:distributionpole}), with the replacement $\epsilon \to 2\epsilon$. 
The renormalization of the singlet detector is then
\be
\mathcal{D}_{J_L,s,R}^{\rm detector}(z)=Z_{J_L,s}^{-1}\mathcal{D}_{J_L}^{\rm detector}(z)\,,\quad Z_{J_L,s}=Z_\phi^{-1}\Big(1+\fft{\mathcal{F}^1(z,p)\Big|_{\epsilon\,{\rm div}}+\mathcal{F}^2(z,p)\Big|_{\epsilon\,{\rm div}}}{V_{J_L}(p,z)}\Big)\,.
\ee
The anomalous dimension of the singlet detector is
\be
\gamma_{L,s}
&=-\mu \fft{d\log Z_{J_L,s}}{d\mu}
\nn\\
&=\fft{1}{N} \frac{2^{d+3} \sin \left(\frac{\pi  d}{2}\right) \Gamma \left(\frac{d-1}{2}\right) \left(J_L-1\right) \left(d+J_L-2\right)}{\pi ^{3/2} (d-2) d  \Gamma \left(\frac{d}{2}-2\right) \left(d+2 J_L-4\right) \left(d+2 J_L-2\right)}\nn\\
& \quad +\fft{1}{N}\frac{4^{d-1} (d-4) \sin \left(\frac{\pi  d}{2}\right) \Gamma \left(\frac{d-1}{2}\right)^2 \csc \left(\pi  \left(d+J_L-4\right)\right)}{\pi   \left(d+2 J_L-4\right) \left(d+2 J_L-2\right) \Gamma \left(-J_L\right) \Gamma \left(d+J_L-3\right)}\,.\label{eq: gammaL sing}
\ee
Upon using the reciprocity relation \eqref{eq: reciprocity relation}, we recover \eqref{eq: leading twist ano}.


\subsection{Pomeron from shadow mixing}
\label{sec: detector operator mixing}

So far, we have renormalized the DGLAP-type detector and reproduced the leading-twist anomalous dimensions for generic $J_L$. However, this procedure does not resolve the singularity at $J_L = \frac{1}{2}(2-d)$, at which DGLAP trajectories mix with their shadows. We now explicitly describe this mixing and locate the Pomeron operator in this theory.

Following \cite{Caron-Huot:2022eqs}, the key observation is that the $1/N$ corrections \eqref{eq: F1} and \eqref{eq: F2}  (considered as distributions) contain poles in $J_L$ even for finite $\epsilon$. The relevant distributional identity (which can easily be verified in the position space) is given by
\be
(-2z\cdot p)^{J_L}(-p^2)^{-d/2-J_L-a}\theta(-p^2)= -\fft{1}{J_L-J_{L,{\rm div}}} (-2z\cdot p)^{J_L}\delta(p^2) + \dots\,,\quad J_{L,{\rm div}}=\tfrac{2-d}{2}-a\,.
\ee
The one-loop diagram in figure~\ref{fig:oneloopcorrectionsubfigure} corresponds to $a=\epsilon$, while the two-loop diagram in figure~\ref{fig:mysubfigure} corresponds to $a=2\epsilon$. The coefficient of this divergence is proportional to the matrix element of the shadow of a tree-level detector \cite{Caron-Huot:2022eqs}. Recall that the shadow transform is defined by \cite{Simmons-Duffin:2012juh}
\be
\bold{S}_J[\mathcal{D}_{J_L}^{\rm detector}]=\int D^{d-2}z' (-2z\cdot z')^{2-d-J_L}\mathcal{D}_{J_L}^{\rm detector}(z')\,.
\ee
It is convenient to define a normalized shadow transform that squares to the identity:
\be
\tl{\bold{S}}_J=\fft{\Gamma(-J_L)}{2\pi^{\fft{d}{2}-1}\Gamma(\fft{2-d}{2}-J_L)}\bold{S}_J\,
\label{eq: norm shadow}
\qquad \tl{\bold{S}}_J^2 = 1.
\ee
Finally, we define the shadow DGLAP detector by
\be
\tl \cD_{J_L,\rho}^\mathrm{detector} \equiv \tl{\bold{S}}_J[\cD_{2-d-J_L,\rho}^\mathrm{detector}],
\ee
and denote its associated vertex by $\tl V_{J_L}=\tl{\bold{S}}_J[V_{2-d-J_L}]$.

In this language, we can write the $J_L$ poles of our two diagrams as
\be
\mathcal{F}^1(z,p)\big|_{J_{L,\rm div}^1}&=\ft{\pi ^{\epsilon -\frac{1}{2}} 2^{d+4 \epsilon-1} \mu ^{2 \epsilon } \Gamma (-\epsilon ) \sin \left(\frac{1}{2} \pi  (d+2 \epsilon -2)\right) \Gamma \left(\frac{d}{2}+\epsilon -\frac{1}{2}\right) \csc \left(\pi  J_L\right) }{N \Gamma \left(\frac{d}{2}-\epsilon -1\right) \Gamma \left(\frac{d}{2}+\epsilon -2\right) \Gamma \left(J_L+1\right) \left(d+2 \left(J_L+\epsilon -1\right)\right) \Gamma \left(-\frac{d}{2}-\epsilon -J_L+3\right)}\tl{V}_{\frac{2-d}{2}-\e}\,,\nn\\
\mathcal{F}^2(z,p)\big|_{J_{L,\rm div}^2}&=-\ft{\pi ^{2 \epsilon -1} 2^{2 d+8 \epsilon -4} \Gamma \left(\frac{d}{2}-1\right) \mu ^{4 \epsilon } \sin ^2\left(\frac{1}{2} \pi  (d+2 \epsilon -2)\right) \csc \left(\frac{1}{2} \pi  (d+4 \epsilon )\right) \Gamma \left(\frac{d}{2}+\epsilon -\frac{1}{2}\right)^2 \csc \left(\pi  \left(d+J_L+2 \epsilon -4\right)\right) }{N \epsilon  (2 \epsilon -1) \Gamma \left(\frac{d}{2}-2 \epsilon -1\right) \Gamma \left(\frac{d}{2}+2 \epsilon -2\right) \left(d+2 J_L+4 \epsilon -2\right) \Gamma \left(-\frac{d}{2}-2 \epsilon -J_L+3\right) \Gamma \left(d+2 \epsilon +J_L-3\right)}\nn\\
&\quad \times \tl{V}_{\frac{2-d}{2}-2\e}\,.
\ee
We see indeed that the bare detectors are divergent in $J_L$ for finite $\epsilon$, and this divergence is proportional to shadow vertices. A key point discussed in \cite{Caron-Huot:2022eqs} is that we should renormalize divergences in both $1/\epsilon$ and $1/(J_L-J_{L,{\rm div}})$, otherwise we will have failed to define finite operators. The full divergence of the matrix elements at $O(1/N)$ can be summarized as
\be
\langle\mathcal{D}_{J_L,\rho}^{\rm detector}\rangle
&\sim \big( \mathcal{F}'^1(z,p)\big|_{J_{\rm div}^1}+\delta_{\rho s}\mathcal{F}'^2(z,p)\big|_{J_{\rm div}^2}\big)\langle\mathcal{D}_{J_L,\rho}^{\rm detector}\rangle_{\rm tree} \nn\\
&\quad+ \fft{1}{\epsilon}\big(Y_1 \mu^{2\epsilon}+ \delta_{\rho,s} Y_2 \mu^{4\epsilon}\big)\langle \mathcal{D}_{J_L,\rho}^{\rm detector}\rangle_{\rm tree}
\nn\\
&\quad+ \epsilon \big( \mathcal{F}'^1(z,p)\big|_{J_{\rm div}^1}+2 \delta_{\rho,s}  \mathcal{F}'^2(z,p)\big|_{J_{\rm div}^2}\big) \langle \mathcal{D}_{J_L,\rho}^{\prime{\rm detector}}\rangle_{\rm tree}\,,\label{eq: bare full div}
\ee
where $\mathcal{F}'$ is $\mathcal{F}$ without the vertex factor $\tl V_{J_L}$, and we follow \cite{Caron-Huot:2022eqs} in choosing a non-degenerate basis of detectors for all $J_L$ given by $\cD^\mathrm{detector}_{J_L}$ and
\be
\mathcal{D}^{\prime{\rm detector}}_{J_L,\rho}&\equiv \fft{\tl{\mathcal{D}}_{J_L,\rho}^{\rm detector}-\mathcal{D}_{J_L,\rho}^{\rm detector}}{J_L-\fft{2-d}{2}}.
\ee
Here, $Y_1$ and $Y_2$ are determined such that the $1/\epsilon$ divergence detailed in the last subsection is correctly reproduced when we take $\epsilon\rightarrow 0$ in \eqref{eq: bare full div}. We should renormalize the detector vector using a $2\x 2$ renormalization matrix $Z_{J_L,\rho}$:
\be
\mathbb{D}_{J_L,\rho}^{\rm detector}=\left(
\begin{array}{c}
 \mathcal{D}_{J_L,\rho}^{\rm detector}  \\
 \mathcal{D}^{\prime{\rm detector}}_{J_L,\rho}  \\
\end{array}
\right)\,,\quad \mathbb{D}_{J_L,\rho,R}^{\rm detector}= Z_{J_L,\rho}^{-1} \mathbb{D}_{J_L,\rho}^{\rm detector}\,.
\ee
We can now evaluate the full dilatation operator up to $1/N$ order
\be
\mathbf{\Delta}_{L,\rho}=\lim_{\epsilon\rightarrow 0}Z_{J_L,\rho}^{-1} (\mathbf{\Delta}_L^{\rm tree}- \fft{\partial}{\partial \log \mu}) Z_{J_L,\rho}\,,\quad
\mathbf{\Delta}_L^{\rm tree}=\left(
\begin{array}{cc}
 d+J_L-2 & 0 \\
 -2 & -J_L \\
\end{array}
\right)\,.
\ee

If we diagonalize $\mathbf{\Delta}_L$ for $J_L\neq \frac{2-d}{2}$, we reproduce the anomalous dimensions in \eqref{eq: gammaL sym} and \eqref{eq: gammaL sing}, together with anomalous dimensions of the shadow trajectories with $J_L\rightarrow 2-d-J_L$. Furthermore, the characteristic equation of ${\rm det}\,\big(\mathbf{\Delta}-\Delta_L I\big)=0$ gives a smooth curve near $J_L=\frac{2-d}{2}$. To identify the Pomeron and its Regge intercept, we simply diagonalize $\mathbf{\Delta}_L(J_L=\frac{2-d}{2})$. This gives two eigenvalues $\Delta_L^{\pm}$. Using $\Delta=d/2$ and $\Delta_L^+(\Delta)=1-J$, we precisely recover the Regge intercept \eqref{eq: intercept}. Meanwhile, $\Delta_L^-$ gives \eqref{eq: intercept} with the sign in front of $c_{1,\rho}/\sqrt{N}$ flipped, providing the subleading intercept.

\subsection{BFKL-type horizontal trajectories}

In this subsection, we follow \cite{Caron-Huot:2022eqs} to understand renormalization of the horizontal trajectory \eqref{eq: detector H} in the $O(N)$ model. For simplicity, we will restrict our attention to $O(N)$ singlet detectors.\footnote{Note that the symmetric and anti-symmetric DGLAP-type anomalous dimensions \eqref{eq: leading twist ano} do not exhibit singularities at $J = -1$, indicating that these detectors don't mix with horizontal trajectories at leading order in $1/N$.}

\begin{figure}[t]
\centering
\begin{tikzpicture}
   \draw[thick] (-3-7,1.5) -- (3-7,1.5);
\draw[thick] (-3-7,-1.5) -- (3-7,-1.5);
\draw[thick,->] (-3-7,-1.5) -- (-1.5-7,-1.5);
\node at (-1.5-7,-2.3+0.5) {$p_1$};
\draw[thick,->] (-7,-2+0.5) -- (1.5-7,-2+0.5);
\node at (1.5-7,-2.3+0.5) {$q_1$};
\draw[thick, ->] (-3-7,2-0.5) -- (-1.5-7,2-0.5);
\draw[thick, ->] (0-7,2-0.5) -- (1.5-7,2-0.5);
\node at (-1.5-7,2+0.3-0.5) {$p_2$};
\node at (1.5-7,2+0.3-0.5) {$q_2$};
    \draw[red, very thick] (-0.05-7,2.2) -- (-0.05-7,-2.2);
    \draw[red, very thick] (0.05-7,2.2) -- (0.05-7,-2.2);
         \fill (-7,2-0.5) circle (0.1);
      \fill (-7,-2+0.5) circle (0.1);
      \node at (-7+0.27,0.2+2-0.5) {$z_2$};
       \node at (-7+0.27,-2-0.2+0.5) {$z_1$};
        \end{tikzpicture}
\caption{Tree-level diagram for the horizontal detector.}
\label{fig: detector operators horizontal tree}
\end{figure}
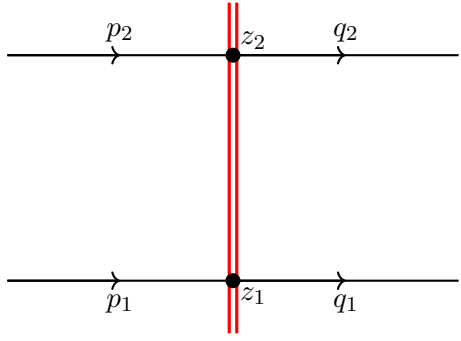

Let us first compute
\be
\langle \phi^i(-q_1)\phi^j(-q_2)|\cH^\mathrm{detector}_{J_{L1}J_{L2}}(z_1,z_2)|\phi^k(p_2)\phi^l(p_1)\rangle\,,
\ee
where
\be
\cH^\mathrm{detector}_{J_{L1}J_{L2}}(z_1,z_2) &\equiv \mathcal{D}_{J_{L1},s}^{\rm detector}(z_1)\mathcal{D}_{J_{L2},s}^{\rm detector}(z_2).
\ee
The tree-level result (see figure \ref{fig: detector operators horizontal tree}) is given by
\be
& \langle \phi_i(-q_1)\phi_j(-q_2) \mathcal{H}_{J_{L1}J_{L2}}^{\rm detector}(z_1,z_2) \phi_k(p_2)\phi_l(p_1)\rangle 
 \nn\\
 &= (2\pi)^d \delta^d(p_2-q_2)V_{J_{L1}}(z_1,q_1) V_{J_{L2}}(z_2,p_2)\delta_{il}\delta_{jk}+{\rm perm}\,,
\ee
where ``perm." denotes permutations $(p_1,l)\leftrightarrow (p_2,k)$ and $(q_1,i)\leftrightarrow (q_2,j)$.
Convolving with a conformal three point function on the celestial sphere to obtain a matrix element of (\ref{eq: detector H}), we find
\be
& \langle \phi_i(-q_1)\phi_j(-q_2) \mathcal{H}_{J_L,J_{L1}J_{L2}}^{\rm detector}(z) \phi_k(p_2)\phi_l(p_1)\rangle
\nn\\
&=(2\pi)^d \delta^d(p_2-q_2)\delta_{il}\delta_{jk}\nn\\
 &\quad \x 4 (-2q_1\cdot p_2)^{\ft{4-2d-J_L-J_{L1}-J_{L2}}{2}}(-2q_1\cdot z)^{\ft{J_L-J_{L1}+J_{L2}}{2}}(-2p_2\cdot z)^{\ft{J_L+J_{L1}-J_{L2}}{2}}\delta(q_1^2)\delta(p_2^2)\nn\\
 &\quad + {\rm perm}\,.
\ee

\begin{figure}[h]
\centering
\begin{tikzpicture}
\draw[thick] (-3-7,0) -- (3-7,0);
\draw[thick] (-3-7,-2) -- (3-7,-2);
\draw[thick,->] (-3-7,-2) -- (-1.5-7,-2);
\node at (-1.5-7,-2.3) {$p_1$};
\draw[thick,->] (-7,-2) -- (1.5-7,-2);
\node at (1.5-7,-2.3) {$q_1$};
\draw[thick, ->] (-3-7,0) -- (-2.25-7,0);
\draw[thick, ->] (-3-7,0) -- (-0.75-7,0);
\draw[thick, ->] (0-7,0) -- (0.75-7,0);
\draw[thick, ->] (0.75-7,0) -- (2.25-7,0);
\node at (-2.25-7,0.3) {$p_2$};
\node at (2.25-7,0.3) {$q_2$};
\node at (-0.75-7,0.3) {$\ell$};
    \draw[red, very thick] (-0.05-7,2.2) -- (-0.05-7,-2.2);
    \draw[red, very thick] (0.05-7,2.2) -- (0.05-7,-2.2);
    \draw[dashed, thick] (1.5-7,0) arc (0:180:1.5);
    \draw[dashed, thick,->] (-1.5-7,0) arc (0:-45:-1.5);
    \node at (-0.75-7-0.7,1.3) {$p_2-\ell$};
     \fill (-7,0) circle (0.1);
      \fill (-7,-2) circle (0.1);
      \node at (-7+0.27,0.2) {$z_1$};
       \node at (-7+0.27,-2-0.2) {$z_2$};
    \node at (-3.5,-1) {$+$};
    \draw[thick] (-3,0) -- (-2,0);
      \draw[thick] (-3,-2) -- (3,-2);
     \draw[dashed,thick] (-2,0) -- (-1,0);
     \draw[thick] (-1,0) -- (1,0);
     \draw[dashed,thick] (1,0) -- (2,0);
     \draw[thick] (2,0) -- (3,0);
    \draw[red, very thick] (-0.05,2.2) -- (-0.05,-2.2);
    \draw[red, very thick] (0.05,2.2) -- (0.05,-2.2);
    \draw[thick] (1,0) arc (0:180:1);
    \draw[thick] (2,0) arc (0:180:2);
    \node at (4,-1) {$+\,\text{perm}$};
     \fill (0,0) circle (0.1);
      \fill (0,-2) circle (0.1);
        \node at (0.27,0.2) {$z_1$};
       \node at (0.27,-2-0.2) {$z_2$};
       \draw[thick,->] (-3,-2) -- (-1.5,-2);
       \draw[thick,->] (0,-2) -- (1.5,-2);
       \node at (-1.5,-2.3) {$p_1$};
\node at (1.5,-2.3) {$q_1$};
\draw[thick,->] (-3,0) -- (-2.25,0);
\node at (-2.25,-0.3) {$p_2$};
\draw[thick,->,dashed] (-1.5,0) -- (-1.25,0);
\node at (-1.25,-0.3) {$\ell_1$};
\draw[thick,->] (-1,0) -- (-0.5,0);
\node at (-0.5,-0.3) {$\ell_2$};
\draw[thick,->] (0,0) -- (0.5,0);
\draw[thick,->,dashed] (1,0) -- (1.5,0);
\draw[thick,->] (2,0) -- (2.5,0);
 \draw[thick,->] (-1,0) arc (0:-45:-1);
  \draw[thick,->] (-2,0) arc (0:-45:-2);
\node at (2.5,-0.3) {$q_2$};
\node at (-1,1.05) {$\ell_1-\ell_2$};
\node at (-2,1.7) {$p_2-\ell_1$};
     \end{tikzpicture}
\caption{The disconnected diagram at $1/N$ order for the horizontal detector.}
\label{fig: detector operators horizontal disc}
\end{figure}
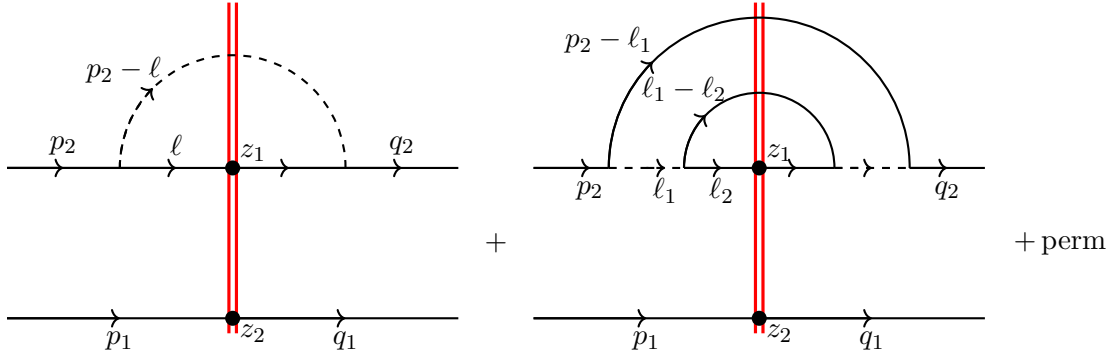

Let us now consider $1/N$ corrections. These come from the connected diagrams shown in figure~\ref{fig: detector operators horizontal connect}  and disconnected diagrams shown in figure~\ref{fig: detector operators horizontal disc}. The connected diagrams will give a $J_L$-dependent anomalous dimension. The disconnected diagrams contribute $J_L$-independent divergences, and their status is less clear, as we discuss futher in appendix~\ref{app: more horizontal}. For now, we focus on the $J_L$-dependent connected contributions.


\begin{figure}[h]
\centering
\begin{tikzpicture}
    \draw[thick] (-3,2) -- (3,2);
      \draw[thick] (-3,-2) -- (3,-2);
    \draw[red, very thick] (-0.05,2.2) -- (-0.05,-2.2);
    \draw[red, very thick] (0.05,2.2) -- (0.05,-2.2);
    \draw[thick, dashed] (-1.5,-2) -- (-0.75,-1);
     \draw[thick, dashed] (1.5,2) -- (0.75,1);
     \draw[thick] (0,0) circle (1.3);
      \node at (4,0) {$+\,\text{perm}$};
       \fill (0,1.3) circle (0.1);
     \node at (0.27,0.2+1.3) {$z_1$};
     \fill (0,-1.3) circle (0.1);
     \node at (0.25,0.2-1.7) {$z_2$};
     \draw[thick,->] (-3,-2) -- (-2.25,-2);
     \draw[thick,->] (0,-2) -- (1.5,-2);
      \draw[thick,->] (-3,2) -- (-1.5,2);
      \draw[thick,->] (1.5,2) -- (2.25,2);
      \draw[thick,->,dashed] (-1.5,-2) -- (-1.125,-1.5);
       \draw[thick,<-,dashed]  (1.125,1.5) -- (0.75,1);
       \draw[->, thick] (0:1.3) arc (0:10:1.3);
       \draw[->, thick] (0:1.3) arc (0:-180:1.3);
       \node at (-2.25,-2.3) {$p_1$};
        \node at (1.5,-2.3) {$q_1$};
        \node at (-1.5,2.3) {$p_2$};
        \node at (2.25,2.3) {$q_2$};
         \node at (-2,-1.5) {$p_1-q_1$};
         \node at (-1.7,0) {$\ell$};
          \node at (2.2,-0.5) {$p_1-q_1-\ell$};
        \end{tikzpicture}
\caption{A connected diagram that contributes at $O(1/N)$ for the matrix element of the horizontal detector.}
\label{fig: detector operators horizontal connect}
\end{figure}
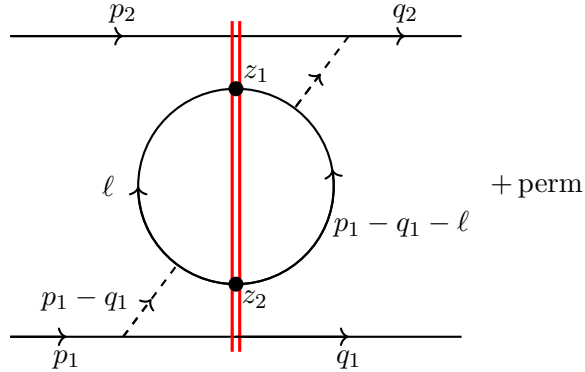

Let us consider the connected diagram in figure~\ref{fig: detector operators horizontal connect}.
\be
& \mathcal{G}_{c}=\fft{1}{N}(2\pi)^2 \delta^+(q_1^2)\delta^+(p_2^2)\nn\\
& \quad \x\int \fft{d^d\ell}{(2\pi)^d} V_{J_{L1}}(z_1,\ell) V_{J_{L2}}(z_2,p_1-q_1-\ell) \fft{1}{p_1^2-i 0}\fft{1}{q_2^2+i 0} G_\sigma((p_1-q_1)^2-i0)^2 
\nn\\
&=\fft{1}{(2\pi)^{d-2}N}\delta^+(q_1^2)\delta^+(p_2^2)\nn\\
& \quad \x\int d\beta_1 d\beta_2\fft{G_\sigma((p_1-q_1)^2-i0)^2}{(p_1^2-i 0)(q_2^2+i 0)}\beta_1^{-J_{L1}-1} \beta_2^{-J_{L2}-1}\delta^d(p_1-q_1-\beta_1 z_1-\beta_2 z_2)\,.
\ee
Let us now locate the divergence in $J_{L1} + J_{L2}$. We first replace $p_1$ and $q_2$ using the delta function $p_1 - q_1 = q_2 - p_2 = \sum \beta_i z_i$.  Then we take the limit of small $\beta_i$ inside the integrand. This leads to
\be
\label{eq:thingthatitleadsto}
\mathcal{G}_c
& =- \fft{X}{N} \delta^d(q_2-p_2)\delta(q_1^2)\delta(p_2^2) K_\alpha(z_1,z_2,p_2,q_1)+\text{regular} \,,
\ee
where $\alpha=(J_{L1}-J_{L2})/2$, the kernel $K_\alpha$ is defined by
\be
K_\alpha(z_1,z_2,z_3,z_4) = (-z_{12})^{4-d-2\epsilon}\int dx \fft{x^{-\alpha}}{(x z_{13}+ z_{23})(x z_{14}+z_{24})}\,,\quad z_{ij}=-2z_i\cdot z_j\,,\label{eq: horizontal K}
\ee
and the constant out front that contains a pole in $J_{L1}+J_{L2}$ is given by
\be
X &= \frac{\mu^{4\epsilon}\pi ^{2 \epsilon -1} 2^{3d+8 \epsilon -4} \sin ^2(\frac{\pi}{2}   (d+2 \epsilon -2)) \Gamma (\frac{d}{2}+\epsilon -\frac{1}{2})^2}{2 d+J_{L1}+J_{L2}+4 \epsilon -6}.
\ee
Above, ``regular" refers to terms that do not contain a pole in $J_{L1}+J_{L2}$ at the same location.
Note that for any function $f(p)$ with homogeneity $2-d-J_L$, we have
\be
 \int D^{d-2}z V_{J_L}(z,p)f(z) = 2 \delta(p^2)f(p).
\ee
The kernel $K_\alpha$ has homogeneity $-1$ in both $z_3$ and $z_4$, so we can rewrite (\ref{eq:thingthatitleadsto}) as
\be
 \mathcal{G}_c &\sim  -\fft{X}{N} \delta^d(q_2-p_2)\int D^{d-2}z_3 D^{d-2}z_4 K_\alpha(z_1,z_2,z_3,z_4)V_{3-d}(z_3,p_2)V_{3-d}(z_4,q_1)\,.
\ee
This is precisely the integral of the kernel $K_\alpha$ against a tree-level matrix element, so we can summarize our result as
\be
\langle \mathcal{H}_{J_{L1},J_{L2}}^{\rm detector}(z_1,z_2)\rangle_{c}
& \sim  -\fft{X}{N} \int D^{d-2}z_3 D^{d-2}z_4   K_\alpha(z_1,z_2,z_3,z_4)\langle \mathcal{H}_{3-d,3-d}^{\rm detector}(z_3,z_4)\rangle_{\rm tree}+\text{regular}\,.
\ee

We still must project onto an irreducible representation of the Lorentz group by integrating against a conformal three-point function. This requires computing the action of the kernel $K_\alpha$ on the three-point function
\be
\int D^{d-2}z_1 D^{d-2}z_2 \, K_\alpha(z_1,z_2,z_3,z_4) \langle \varphi_{-J_L}(z) \varphi_{d-2+J_{L1}}(z_1)\varphi_{d-2+J_{L2}}(z_2)\rangle\,.
\ee
We can think of $K_\alpha$ as a conformal four-point function in CFT$_{d-2}$. By integrating against a three-point function, we are essentially computing the OPE data of that four point function using a Euclidean partial wave expansion, see e.g.\ \cite{Karateev:2018oml}. In appendix~\ref{app: diagonalize K}, we determine this data using the Lorentzian inversion formula \cite{Caron-Huot:2017vep}, providing the diagonalization of $K_\alpha$. The result is
\be
& \int D^{d-2}z_1 D^{d-2}z_2 \, K_\alpha(z_1,z_2,z_3,z_4) \langle \varphi_{-J_L}(z) \varphi_{d-2+J_{L1}}(z_1)\varphi_{d-2+J_{L2}}(z_2)\rangle\nn\\
& = \pi^{d-2}\kappa_\alpha(J_L)\langle \varphi_{-J_L}(z)\varphi_1(z_3)\varphi_1(z_4)\rangle\,,\label{eq: dia1}
\ee
where the eigenvalue is given by
\be
\kappa_\alpha(J_L)=\frac{\sin (\pi  J_L) \Gamma (\frac{J_L}{2}+1) \Gamma (-\frac{J_L}{2})^2 \Gamma (J_L+2) \Gamma (\frac{d+2 \alpha +J_L-2}{2}) \Gamma (\frac{d-2 (\alpha +1)+J_L}{2})}{\sin (\frac{\pi}{2}   (d+J_L)) \Gamma (\frac{d}{2}-1) (J_L+1) \Gamma (\frac{d+J_L-2}{2}) \Gamma (d+J_L-2)}\,.\label{eq: dia2}
\ee
This is a generalization of the result in \cite{Caron-Huot:2022eqs} to generic dimensions. We thus find
\be
\langle \mathcal{H}_{J_{L1},J_{L2},J_L}^{\rm detector}(z)\rangle_c  &=  -\fft{\pi^{d-2} X}{N} \kappa_\alpha(J_L) \langle \mathcal{H}_{3-d,3-d,J_L}^{\rm detector}(z)\rangle_{\rm tree}+\text{regular}\,.
\ee

Note that the divergence is proportional to a tree-level matrix element of a detector with $J_{L1}=J_{L2}=3-d$. In other words, only detectors with $J_{L1}=J_{L2}=3-d$ get multiplicatively renormalized and acquire anomalous dimensions, while the other detectors in this class get additively renormalized. The multiplicatively-renormalized detectors are what we identify as BFKL-type:
\be
\mathcal{H}_{J_L,{\rm BFKL}}(z)= \mathcal{H}_{3-d,3-d,J_L}^{\rm detector}(z)\,.
\ee

Setting $J_{L1}=J_{L2}=3-d$, the pole in $J_{L1}+J_{L2}$ becomes a pole in $1/\epsilon$. We can thus follow the standard renormalization procedure to compute its anomalous dimension \eqref{eq: RG BFKL detector}. We obtain
\be
\gamma_\mathrm{BFKL}(J_L)= \fft{1}{N}\frac{2^{2d-5} (1-\cos (\pi  d)) \Gamma (\frac{d-1}{2})^2}{\pi ^3} \kappa_0(J_L)=\fft{1}{N}\fft{C_\sigma}{(4\pi)^d}\kappa_0(J_L)\,.\label{eq: gammaL}
\ee

Note that the disconnected diagrams in figure~\ref{fig: detector operators horizontal disc} only contribute to additive renormalization, mixing horizontal detector operators with different $J_{L_i}$. We leave discussion of these contributions to appendix \ref{app: more horizontal}.


\section{Renormalization of distribution operators}
\label{sec: RG distribution operator}

In this section, we study renormalization of distribution operators. Since we work in a CFT, distribution operators are simply detectors in a different conformal frame, and thus are described by precisely the same Regge trajectories. Nevertheless, their physical interpretation, and the technical details of their renormalization are different. We find it instructive to understand these details in both frames and compare them.
%

\subsection{Regarding leading-twist distribution operators}

Consider the leading-twist distribution operators in the critical O$(N)$ model, defined as
\be
\mathcal{D}_{J}^{{\rm dis},k,ij}(-\infty z)\propto \int d\alpha_1 d\alpha_2 |\alpha_{12}|^{-1-J} (\mathrm{sign}\,\alpha_{12})^k :\phi^i(\alpha_1 z)\phi^j(\alpha_2 z):\,.
\ee
For simplicity and to make contact with PDFs, we consider the forward in-in correlator
\be
\langle p^i|\mathcal{D}_{J,\rho}^{\rm dis}(-\infty z,x^\perp)|p^j\rangle\,,
\ee
where the external state $|p^i\>$ denotes an on-shell $\phi$-particle with $O(N)$ index $i$. In particular, unlike our discussion of detector operators, we now choose $p^2=0$ for the incoming/outgoing state. As a consequence, these matrix elements will contain collinear divergences coming from the (ill-defined) initial and final states \cite{Collins:1989gx}. Our focus will be on UV divergences associated to the detector, so these collinear divergences will not play a role.


The tree-level matrix elements are (ommitting the argument $(-\infty z,x^\perp=0)$ for brevity):
\be
\langle p^i|\mathcal{D}_{J,\rho}^{\rm dis}|p^j\rangle=\fft{1}{2\pi} (-2p\cdot z)^{J}T^{ij}_\rho\,,
\ee
which using \eqref{eq: from D to f} corresponds to the PDF:
\be
\quad f_{\rho}^{ij}(\xi)= \delta(1-\xi)T_{\rho}^{ij}\,.
\ee
This tree-level PDF has the interpretation that a parton propagates unmodified in the absence of interactions. 

\subsubsection{The Drell-Levi-Yan relation}

As a quick comment, we note the tree-level matrix element of distribution detectors is proportional to the matrix element of a shadow-transformed detector 
\be
\langle p^i|\mathcal{D}_{4-d-J,\rho}^{\rm dis}|p^j\rangle \propto \langle \phi^i(p)|\bold{S}_J[\mathcal{D}_{2-d-J_L,\rho}^{\rm detector}]|\phi^j(p)\rangle \times (-2p\cdot z)\sim (-2p\cdot z)^{4-d-J}\,.
\ee
If we define $D(\xi)$ as the inverse Mellin transform of the shadow detector matrix element,
\be
D(\xi)\sim \oint dJ_L\, \langle\phi^i(p)|\bold{S}_J[\mathcal{D}_{2-d-J_L,\rho}^{\rm detector}]|\phi^j(p)\rangle (-2\xi p\cdot z)^{-J_L} \,,
\ee
then at tree-level in $d=4$ we find the relation
\be
D(\xi)\propto \xi f\left(\fft{1}{\xi}\right)\,.\label{eq: Drell-Levi-Yan}
\ee
This is reminiscent of the Drell-Levi-Yan relation between spacelike and timelike splitting functions \cite{Blumlein:2000wh} (which is however violated at loop levels)
\be
P_T(x)\sim x P_S\left(\fft{1}{x}\right)\,.
\ee
This suggests that in general the Drell-Levi-Yan relation can be understood as a consequence of shadow symmetry and reciprocity.\footnote{We are grateful to Ian Moult for suggesting this idea.} The loop-level modifications of this relation may be related to perturbative corrections to the shadow symmetry. We leave further investigation of this idea to future work.



\subsubsection{Renormalizing distribution operators and PDFs}

We now proceed to compute $1/N$ corrections to distribution operator matrix elements. The relevant diagrams are essentially the same as in figure~\ref{fig: detector operators leading}, but the details of the calculations are slightly different.

The $1/N$ diagrams give
\be
& \mathcal{F}_J^1(p^+)= \fft{1}{N} \int \fft{d^d \ell}{(2\pi)^d} \fft{1}{\ell^2+i 0}\fft{1}{\ell^2-i0} (-2\sin(\pi \delta)G_\sigma(-(p-\ell)^2)\theta(p-\ell)) \fft{(\ell^+)^{J}}{2\pi}\,,\nn\\
& \mathcal{F}_J^2(p^+)= \fft{1}{N} \int \fft{d^d\ell_1 d^d\ell_2}{(4\pi^2)^{d-1}}\fft{1}{\ell_2^2+i 0}\fft{1}{\ell_2^2-i0}G_\sigma(\ell_1^2-i 0)G_{\sigma}(\ell_1^2+i 0) \delta^+((p-\ell_1)^2) \delta^+((\ell_1-\ell_2)^2) \fft{(\ell_2^+)^J}{2\pi}\,.\label{eq: distribution operator 1/N}
\ee
(These integrals are slightly more involved than directly computing the PDF, which replaces $(\ell^+)^J$ by $\delta(\ell^+ - k^+)$, thus trivializing the integral along $\ell^+$.) As is standard for perturbative computations of PDFs, these integrals are actually scaleless, proportional to $\int d^{d-2} \ell^\perp \, 1/(\ell^\perp)^\#$, which gives $1/\epsilon_{\rm UV} - 1/\epsilon_{\rm IR}$. The IR collinear singularity exists because we have chosen an IR-divergent initial state. On the other hand, the UV divergence must be renormalized.  This is in contrast to detector operators, where we renormalize the IR divergence.

We thus focus on the UV divergence, which gives
\be
& \mathcal{F}_J^1(p^+)\Big|_{\epsilon\,{\rm div}}=\fft{1}{N}\frac{(2-d) \pi ^{\epsilon -\frac{3}{2}} 2^{d+4 \epsilon -2} (d+2 \epsilon -4) \mu ^{2 \epsilon } \sin \left(\frac{1}{2} \pi  (d+2 \epsilon -2)\right) \Gamma \left(\frac{d}{2}+\epsilon -\frac{1}{2}\right)}{\epsilon  \Gamma \left(\frac{d}{2}\right) (d+2 (J+\epsilon -2)) (d+2 (J+\epsilon -1))} \fft{(p^+)^J}{2\pi}\,,\nn\\
& \mathcal{F}_J^2(p^+)\Big|_{\epsilon\, {\rm div}}=\fft{1}{N}\ft{\pi ^{2 \epsilon -2} 4^{d+4 \epsilon -3} \csc \left(\frac{\pi  d}{2}\right) \Gamma (J+1) \mu ^{4 \epsilon } \Gamma \left(\frac{d}{2}+J-2\right) (1-\cos (\pi  (d+2 \epsilon ))) \Gamma \left(\frac{d-1}{2}+\epsilon \right)^2 \Gamma \left(\frac{d}{2}+2 \epsilon -1\right)}{\epsilon  \Gamma \left(\frac{d}{2}-2\right) \Gamma (d+J-3) \Gamma \left(\frac{d}{2}+J+2 \epsilon \right)}\fft{(p^+)^J}{2\pi}\,.\label{eq: distribution div}
\ee
We can directly see singularities in both $\epsilon$ and $J$. We can renormalize the UV $1/\epsilon$ divergences to obtain the anomalous dimension 
\be
\gamma_\rho(J)=-\mu \fft{d}{d\mu}\log Z_{J,\rho}\,,\quad Z_{J,\rho}=Z_\phi^{-1}\Big(1+(4\pi)^2 \fft{\mathcal{F}_J^1(p^+)\Big|_{\epsilon\,{\rm div}}+\delta_{\rho s}\mathcal{F}_J^2(p^+)\Big|_{\epsilon\,{\rm div}}}{(p^+)^J}\Big)\,.
\ee
This reproduces \eqref{eq: leading twist ano}.

Let's now elaborate on the details of relating the distribution operators to PDFs and splitting functions at the $1/N$ order. We can simply redo the computations in \eqref{eq: distribution operator 1/N} by replacing $(\ell^+)^J$ with $\delta(\ell^+ - k^+)$. This gives the same answer as if we directly take the inverse Mellin transform of \eqref{eq: distribution div}. This can be easily done by taking the residues of $J$. To get the full answer for the corrected PDF, we need to include $O(1/N)$ wavefunction renormalization corrections to the external states. Overall, we find 
\be
f^0(\xi)\big|_{\epsilon\,{\rm div}}&=-\fft{1}{N}\frac{\pi ^{\epsilon -\frac{1}{2}} 2^{d+4 \epsilon -2} \Gamma \left(\frac{d}{2}-1\right) \mu ^{2 \epsilon } \csc (\pi  \epsilon ) \sin \left(\frac{1}{2} \pi  (d+2 \epsilon )\right) \Gamma \left(\frac{d-1}{2}+\epsilon \right)}{\Gamma \left(\frac{d}{2}-\epsilon +1\right) \Gamma \left(\frac{d}{2}+\epsilon -2\right)}\xi \delta (1-\xi)\,,\nn\\
 f^1(\xi)\big|_{\epsilon\,{\rm div}}&=\fft{1}{N}\frac{(1-\xi ) \pi ^{\epsilon -\frac{3}{2}} 2^{d+4 \epsilon -3} (d+2 \epsilon -4) \mu ^{2 \epsilon } \xi ^{\frac{d}{2}+\epsilon -2} \sin \left(\frac{1}{2} \pi  (d+2 \epsilon -2)\right) \Gamma \left(\frac{d}{2}+\epsilon -\frac{1}{2}\right)}{\epsilon  \Gamma \left(\frac{d-2}{2}\right)}\,,\nn\\
 f^2(\xi)\big|_{\epsilon\,{\rm div}}&=\fft{1}{N} \ft{(4-d) \pi ^{2 \epsilon -2} \xi ^{\frac{d}{2}-2} 2^{2 (d+4 \epsilon -3)} (d+2 \epsilon -2) \csc \left(\frac{\pi  d}{2}\right) \mu ^{4 \epsilon } (1-\xi )^{d+2 \epsilon -3} \sin ^2\left(\frac{1}{2} \pi  (d+2 \epsilon -2)\right) \Gamma \left(\frac{d}{2}+\epsilon -\frac{1}{2}\right)^2 \Gamma \left(\frac{d}{2}+2 \epsilon -1\right) }{\epsilon  \Gamma \left(\frac{d-2}{2}\right) \Gamma (d+2 \epsilon -1)}\nn\\
& \times \,_2F_1\left(d-4,\frac{d}{2}+2 \epsilon -1;d+2 \epsilon -2;1-\xi \right)\,,
\ee
where $f^0$ is the contribution from wavefunction renormalization. These divergences have to be renormalized by the DGLAP-type splitting in \eqref{eq: DGLAP evol}, giving the space-like splitting function
\be
& P_\rho(x)=-2\gamma_\phi \, x \delta(1-x)+\frac{2^{d-2} (d-4) (1-x) x^{\frac{d}{2}-2} \sin \left(\frac{1}{2} \pi  (d-2)\right) \Gamma \left(\frac{d-1}{2}\right)}{\pi ^{3/2} \Gamma \left(\frac{d-2}{2}\right)}\nn\\
& -\ft{2^{2 (d-2)} (d-4) (d-2) (1-x)^{d-3} x^{\frac{d}{2}-2} \sin ^2\left(\frac{1}{2} \pi  (d-2)\right) \csc \left(\frac{\pi  d}{2}\right) \Gamma \left(\frac{d}{2}-1\right) \Gamma \left(\frac{d-1}{2}\right)^2 }{\pi ^2 \Gamma \left(\frac{d-2}{2}\right) \Gamma (d-1)}\, _2F_1\Big(\frac{d}{2}-1,d-4,d-2,1-x\Big) \delta_{\rho s}\,.
\ee
This is indeed the result that we can reproduce by directly taking the inverse Mellin transform of \eqref{eq: leading twist ano}.

\subsubsection{Comments on shadow mixing for distribution operators}

We have computed the leading-order renormalization of distribution operators and their associated PDFs and reproduced the associated leading-twist anomalous dimensions. It is natural to ask whether we can redo the analysis in section \ref{sec: detector operator mixing} to resolve the singularity at $ J = (4-d)/2 $ in the anomalous dimensions and determine the pomeron spin in the distribution frame.

However, we should note that the exact quantum number in the distribution frame is the spin $J$. The singularity at $J\sim (4-d)/2$ comes from mixing between a distribution operator and its shadow, but in this context the shadow transform is a conformally-invariant integral transform in the transverse plane
%
\be
\bold{S}_\Delta[\mathcal{D}_J^{\rm dis}(-\infty z,y^\perp)]\sim \int d^{d-2}y'^\perp |y^\perp-y'^\perp |^{2(2-d-\Delta)}\mathcal{D}_J^{\rm dis}(-\infty z,y'^\perp)\,.
\ee
Because conformal invariance is not manifestly respected at each order in perturbation theory, in general we must redefine the shadow transform at each order in perturbation theory, taking into account perturbative shifts in $\Delta$. Furthermore, we cannot use analytic continuation in $J_L$ as a regulator, as we did in the distribution frame. Nonetheless, it should be possible to perform this analysis, and we leave this exercise for future work.

More subtle issues arise away from the critical point, where there is no nonperturbative shadow symmetry in the distribution frame. In this case, conformal symmetry of the transverse plane does not constrain the pattern of mixing in a simple way. The resulting mixing matrix could be a general kernel in transverse momentum, which requires understanding transverse-momentum-dependent PDFs, see e.g.\ \cite{Collins:1981uk,Collins:1984kg}. Furthermore, one cannot use analytic continuation in $J_L$ to regulate divergences, as in the detector frame.

\subsection{Collinear function and rapidity renormalization}

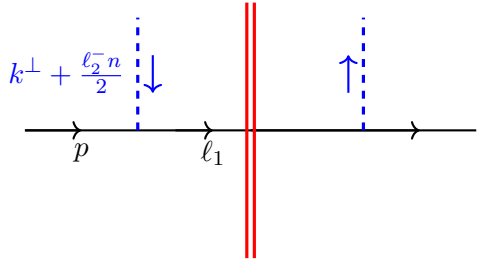
\begin{figure}
    \centering
    \begin{tikzpicture}
    \draw[thick] (-3,0) -- (3,0);
    \draw[->, thick] (-3,0) -- (-2.25,0);
    \draw[->, thick] (0,0) -- (2.25,0);
    \draw[very thick, dashed,blue] (-1.5,0) -- (-1.5,1.5);
    \draw[very thick, dashed,blue] (1.5,0) -- (1.5,1.5);
    \draw[red, very thick] (-0.05,1.7) -- (-0.05,-1.7);
    \draw[red, very thick] (0.05,1.7) -- (0.05,-1.7);    
    \node at (-2.25,-0.3) {$p$};
    \draw[->, thick,blue] (-0.8-0.5,1) -- (-0.8-0.5,0.5);
    \draw[->, thick,blue] (0.8+0.5,0.5) -- (0.8+0.5,1);
    \node at (-0.45-2,0.8) {${\color{blue} k^\perp+\fft{\ell_2^- n}{2}}$};
    \draw[->, thick] (-1,0) -- (-0.5,0);
    \node at (-0.5,-0.3) {$\ell_1$};
\end{tikzpicture}
\caption{The leading order contribution of the collinear function of $\sigma$.}
\label{fig: tree collinear}
\end{figure}

Now let us discuss BFKL-type horizontal trajectories in the distribution frame. These operators are closely related to collinear functions that appear in the study of Regge scattering and factorization at small-$x$  \cite{Rothstein:2016bsq,Gao:2024qsg}. The collinear function \eqref{eq: collinear} in our context is
\be
C(p^+,k^\perp)=\fft{1}{{\rm Vol}\,\mathbb{R}^{d+1}}\fft{1}{(k^\perp)^{d-2}}\langle \phi(p)|\big(\int dx^-\sigma(x^+=0,x^-,k^\perp)\big)\big(\int dx^- \sigma(x^+=0,x^-,0)\big)|\phi(p)\rangle\,,
\ee
where $p=(p^+,0,0)$. Th leading order collinear function is illustrated in figure~\ref{fig: tree collinear}. It is given by
\be
C^{(0)}(p^+,k^\perp)=-\fft{1}{2N}\fft{1}{(2\pi)^{d-1}}\fft{1}{p^+ (k^\perp)^{d-2}}\,.
\ee

\begin{figure}
    \centering
    \begin{tikzpicture}
    \draw[->, thick] (-3,-2.5) -- (-2.25,-2.5);
     \draw[->, thick] (-1.5,-2.5) -- (-0.75,-2.5);
    \draw[->, thick] (1.5,-2.5) -- (2.25,-2.5);
    \draw[very thick, dashed,blue] (-1-0.5,0) -- (-1-0.5,1.5);
    \draw[very thick, dashed,blue] (1+0.5,0) -- (1+0.5,1.5);
    \draw[very thick, red] (-0.15,-2.7) -- (-0.15,1.7);
    \draw[very thick, red] (-0.25,-2.7) -- (-0.25,1.7);
    \draw[->, thick,blue] (-0.8-0.5,1) -- (-0.8-0.5,0.5);
    \draw[->, thick,blue] (0.8+0.5,0.5) -- (0.8+0.5,1);
    \node at (-0.45-2,0.8) {${\color{blue} k^\perp+\fft{\ell_3^- n}{2}}$};
    \draw[thick] (-1-0.5,0)--(1+0.5,-1.5);
     \draw[thick, ->] (-1-0.5,0) -- (-0.5,-0.5);
       \draw[thick, ->] (-1-0.5,-1.5) -- (0.5,-0.5);
     \draw[thick] (-1-0.5,0)--(-1-0.5,-1.5);
      \draw[thick,->] (-1-0.5,0)--(-1-0.5,-0.75);
      \draw[thick] (1+0.5,0)--(1+0.5,-1.5);
       \draw[thick,dashed,->] (-1-0.5,-1.5)--(-1-0.5,-2);
       \draw[thick] (1+0.5,0)--(-1-0.5,-1.5);
       \draw[thick,dashed] (-1-0.5,-1.5)-- (-1-0.5,-2.5);
        \draw[thick,dashed]  (1+0.5,-2.5)--(1+0.5,-1.5);
        \draw[thick,dashed,->]  (1+0.5,-2.5)--(1+0.5,-2);
        \draw[thick,->]  (1+0.5,-1.5)--(1+0.5,-0.75);
         \draw[thick] (-3,-2.5) -- (3,0-2.5);
         \node at (-2.25,-2.8) {$p$};
         \node at (-0.75,-2.8) {$p+\ell_1$};
          \node at (-1.5+0.3,-2) {$\ell_1$};
           \node at (-1.5+0.3,-2+1.25) {$\ell_2$};
            \node at (-1.5+0.6,-2+2.2) {$\ell_3-\ell_2$};
            \node at (-1.5+0.6+1.5,-2+2) {$\ell_2-\ell_1$};
             \node at (-1.5+0.3+3.8,-2+1.25) {$\ell_1-\ell_2+\ell_3$};
\end{tikzpicture}
\caption{The nontrivial $1/N$ correction to the collinear function of $\sigma$.}
\label{fig: BFKL collinear}
\end{figure}
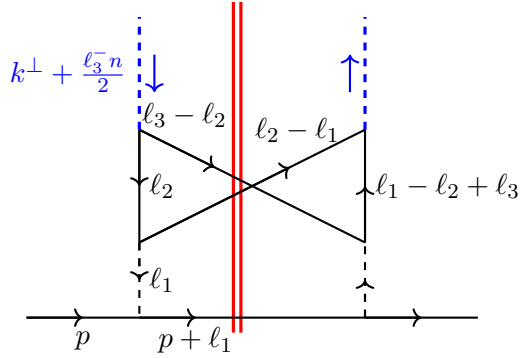

The only nontrivial next-to-leading order diagram is figure~\ref{fig: BFKL collinear}, which, as expected, has precisely the topology that appears in the  horizontal detector calculation in figure~\ref{fig: detector operators horizontal connect}. 
We have
\be
C^{(1)}(p^+,k^\perp)&=\fft{1}{N^2  (k^\perp)^{d-2}}\int \fft{d^d\ell_1 d^d\ell_2 d^d\ell_3}{(2\pi)^{3d}}\fft{G_\sigma(\ell_1^2-i 0)G_\sigma(\ell_1^2+i0)G_\sigma(\ell_3^2-i 0)G_\sigma(\ell_3^2+i0)}{(\ell_2^2-i0)((\ell_1-\ell_2+\ell_3)^2+i 0)}\nn\\
& \times (2\pi)^3 \delta^+((p+\ell_1)^2) \delta^+((\ell_2-\ell_1)^2) \delta^+((\ell_3-\ell_2)^2) \delta(\ell_3^+)\delta^{(d-2)}(k^\perp-\ell_3^\perp)\,.\label{eq: do Cs}
\ee
It is useful to parametrize $\ell_2^+ = \xi p^+$ and $\ell_1^+ = z p^+$. We see that a divergence arises from the limit $\ell_2^+ \to 0$, i.e., $\xi \to 0$. Note it is not regulated by adjusting the $\sigma$ scaling dimension. This is a rapidity divergence \cite{Collins:1981uk,Ji:2004wu,Chiu:2011qc,Chiu:2012ir} arising from integrating the $\sigma$ operators along exactly null lines. To extract it, we can take the limit $0 < \xi < z \ll 1$ 
\be
&C^{(1)}(p^+,k^\perp)= \frac{1}{8N^2 (2\pi)^{3(d-1)}}\fft{1}{p^+  (k^\perp)^{d-2}} \int_0^1 \fft{d\xi}{\xi} \int d^{d-2}\ell_1^\perp d^{d-2}\ell_2^\perp G_{\sigma}^2((k^\perp)^2)\nn\\
&\times G_{\sigma}^2((\ell_1^\perp)^2) \int_0^\infty dx \fft{1}{\Big(x (\ell_1^\perp)^2+(\ell_1^\perp-\ell_2^\perp)^2\Big)\Big(x (k^\perp-\ell_1^\perp)^2+(k^\perp-\ell_1^\perp+\ell_2^\perp)^2\Big)}\,,\label{eq: C1}
\ee
where $x=z/\xi -1$. Note that the kernel in the second line is precisely $C_\sigma^2 K_0(z_1,z_2;z_3,z_4)$ with
\be
z_{13}=(\ell_1^\perp)^2\,,\quad z_{23}=(\ell_1^\perp-\ell_2^\perp)^2\,,\quad z_{14}=(k^\perp-\ell_1^\perp)^2\,,\quad z_{24}=(k^\perp-\ell_1^\perp+\ell_2^\perp)^2\,.
\ee
However, this kernel is now defined on the transverse momentum plane, and we will call it $K_0^\perp(\ell_1^\perp,\ell_2^\perp;k^\perp)$. It can be embedded in the lightcone of $d$-dimensional space by stereographic projection $z = (1, (\ell^\perp)^2, \ell^\perp)$, where we translate one point to the origin, e.g.,
\be
z_1=(1,(\ell_1^\perp)^2,\ell_1^\perp)\,,\quad z_2=(1,(\ell_1^\perp-\ell_2^\perp)^2,\ell_1^\perp-\ell_2^\perp)\,,\quad z_3=(1,0,0)\,,\quad z_4=(1,(k^\perp)^2,k^\perp)\,.
\ee 
The first line of \eqref{eq: C1} exhibits the rapidity divergence $\int \frac{d\xi}{\xi}$. To regulate it, we follow \cite{Chiu:2011qc,Chiu:2012ir} and introduce a rapidity regulator 
\be
\int \frac{d\xi}{\xi} &\to \int \frac{d\xi}{\xi}\Big|\fft{p^+ \xi}{\nu}\Big|^{-\eta},\label{eq: rapidity regulator}
\ee
where $\eta$ is a small parameter that regulates the divergence, and $\nu$ is a rapidity renormalization scale.
Conceptually, this can be thought of as smearing the light-ray operator over a small window in $x^+$ 
\be
-\int \fft{d\delta x^+}{\nu} \sigma(\delta x^+,x^-,x^\perp)\left(\fft{\delta x^+}{\nu}\right)^{-1-\eta}\eta\, \theta(\delta x^+)\,,
\ee
see appendix \ref{app: rapidity regulator}.

We can then renormalize the rapidity divergence to find
\be
\nu \fft{d}{d\nu} C(p^+,k^\perp)=-\fft{C_\sigma^2}{N( 4\pi)^d \pi^{d-2}}\int d^{d-2}\ell_1^\perp d^{d-2}\ell_2^\perp \left(\fft{\ell_1^\perp}{k^\perp}\right)^{d-2} K_0^\perp(\ell_1^\perp,\ell_2^\perp;k^\perp) C(p^+,\ell_1^{\perp})\,.\label{eq: BFKL eq}
\ee
This is our BFKL-type evolution equation for the collinear function associated with horizontal detectors in the distribution frame. As is standard in BFKL physics, the leading order evolution kernel is diagonalized by restricting to irreducible representations of the transverse conformal group, which is equivalent to integrating it against a conformal three-point function $\<\varphi_{\Delta-1}(x^\perp)\varphi_1(x_1^\perp)\varphi_1(x_2^\perp)\>$. By conformal invariance, we can set $x^\perp=\oo$ without loss of generality, and then pass to Fourier space in $x_1^\perp,x_2^\perp$. This is equivalent to integraing our momentum-space kernel against a pure power of $k^\perp$:
%
%
\be
\langle\phi(p)|\mathcal{H}^{\rm dis}_{1-\Delta,{\rm BFKL}}|\phi(p)\rangle&=\int d^{d-2}k^\perp (k^\perp)^{5-\Delta-d} \langle\phi(p)|\bold{L}[\sigma](-\infty,k^\perp)\bold{L}[\sigma](-\infty z,-k^\perp)|\phi(p)\rangle\nn\\
&=\int d^{d-2}k^\perp (k^\perp)^{3-\Delta} C(p^+,k^\perp):=H_{\Delta,{\rm BFKL}}(p^+)\,.
\ee
The collinear function only depends on $|k^\perp|$, thus we have
\be
 (k^\perp)^{d-2}C(k^\perp)=\fft{1}{\Omega_{d-3}}\oint \fft{d\Delta}{2\pi i} H_{\Delta,{\rm BFKL}}(p^+) (k^\perp)^{\Delta-3}\,.
\ee
We can then diagonalize the BFKL-type equation \eqref{eq: BFKL eq} using \eqref{eq: dia1} with a particular choice of $z_i$
\be
\int d^{d-2}\ell_1^\perp K_0^{\perp}(\ell_1^\perp,\ell_2^\perp;k^\perp) (\ell_1^{\perp})^{\Delta-3}=\pi^{d-2}\kappa_0(1-\Delta) (k^\perp)^{\Delta-3}\,.
\ee
This indeed leads to the expected result (after regulating the integral over $k^\perp$, to account for IR divergences in the initial state)
\be
\nu \fft{d}{d\nu} H_{\Delta,{\rm BFKL}}(p^+)=-\fft{C_\sigma^2}{N(4\pi)^d}\kappa_0(1-\Delta) H_{\Delta,{\rm BFKL}}(p^+)\,.\label{eq: rapidity ano}
\ee

It is worth emphasizing that this diagonalization procedure applies to the particular case $x^\perp=\oo$ in \eqref{eq: BFKL dis}. This method of projecting onto irreducible representations of the conformal group is equivalent to the standard approach of solving the forward BFKL equation starting from a Mellin space ansatz $C(p^+,k^\perp)$ as (see e.g.\ \cite{Sterman:1995fz})
\be
C(p^+,k^\perp)=\oint \fft{d\Delta\, dJ}{(2\pi i)^2} \fft{h(\Delta,J)}{p^+} \left(\fft{p^+}{\nu}\right)^{J+1} \left(\fft{k^\perp}{\mu}\right)^{\Delta-d-1}\,.
\ee

\section{From Bethe-Salpeter equations to detectors}
\label{sec: BS equation}

So far, we have explicitly constructed light-ray operators in the $O(N)$ model and renormalized them to find their quantum numbers. However, it should also be possible to determine light-ray quantum numbers by computing correlation functions and extracting conformal data using the conformal block decomposition and analytic continuation in spin.


In this section, we study correlation functions in the critical O$(N)$ model, not at a fixed perturbative order but by resumming subsets of diagrams to all orders in $1/N$, in order to extract contributions of particular light-ray operators. In other words, we will study the large-$N$ limit at finite $1/N \log x$, where $x \to 0$ in either the lightcone limit or the Regge limit. This resummation of leading logarithms is achieved by using the Bethe-Salpeter equation. We will leverage conformal symmetry and harmonic analysis to perform the necessary loop integrals.


\subsection{Review: Bethe-Salpeter equation}

Recall that the Dyson equation resums self-energy diagrams
\be
G_2= G_2^{(0)}+  G_2\circ K_2\circ G_2^{(0)} \,,
\ee
where $\circ$ refers to the convolution with proper integral measure, $G^{(0)}$ refers to the Green function without interactions. We can graphically illustrate this equation as
\be
    \begin{tikzpicture}
    \draw[thick] (-3.5,0) -- (-2,0);
    \draw[thick] (-2,0) -- (-0.5,0);
   \node[draw, thick, circle, shading=radial, inner color=gray!30, outer color=white, minimum size=1cm, inner sep=0pt] (vertex) at (-2, 0) {$G_2$};
   \node at (0,0) {$=$};
   \draw[thick] (0.5,0) -- (2,0);
   \node at (2.5,0) {$+$};
   \draw[thick] (3,0) -- (3.5,0);
   \node[draw, thick, circle,  inner color=red!30, outer color=white, minimum size=1cm, inner sep=0pt] (vertex) at (5.5, 0) {$K_2$};
   \draw[thick] (4.5,0) -- (5,0);
   \node[draw, thick, circle, shading=radial, inner color=gray!30, outer color=white, minimum size=1cm, inner sep=0pt] (vertex) at (4, 0) {$G_2$};
    \draw[thick] (6,0) -- (7,0);
    \node at (1.25,0.4) {$G_2^{(0)}$};
    \node at (6.5,0.4) {$G_2^{(0)}$};
\end{tikzpicture}
\nn
\ee

The Bethe-Salpeter equation is essentially a generalization of the Dyson equation to a four-point function. It resums a subset of diagrams associated with two-particle exchange. Schematically, it is given by
\be
G_4=G_4^{(0)}+ G_4\circ K_4\circ G_4^{(0)}\,,
\ee
where $G_4$ is a resummation of two-particle exchange diagrams, and the kernel $K_4$ consists of 2-particle irreducible diagrams. In pictures:
\be
    \begin{tikzpicture}
   \node[draw, thick, circle, shading=radial, inner color=gray!30, outer color=white, minimum size=1cm, inner sep=0pt] (vertex) at (0, 0) {$G_4$};
   \draw[thick] (-0.35,0.35)--(-1,1);
   \draw[thick] (0.35,0.35)--(1,1);
   \draw[thick] (-0.35,-0.35)--(-1,-1);
   \draw[thick] (0.35,-0.35)--(1,-1);
   \node at (1.25,0) {$=$};
   \draw[thick] (-0.5+2+0.25,1)--(-0.5+2+0.25,-1);
    \draw[thick] (0.5+2+0.25+0.3,1)--(0.5+2+0.25+0.3,-1);
     \node at (2.1+0.3,0.5) {$G_4^{(0)}$};
     \node at (3.5,0) {$+$};
        \node[draw, thick, circle, shading=radial, inner color=gray!30, outer color=white, minimum size=1cm, inner sep=0pt] (vertex) at (0+4.5, 0+0.6) {$G_4$};
   \draw[thick] (-0.35+4.5,0.35+0.6)--(-1+4.5,1+0.6);
   \draw[thick] (0.35+4.5,0.35+0.6)--(1+4.5,1+0.6);
   \draw[thick] (-0.35+4.5,-0.35+0.6) to[out=210, in=150] (-0.35+4.5,-0.85+0.6);
\draw[thick] (0.35+4.5,-0.35+0.6) to[out=-30, in=30] (0.35+4.5,-0.85+0.6);
 \node[draw, thick, circle,  inner color=red!30, outer color=white, minimum size=1cm, inner sep=0pt] (vertex) at (4.5, -1.2+0.6) {$K_4$};
\draw[thick] (-0.35+4.5,-1.5-0.07+0.6)-- (-0.35+4.5,-2+0.08-0.25+0.6);
\draw[thick] (0.35+4.5,-1.5-0.07+0.6) -- (0.35+4.5,-2+0.08-0.25+0.6);
\node at (0.35+4.15,-2+0.08-0.5-0.25+0.6+0.2) {$G_4^{(0)}$};
\end{tikzpicture}
\nn
\ee

In the context of the critical $O(N)$ model, the Bethe-Salpeter equation will describe contributions to the data of the leading-twist operators $[(\phi^i\phi^j)_\rho]_J$. There also exists a Bethe-Salpeter equation that can describe contributions to horizontal trajectories like $\mathbf{L}[\sigma] \mathbf{L}[\sigma]$ \cite{Caron-Huot:2022eqs}; from a perturbative perspective, this corresponds to studying ``two-Reggeon bound states'' \cite{Gribov:2003nw}.

\subsection{Bethe-Salpeter equations for leading-twist anomalous dimensions}

Let us consider the four-point function $\langle\phi^i\phi^j\phi^k\phi^l\rangle$, which can be organized in terms of irreps of O$(N)$ in the $ij\to kl$ channel as follows:
\be
\langle \phi^i\phi^j\phi^k \phi^l\rangle = (G_{\rm sym}+G_{\rm asym})\delta^{il}\delta^{jk} + (G_{\rm sym}-G_{\rm asym})\delta^{ik}\delta^{jl} + (G_{\rm sing}-\fft{2}{N}G_{\rm sym})\delta^{ij}\delta^{kl}\,,\label{eq: correlator str}
\ee
where $G_{\rm sing}, G_{\rm sym}, G_{\rm asym}$ refer to the singlet, symmetric and antisymmetric representations, respectively. Let us discuss the (anti)symmetric and singlet reps separately.

\subsubsection{(Anti)symmetric sector}

Let us start by considering $1/N$ corrections to the symmetric and antisymmetric sectors. We can think of the Bethe-Salpeter equation in terms of a composition of conformally-invariant kernels. It is efficient to analyze this composition by diagonalizing the kernels, i.e. using harmonic analysis with respect to the conformal group. Thus, our initial goal will be to consider two-particle irreducible diagrams and decompose them into conformal partial waves.

\begin{figure}[h]
\centering
\begin{subfigure}[b]{0.45\textwidth}
\centering
 \begin{tikzpicture}
    \draw[thick] (-1.5,1.5) -- (-1.5,-1.5);
    \draw[thick] (1.5,1.5) -- (1.5,-1.5);
    \draw[thick,dashed] (-1.5,0) -- (1.5,0);
    \node at (-2.2+0.5,2-0.5) {$\phi^i$};
    \node at (2.25-0.5,2-0.5) {$\phi^j$};
    \node at (-2.2+0.5,-2+0.5) {$\phi^l$};
    \node at (2.25-0.5,-2+0.5) {$\phi^k$};
\end{tikzpicture}
\caption{}
\end{subfigure}
\begin{subfigure}[b]{0.45\textwidth}
\centering
 \begin{tikzpicture}
    \draw[thick] (-1.5,1.5) -- (-1.5,0);
    \draw[thick] (-1.5,0) -- (1.5,-1.5);
    \draw[thick] (1.5,1.5) -- (1.5,0);
    \draw[thick] (1.5,0) -- (-1.5,-1.5);
        \draw[thick,dashed] (-1.5,0) -- (1.5,0);
     \node at (-2.2+0.5,2-0.5) {$\phi^i$};
    \node at (2.25-0.5,2-0.5) {$\phi^j$};
    \node at (-2.2+0.5,-2+0.5) {$\phi^l$};
    \node at (2.25-0.5,-2+0.5) {$\phi^k$};
\end{tikzpicture}
\caption{}
\end{subfigure}
\caption{$t$-channel (a) and $u$-channel (b) leading contributions to four-point function $\langle\phi^i\phi^j\phi^k\phi^l\rangle$.}
\label{eq: signa sigma diagram}
\end{figure}
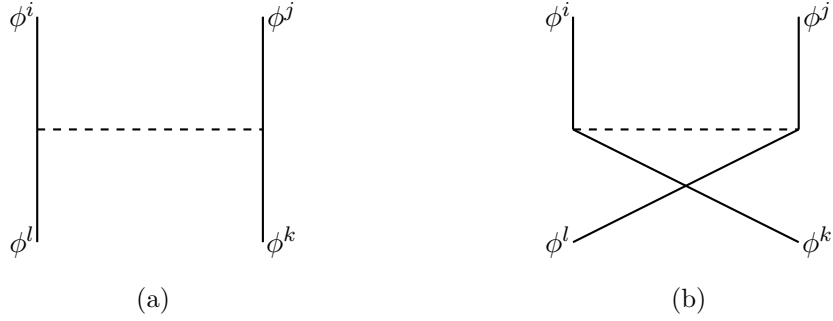

Let us start with diagrams contributing to $\langle \phi^i\phi^j\phi^k \phi^l\rangle$ with a single $\sigma$ exchange in the $t$-channel or $u$-channel (see Fig.~\ref{eq: signa sigma diagram}). 

Their sum contributes to the kernel $K_4$ for the symmetric sector, while their difference contributes to the antisymmetric sector. Note that the swap $1\leftrightarrow 2$ corresponds to an insertion of the phase $(-1)^J$ in the conformal partial wave decomposition. Thus it suffices to compute the partial wave decomposition of one of the diagrams.


Stripping off $O(N)$ indices, the $t$-channel diagram gives\footnote{We choose the normalization of $\langle\phi\phi\rangle$ as $N_\phi = \frac{1}{4} \pi^{-d/2} \Gamma\left(\frac{1}{2} (d-2)\right)$ so that, in momentum space, it is  $1/p^2$.}
\be
& G(x_1,x_2,x_3,x_4)= \fft{1}{N} \int d^dy_1 d^dy_2 \fft{N_\phi^4}{|x_1-y_1|^{d-2}|x_4-y_1|^{d-2}|x_2-y_2|^{d-2}|x_3-y_2|^{d-2}}\fft{N_\sigma}{|y_{12}|^4}.\nn\\
\label{eq: sym leading correlator}
\ee
We would like to express $G$ in terms of a conformal partial wave decomposition 
\be
G(x_1,x_2,x_3,x_4)= \sum_J \int_{\frac d 2}^{\frac d 2 + i\oo} \fft{d\Delta}{2\pi i} I_{{\rm (a)sym}}^{(1)}(\Delta,J)\Psi^{\Delta_\phi\Delta_\phi\Delta_\phi\Delta_\phi}_{\Delta,J}(x_1,x_2,x_3,x_4)\,.\label{eq: OPE leading-twist decomp}
\ee
To do so, we recognize the integral as a composition of two kernels
\be
F_1(x_1,x_2,y_1,y_2) &= \fft{1}{|y_{12}|^4 |x_1-y_1|^{d-2}|x_2-y_2|^{d-2}}\,.
\nn\\
F_2(y_1,y_2,x_3,x_4) &= \fft{1}{ |x_4-y_1|^{d-2}|x_3-y_2|^{d-2}}.
\ee
These kernels are products of two-point functions, and are closely related to the four-point function in free field theory, for which the harmonic decomposition is known \cite{Karateev:2018oml}. Using this, together with the composition formula for partial waves in terms of conformal bubble diagrams \cite{Karateev:2018oml}, we find
\be
I_{\rm (a)sym}^{(1)}(\Delta,J)=\fft{N_\phi^4 N_\sigma}{N}  
B_{\Delta,J} I_\mathrm{GFF}^{\Delta_\phi\Delta_\phi\Delta_\phi\Delta_\phi}(\Delta,J)^2 \frac{S^{\Delta_\phi\Delta_\phi}_{\tl \Delta,J}}{S^{\tl \Delta_\phi \tl \Delta_\phi}_{\tl \Delta,J}}\,,
\ee
where 
\be
& I_\mathrm{GFF}^{\Delta_1 \Delta_2 \Delta_1 \Delta_2}(\Delta,J) =\frac{\pi ^{-\frac{d}{2}} \Gamma (\Delta -1)^2 \Gamma \left(\frac{d}{2}+J\right)}{2 \Gamma (\beta -1)^2 \Gamma (J+1)}\nn\\
&\times \ft{\Gamma \left(\frac{1}{2} \left(\beta -\Delta _{12}\right)\right){}^2 \Gamma \left(\frac{1}{2} \left(\beta +\Delta _{12}\right)\right){}^2 \Gamma \left(\frac{d}{2}-\Delta _1\right) \Gamma \left(\frac{d}{2}-\Delta _2\right) \Gamma \left(\frac{1}{2} \left(-\tau +\Delta _1+\Delta _2\right)\right) \Gamma \left(\frac{1}{2} \left(-d+\beta +\Delta _1+\Delta _2\right)\right)}{\Gamma \left(\Delta _1\right) \Gamma \left(\Delta _2\right) \Gamma \left(\Delta -\frac{d}{2}\right)^2 \Gamma \left(\frac{1}{2} \left(d+\beta -\Delta _1-\Delta _2\right)\right) \Gamma \left(\frac{1}{2} \left(2 d-\tau -\Delta _1-\Delta _2\right)\right) \Gamma \left(\frac{1}{2} \left(d-\tau -\Delta _{12}\right)\right){}^2 \Gamma \left(\frac{1}{2} \left(d-\tau +\Delta _{12}\right)\right)^2}\,,
\ee
encodes the OPE data of generalized free field theory \cite{Fitzpatrick:2012yx,Karateev:2018oml} (where $\beta=\Delta+J, \tau=\Delta-J, \Delta_{12}=\Delta_1-\Delta_2$), the shadow coefficients $S^{\Delta_1\Delta_2}_{\Delta_3,J}$ are defined in appendix~\eqref{eq: shadow coefficients}, and the ``bubble coefficient" $B_{\Delta,J}$ is given by
\be
B_{\Delta,J}=\ft{2 \pi ^{\frac{3 d}{2}} \Gamma (d-1) \Gamma (J+1) \Gamma \left(\frac{d}{2}-\Delta \right) \Gamma \left(\Delta -\frac{d}{2}\right) (d-2)_J}{(d+2 J-2) \Gamma \left(\frac{d}{2}\right) \Gamma (\Delta -1) (\Delta +J-1) \Gamma (d-\Delta -1) (d-\Delta +J-1)  \Gamma (d+J-2)\left(\frac{d-2}{2}\right)_J}\,.
\ee

Let us also consider self-energy diagrams that modify the scaling dimension of $\phi$ to $1/2(d-2)+\gamma_\phi$. Their contribution to the partial wave decomposition is given by
\be
I_{{\rm (a)sym}}^{(2)}(\Delta,J)=I_{\rm GFF}^{\Delta_\phi=1/2(d-2)+\gamma_\phi}(\Delta,J)-I^{\Delta_\phi=1/2(d-2)}_{\rm GFF}(\Delta,J)\,.
\ee
We thus summarize the OPE data for (anti)symmetric leading-twist operators as $I_{\rm (a)sym}=I_{\rm (a)sym}^{(1)}+I_{\rm (a)sym}^{(2)}$.

This function contains both a single pole and a double pole in $1/(\Delta-(d-2+J))$, where the former encodes leading corrections to OPE coefficients and the latter encodes anomalous dimensions. This structure arises from perturbatively expanding poles in $\Delta$ in the full function $I^\mathrm{full}(\Delta,J)$ that encodes the OPE data:
\be
I^{\rm full}(\Delta,J)=\fft{(c^{(0)}_{\Delta,J}+\delta c_{\Delta,J})}{\Delta-(d-2+J+\gamma_{\Delta,J})}\simeq \fft{(c^{(0)}_{\Delta,J}+\delta c_{\Delta,J})}{\Delta-(d-2+J)}+\fft{c^{(0)}_{\Delta,J}\gamma_{\Delta,J}}{(\Delta-(d-2+J))^2}+\cdots\,.
\ee
The anomalous dimensions can be easily extracted by taking the coefficients of $1/\epsilon^2$ at $\Delta=d-2+J+2\epsilon$ as $\epsilon\rightarrow 0$, and dividing by the GFF OPE coefficients. Doing so, we recover the correct $\gamma_{{\rm (a)sym},J}$ for leading-twist operators in \eqref{eq: leading twist ano}. The corrections to OPE coefficients are encoded in the single pole $1/\epsilon$. We find
\be
& \de c_{\rm (a)sym}=\nn\\
& \ft{\pi ^{-d-1} \Gamma (d-2) \Gamma \left(\frac{d}{2}+J-1\right)^2 \Gamma (d+J-3)}{4N \Gamma \left(\frac{d}{2}-2\right) \Gamma \left(\frac{d}{2}+1\right) \Gamma (J+1) \Gamma (d+2 J-3)}\Big(H_{\frac{d}{2}+J-2}+\ft{4 (J-1) (d+J-2) \left(H_{d+J-4}-H_{d+2 J-4}\right)}{(d+2 J-4) (d+2 J-2)}-H_{\frac{d}{2}-2}+\ft{2 (d-2) d}{(d+2 J-4) (d+2 J-2)^2}\Big)\,.\label{eq: OPE coeff leading-twist}
\ee
which matches the known result in \cite{Lang:1992zw,Dey:2016mcs,Manashov:2017xtt,Alday:2019clp}.

The Bethe-Salpeter equation essentially sums these corrections to OPE data as a geometric series in the harmonic representation, leading to
\be
I^{\rm BS-resummed}_{\rm (a)sym}(\Delta,J)=\fft{I^{\Delta_\phi\Delta_\phi\Delta_\phi\Delta_\phi}_{\rm GFF}(\Delta,J)}{1-\fft{I_{\rm (a)sym}^{(1)}(\Delta,J)+I_{\rm (a)sym}^{(2)}(\Delta,J)}{I^{\Delta_\phi\Delta_\phi\Delta_\phi\Delta_\phi}_{\rm GFF}(\Delta,J)}}\,.
\ee
This resummation now exhibits a single pole located at the shifted location $\Delta=d-2+J+\gamma_{{\rm (a)sym},J}$, and its residue agrees with \eqref{eq: OPE coeff leading-twist} to leading order in $1/N$.

\subsubsection{Singlet sector: a single box diagram}

Let's now consider the singlet data. First, from the global structure of the correlator \eqref{eq: correlator str}, we easily deduce that \cite{Lang:1992zw,Dey:2016mcs,Manashov:2017xtt,Alday:2019clp}
\be
\de c_{\rm sing}=\fft{2c_{\rm (a)sym}^{(0)}}{N}\,.\label{eq: OPE sing}
\ee
This comes from the $O(1/N)$ diagram given by $\sigma$ exchange in the $s$-channel, which doesn't contribute to higher-spin leading-twist operators in the $s$-channel OPE. Such a diagram gives a contribution proportional to $\delta_{ij} \delta_{jk}$, which immediately implies \eqref{eq: OPE sing}.

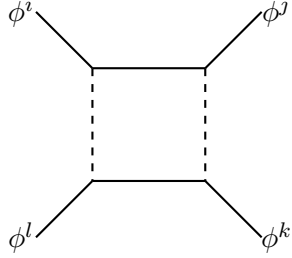
\begin{figure}[h]
\centering
 \begin{tikzpicture}
   \draw[thick] (-1.5,1.5)--(-0.75,0.75);
   \draw[thick] (1.5,1.5)--(0.75,0.75);
   \draw[thick] (-1.5,-1.5)--(-0.75,-0.75);
   \draw[thick] (1.5,-1.5)--(0.75,-0.75);
   \draw[thick] (-0.75,0.75)--(0.75,0.75);
    \draw[thick] (-0.75,-0.75)--(0.75,-0.75);
     \draw[thick,dashed] (-0.75,0.75)--(-0.75,-0.75);
     \draw[thick,dashed] (0.75,0.75)--(0.75,-0.75);
     \node at (-1.5-0.2,1.5) {$\phi^i$};
     \node at (1.5+0.2,1.5) {$\phi^j$};
     \node at (-1.5-0.2,-1.5) {$\phi^l$};
     \node at (1.5+0.2,-1.5) {$\phi^k$};
   \end{tikzpicture}
\caption{The box diagram that computes the additional anomalous dimension of the singlet leading-twist operator $[(\phi^i\phi^j)_s]_J$.}
\label{eq: singlet double pole}
\end{figure}

The global structure mixes different order in $1/N$ \eqref{eq: correlator str}, therefore it is not obvious to have a well-organized Bethe-Salpeter equation. The singlet anomalous dimension is encoded in the double pole of Fig \ref{eq: singlet double pole} in the structure of $c_{\rm sing} \gamma_{\rm sing}$. 

Explicitly, this diagram is 
\be
G^{s}(x_1,x_2,x_3,x_4)&=\fft{N_\phi^6}{N^2} \int d^dy_1 d^dy_2 d^dy_3 d^dy_4 \fft{1}{|y_{14}|^4 |y_{23}|^4}\times\nn\\
& \fft{1}{|x_1-y_1|^{d-2}|x_2-y_2|^{d-2}|y_{12}|^{d-2}}\fft{1}{|x_3-y_3|^{d-2}|x_4-y_4|^{d-2}|y_{34}|^{d-2}}\,.
\ee
Using harmonic analysis leads to 
\be
I^s(\Delta,J)= N_\phi^6 N_\sigma \left(I_{\rm GFF}^{\Delta_\phi \Delta_\phi \Delta_\phi\Delta_\phi}(\Delta,J)\fft{S^{\Delta_\phi \Delta_\phi}_{\tilde{\Delta},J}}{S^{2\Delta_\phi,2\Delta_\phi}_{\tilde{\Delta},J}}\right)^2 I_{\rm GFF}^{2222}(\Delta,J)B_{\Delta,J}^2\,.
\ee
The coefficient of its double pole $1/(\Delta-(d-2+J))^2$ precisely gives the correct result $\gamma_{{\rm sing},J}-\gamma_{{\rm (a)sym}}$.

\subsection{A Bethe-Salpeter equation for BFKL anomalous dimension}

We now use a Bethe-Salpeter equation to reproduce the BFKL-type anomalous dimension for the horizontal Regge trajectory, in the spirit of \cite{Caron-Huot:2022eqs} in the Wilson Fisher theory. The same method was also used to understand BFKL-type physics for $\phi^3$ theory \cite{Kirschner:1989pw}.

As we will justify later (see also the horizontal detector Fig \ref{fig: detector operators horizontal connect}), BFKL physics here is captured by the topology of crossed ladders. We can thus write down the Bethe-Salpeter equation for it, as also illustrated here

\be
 \begin{tikzpicture}
   \node[draw,thick, circle, shading=radial, inner color=gray!30, outer color=white, minimum size=1cm, inner sep=0pt] (vertex) at (0, 0) {$G^{\rm BFKL}$};
   \draw[thick,dashed] (-0.4,0.4)--(-1,1);
   \draw[thick,dashed] (0.4,0.4)--(1,1);
   \draw[thick,dashed] (-0.4,-0.4)--(-1,-1);
   \draw[thick,dashed] (0.4,-0.4)--(1,-1);
   \node at (1.25,0) {$=$};
   \draw[thick,dashed] (-0.5+2+0.25,1)--(-0.5+2+0.25,-1);
    \draw[thick,dashed] (0.5+2+0.25+0.3,1)--(0.5+2+0.25+0.3,-1);
     \node at (3.75,0) {$+$};
        \node[draw, thick, circle, shading=radial, inner color=gray!30, outer color=white, minimum size=1cm, inner sep=0pt] (vertex) at (0+4.5+0.5, 0+0.6-0.6+0.5) {$G^{\rm BFKL}$};
   \draw[thick,dashed] (-0.35+4.5+0.5-0.05,0.35+0.6-0.6+0.05+0.5)--(-1+4.5+0.5,1+0.6-0.6+0.5);
   \draw[thick,dashed] (0.4+4.5+0.5,0.4+0.6-0.6+0.5)--(1+4.5+0.5,1+0.6-0.6+0.5);
\draw[thick,dashed] (-0.45+4.5+0.5,-1.5-0.07+0.6+1.2-0.6+0.5)-- (-0.45+4.5+0.5,-1+0.5);
\draw[thick,dashed] (0.45+4.5+0.5,-1.5-0.07+0.6+1.2-0.6+0.5) -- (0.45+4.5+0.5,-1+0.5);
\draw[thick] (-0.45+4.5+0.5,-1+0.13+0.5)-- (-0.45+4.5+0.5,-1+0.13-0.5+0.5);
\draw[thick] (-0.45+4.5+0.5,-1+0.13+0.5)-- (0.45+4.5+0.5,-1+0.13-0.5+0.5);
\draw[thick] (-0.45+4.5+0.5,-1+0.13-0.5+0.5)-- (0.45+4.5+0.5,-1+0.13+0.5);
\draw[thick] (0.45+4.5+0.5,-1+0.13+0.5)-- (0.45+4.5+0.5,-1+0.13-0.5+0.5);
\draw[thick,dashed] (-0.45+4.5+0.5,-1+0.13-0.5+0.5)-- (-0.45+4.5+0.5,-1+0.13-1+0.5);
\draw[thick,dashed] (0.45+4.5+0.5,-1+0.13-0.5+0.5)-- (0.45+4.5+0.5,-1+0.13-1+0.5);
\end{tikzpicture}
\nn
\ee
Explicitly, we have 
\be
& G^{\rm BFKL}(x_1,x_2,x_3,x_4)=\fft{1}{|x_{14}|^4 |x_{23}|^4}\nn\\
& +\fft{1}{N}\int d^dy_1d^dy_2 d^dy_3 d^dy_4 G^{\rm BFKL}(x_1,x_2,y_1,y_2)\fft{1}{|x_3-y_3|^4 |x_4-y_4|^4} K^{\rm BFKL}(y_1,y_2,y_3,y_4)\,,\label{eq: Kernal BFKL}
\ee
where the BFKL-type kernel is
\be
K^{\rm BFKL}(y_1,y_2,y_3,y_4)
&=\fft{N_\phi^4}{|y_{14}|^{d-2}|y_{13}|^{d-2}|y_{23}|^{d-2}|y_{24}|^{d-2}}
\nn\\
&=\sum_J \int_{\frac d 2}^{\frac d 2 + i \oo} \fft{d\Delta}{2\pi i} I_{K}(\Delta,J)\Psi^{2\Delta_\phi,2\Delta_\phi,2\Delta_\phi,2\Delta_\phi}_{\Delta,J}(y_1,y_2,y_3,y_4)\,.
\ee
The harmonic decomposition of $K^\mathrm{BFKL}$ cannot be expressed in a simple way in terms of GFF OPE data. Instead, we computed it using the Lorentzian inversion formula \cite{Caron-Huot:2017vep}, $I_K(\Delta,J)$; see appendix \ref{app: sole BS} for details. The result is
\begin{align}
& I_K(\Delta,J)=\small{(d-\Delta +J-1) \cos \left(\frac{1}{2} \pi  (2 d+J)\right) 
\Gamma \left(\frac{1}{2} (-J+\Delta -2)\right) 
\Gamma \left(\frac{1}{2} (d-J-\Delta -2)\right)}\nn\\
&  \times
\ft{\pi ^{\frac{3}{2}-\frac{5 d}{2}} 2^{-J-10} (d-2 \Delta ) \cot \left(\frac{\pi  d}{2}\right) 
\sec \left(\frac{\pi  \Delta }{2}\right) (\Delta +J-1) 
\sin \left(\frac{1}{2} \pi  (d-2 \Delta )\right) 
\Gamma \left(\frac{1}{2} (d-\Delta -1)\right)\Gamma \left(d+\frac{J}{2}-\frac{\Delta }{2}-2\right) 
\Gamma \left(\frac{1}{2} (d+J+\Delta -4)\right)}{\Gamma \left(\frac{3}{2}-\frac{\Delta }{2}\right) 
\Gamma \left(\frac{J}{2}+1\right) \left(\sin \left(\frac{3 \pi  d}{2}+\pi  J\right)-\sin \left(\frac{\pi  d}{2}\right)\right) 
\Gamma \left(-\frac{d}{2}-J+1\right) \Gamma \left(\frac{1}{2} (d+J-1)\right)}\,.\label{eq: OPE IK}
\end{align}
Thus, the solution of this Bethe-Salpeter equation is
\be
I^{\rm BFKL}(\Delta,J)=\fft{N_\sigma^2 I^{2222}_{\rm GFF}(\Delta,J)}{1-\fft{1}{N}B_{\Delta,J}^2 N_\sigma^2 I^{2222}_{\rm GFF}(\Delta,J) I_K(\Delta,J)}\,.
\ee
This OPE data shows a singularity at
\be
J=-1+\fft{C_\sigma^2}{N (4\pi)^d}\kappa_0(1-\Delta)\,,
\ee
which indicates that these diagrams capture the same physics as the detector calculations \eqref{eq: gammaL} and \eqref{eq: rapidity ano}.

\subsection{Comments: from correlators to detectors}

We end this section by commenting on the intuitive picture of how detectors relate to correlation functions, closely following \cite{Caron-Huot:2022eqs}.

\begin{figure}[t]
\centering
\begin{subfigure}[b]{0.45\textwidth}
\centering
 \begin{tikzpicture}
   \draw[thick] (-1.5,-1.5)--(1.5,1.5);
   \fill (0,0) circle (0.075);
   \fill (-1.5,-1.5) circle (0.075);
   \fill (1.5,1.5) circle (0.075);
  \draw[thick, dashed] (0,0)--(-0.5,0.5);
  \node at (-1.5-0.2,-1.5) {$1$};
  \node at (1.5+0.2,1.5) {$2$};
  \node at (0.4,0) {$\fft{1}{\sqrt{N}}$};
\end{tikzpicture}
\caption{}
\label{eq: detector from correlator a}
\end{subfigure}
\begin{subfigure}[b]{0.45\textwidth}
\centering
 \begin{tikzpicture}
   \draw[thick,dashed] (-1.5,-1.5)--(-0.71,-0.71);
    \draw[thick,dashed] (1.5,1.5)--(0.71,0.71);
     \draw[thick] (0,0) circle (1);
   \fill (-0.71,0.71) circle (0.075);
    \fill (0.71,-0.71) circle (0.075);
   \fill (-1.5,-1.5) circle (0.075);
   \fill (1.5,1.5) circle (0.075);
  \draw[thick, dashed] (-0.71,0.71)--(-0.71-0.5,0.71+0.5);
  \draw[thick, dashed] (0.71,-0.71)--(0.71+0.5,-0.71-0.5);
  \node at (-1.5-0.2,-1.5) {$1$};
  \node at (1.5+0.2,1.5) {$2$};
\end{tikzpicture}
\caption{}
\label{eq: detector from correlator b}
\end{subfigure}
\caption{(a) Two scalars $\phi^i$ create a $\sigma$ that is smeared over a null line to create a light-ray operator when two $\phi^i$ are boosted apart. (b) When two $\sigma$ are boosted, the same mechanism creates a horizontal detector $\bold{L}[\sigma]\bold{L}[\sigma]$.}
\label{eq: detector from correlator}
\end{figure}
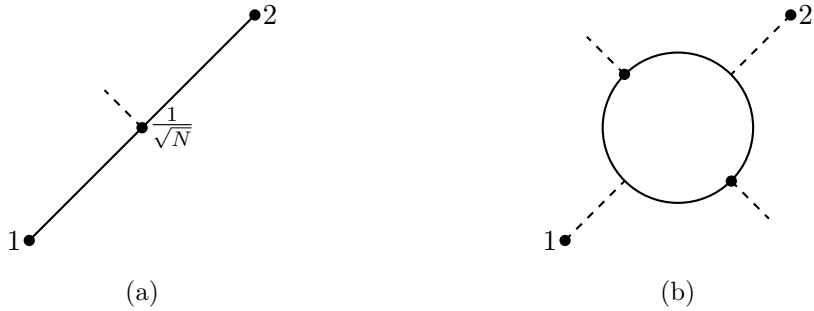

Consider operators $\phi^i(x_1)$ and $\phi^j(x_2)$. They can create a $\sigma$ operator in their causal diamond via a $1/\sqrt{N} \sigma \phi^i \phi^i$ interaction. As we boost them apart to the lightcone limit, this $\sigma$ operator is constrained to exist only on the null line, where it gets smeared. This mechanism creates a detector $\bold{L}[\sigma]$. See Fig.\ref{eq: detector from correlator a} for an illustration. One can also intuitively think of this picture as a particle created by $\phi$ splitting into $\sigma$, which is later detected. This picture receives corrections from interactions that trigger recurrent collinear emissions, all of which can be incorporated into the four-point function $\langle\phi^l\phi^i\phi^j\phi^k\rangle$ by taking $\phi^i$ and $\phi^j$ to the lightcone limit. From analytic conformal bootstrap analysis \cite{Fitzpatrick:2012yx,Komargodski:2012ek}, we then know that this procedure is essentially controlled by the $\phi^i\phi^j$ OPE and thus produces the leading-twist operators $[(\phi^i\phi^j)_\rho]_J$ smeared over the null line. This can also be embedded into the DIS picture, where we describe $1/N \langle\phi_l(p)|(\sigma\phi_i)(q)(\sigma\phi_j)(-q)|\phi_k(p)\rangle$ and focus on the contribution of leading-twist detectors.

Horizontal detectors are created in a similar way. We now take two $\sigma$ operators and boost them away, which creates a pair of detectors $\bold{L}[\sigma]\bold{L}[\sigma]$; see Fig.\ref{eq: detector from correlator b}. It is now clear that this topology can be embedded inside the four-point function $\langle\sigma\sigma\sigma\sigma\rangle$ and soft emissions result in repeated nesting of this diagram.

\section{Summary}
\label{sec: summary}

We have explored light-ray operators in the $O(N)$ model, focusing on the leading-twist trajectory (DGLAP-type), its shadow trajectory, and a leading horizontal trajectory (BFKL-type) in various dimensions. We furthermore explored these operators in two different frames: the detector frame, where light-ray operators measure fluxes in final states, and the distribution frame, where matrix elements of light-ray operators define PDFs, collinear functions, and their generalizations. We renormalized light-ray operators to leading order in $1/N$, extracting spacelike and timelike anomalous dimensions, related by reciprocity.

The complete renormalization of leading-twist detectors at order $1/N$ allowed us to resolve its mixing with its shadow trajectory and extract the Regge intercept of the theory. Our results straightforwardly generalize those of the Wilson-Fisher theory ($O(1)$ model) in $d = 4 - \epsilon$ \cite{Caron-Huot:2022eqs} to the critical $O(N)$ model in general dimensions.

We also identified matrix elements of distribution operators with observables commonly studied in perturbative QCD, including parton distribution functions and collinear functions, allowing us to meaningfully interpret these objects within CFT. We established this identification at the level of bare operators and further analyzed RG evolution after renormalization, deriving leading-twist splitting functions and BFKL-type evolution kernels that align with the results obtained from detector operators. In this way, we built a bridge linking the language of perturbative QCD to formal structures in CFT, with the critical $O(N)$ model as an example.

Finally, we leveraged the full power of conformal symmetry and showed that DGLAP-type anomalous dimensions and BFKL-type anomalous spins are encoded in conformal correlators, which we extracted using harmonic analysis and the Lorentzian inversion formula.

Our study of the critical $O(N)$ model has been limited to the leading-twist trajectories and a single horizontal trajectory in the singlet sector. We expect that our approach can be extended to other families of trajectories, such as higher-twist operators of the form $[\sigma\, (\phi^i \phi^j)_\rho]_J$ (following the construction of \cite{Henriksson:2023cnh}), $O(N)$ vector families like $[\sigma\, \phi^i]_J$, and additional singlet families such as $[\sigma \sigma]_J$. Understanding these trajectories is essential for further clarifying the Chew-Frautschi plot of the critical $O(N)$ model. For example, there is a triple intersection involving the BFKL-type trajectory, the shadow of the leading DGLAP-type trajectory, and the $[\sigma \sigma]_J$ trajectory, which it would be interesting to resolve. This situation is better than in the Wilson-Fisher case, where the BFKL-type trajectory mixes with the shadow of the DGLAP-type and several subleading multi-twist families \cite{Caron-Huot:2022eqs}.

The singlet sector of the critical $O(N)$ model is conjectured to be dual to higher-spin gravity in AdS \cite{Klebanov:2002ja,Sezgin:2003pt,Leigh:2003gk}, a correspondence that has passed several checks at the level of correlation functions and parition functions, e.g., \cite{Giombi:2009wh,Giombi:2012ms,Giombi:2013fka,Giombi:2014iua,Giombi:2011ya}. It would be interesting to explore Regge trajectories of light-ray operators in this context.  This may offer new insights into higher-spin holography, particularly for testing its locality through the lens of Regge theory \cite{Caron-Huot:2022lff} using conformal four-point functions \cite{Leonhardt:2003du,Turiaci:2018nua}.

The $O(N)$ model and its variants may offer valuable windows into understanding the RG flow of Regge trajectories, detectors, and distribution operators from a UV CFT to an IR CFT. The structure of such an RG flow may take the form $\mathcal{D}_{\rm IR} = \sum_{\cD_{\rm UV}}C_{\cD_{\rm UV}}\, \mathcal{D}_{\rm UV}$, where the Wilson coefficients $C_{\cD_{\rm UV}}$ must satisfy matching conditions at intermediate scales along the flow. This calls for repeating the analysis presented in this paper in perturbative studies of $O(N)$ models, with or without cubic interactions \cite{Fei:2014yja,Fei:2014xta,Gracey:2015tta}, using the $4 - \epsilon$ and $6 - \epsilon$ expansions. It has been found that the $O(N)$ model develops non-unitarity for $4 < d < 6$ due to instanton effects \cite{Giombi:2019upv}. It is therefore also interesting to explore whether this non-unitarity can be directly reflected in Regge trajectories and non-local operators. Additionally, it would be interesting to generalize these investigations to other models, such as the Gross-Neveu-Yukawa model.

The most exciting future direction is to broadly understand how detector operators and distribution operators, and more generally light-ray operators, are related to observables in QCD. We have taken a first step in this direction by focusing on a simple scalar model, establishing connections to parton distribution functions and collinear functions for scalars. A natural next step would be to explore this bridge to perturbative QCD.

There are many other collider observables, such as TMD PDFs, various jet functions, and soft functions, which play important roles in the study of small-$x$ physics, Regge physics, and energy correlators. It would be valuable to develop precise and rigorous non-local operator definitions for these quantities. Such a bridge may offer a cleaner computational framework for understanding these phenomenologically relevant observables, even in non-perturbative regimes. This may also provide insights into high-energy gravitational observables, as recently explored in \cite{Rothstein:2024nlq,Herrmann:2024yai}.

It is also important to understand how detector and distribution operators can eventually exponentiate, in the spirit of defect operators \cite{Andrei:2018die}, which may provide insights into saturation physics \cite{Iancu:2003xm}. For example, the detectors renormalized by Balitsky-Kovchegov equation in saturation physics might be understood as the exponentiated detectors renormalized by the BFKL equation \cite{Caron-Huot:2013fea}.

\acknowledgments We thank Cyuan-Han Chang, Hao Chen, Gabriel Cuomo, Simone Giombi, Igor Klebanov, Petr Kravchuk, Simon Caron-Huot, Ian Moult, Sebastian Mizera, and Hua Xing Zhu for useful discussions. We are also grateful to Simon Caron-Huot and Sven Moch for their comments on an earlier version of the draft. The work of Y.Z.L is supported in part by the US National Science Foundation under Grant No. PHY- 2209997, and in part by Simons Foundation grant No. 917464. DSD is supported in part by Simons Foundation grant 488657 (Simons Collaboration on the Nonperturbative Bootstrap) and the U.S. Department of Energy, Office of Science, Office of High Energy Physics, under Award Number DE-SC0011632.

\newpage

\appendix 

\section{Wavefunction renormalization in the critical O$(N)$ model}
\label{app: wavefunction RG}

In this appendix, we revisit the scaling dimension regulator for the critical O$(N)$ model by re-deriving the anomalous dimension of a single scalar $\phi_i$.

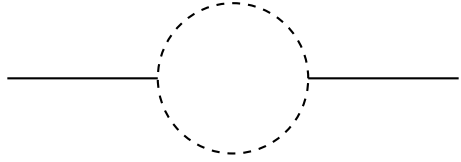
\begin{figure}[h]
\centering
 \begin{tikzpicture}
 \draw[thick] (-2,0) -- (0,0);
 \draw[thick,dashed] (1,0) circle (1);
  \draw[thick] (2,0) -- (4,0);
 \end{tikzpicture}
 \caption{One-loop self-energy diagram that computes the anomalous dimension of $\phi^i$.}
 \label{eq: self-energy}
\end{figure}

The strategy is to study the self-energy diagram of the O$(N)$ scalar $\phi_i$. At order $1/N$, we have a simple diagram, as shown in Fig \ref{eq: self-energy}. 

The self-energy diagram is divergent, which cannot be cured by dimensional regularization, as the entire calculation is performed in a generic dimension $d$. It is therefore useful to consider a regulator that shifts $\Delta_\sigma\rightarrow 2-\epsilon$ and then takes $\epsilon\rightarrow 0$. This procedure then gives.
\be
\Pi_\phi(p^2)&=\fft{1}{p^4} \int \fft{d^d\ell}{(2\pi)^d} \fft{G_\sigma((p-\ell)^2)}{\ell^2}\nn\\
&=\frac{\pi ^{\epsilon -\frac{1}{2}} 2^{d+4 \epsilon -2} \Gamma \left(\frac{d}{2}-1\right) p^{-2-2 \epsilon } \mu ^{2 \epsilon } \csc (\pi  \epsilon ) \sin \left(\frac{1}{2} \pi  (d+2 \epsilon -2)\right) \Gamma \left(\frac{d}{2}+\epsilon -\frac{1}{2}\right)}{N \Gamma \left(\frac{d}{2}-\epsilon +1\right) \Gamma \left(\frac{d}{2}+\epsilon -2\right)}\,.\label{eq: pert Pi}
\ee
This result contains a UV divergence when taking $\epsilon \rightarrow 0$, requiring wavefunction renormalization
\be
\phi_{R}=Z^{-\fft{1}{2}}\phi\,,\quad Z=1+\Pi(p=1)\simeq 1-\fft{1}{ \epsilon}\frac{3\ 2^{d-5} (d-4) \sin \left(\frac{\pi  d}{2}\right) \Gamma \left(\frac{d-1}{2}\right)}{\pi ^{3/2} N  \Gamma \left(\frac{d}{2}+1\right)}\,,\label{eq: Z phi}
\ee
to absorb the divergence. Following the standard renormalization procedure, the anomalous dimension of $\phi$ is given by $\gamma_\phi= - \mu \fft{dZ_\phi}{d\mu}$. Thus, we find
\be
\gamma_\phi= \fft{1}{N}\frac{2 \sin \left(\frac{\pi  d}{2}\right) \Gamma (d-2)}{\pi  \Gamma \left(\frac{d}{2}-2\right) \Gamma \left(\frac{d}{2}+1\right)}\,.\label{eq: ano phi}
\ee
which agrees with \cite{Giombi:2016hkj}.

An alternative way to find this answer is to use the ``EFT'' matching method. Recall that we know the exact expression for the scalar two-point function $\langle \phi(p)\phi(-p)\rangle$ in CFT.
\be
\langle \phi(p)\phi(-p)\rangle\sim p^{2\Delta-d}=p^{\gamma_\phi-2}\,.
\ee
We can thus expand the two-point function perturbatively in $\gamma_\phi$ and then match it to the perturbative calculation in \eqref{eq: pert Pi}, obtaining \eqref{eq: ano phi}.

\section{Lorentzian inversion formula and diagonalizing kernels}
\label{app: Lorentzian inversion}

In this appendix, we review the Lorentzian inversion formula \cite{Caron-Huot:2017vep} and use it to extract the OPE data that appear in the main text.

\subsection{Revisiting Lorentzian inversion formula}

Conformal four-point functions can be decomposed into single-valued conformal partial waves
\be
 \langle\phi_1(x_1)\phi_2(x_2)\phi_3(x_3)\phi_4(x_4)\rangle
 &=\fft{\left(\fft{x_{24}}{x_{14}}\right)^{\Delta_{12}}\left(\fft{x_{14}}{x_{13}}\right)^{\Delta_{34}}}{x_{12}^{\Delta_1+\Delta_2}x_{34}^{\Delta_3+\Delta_4}}\mathcal{G}(z, \bar z)
 \nn\\
 &=\sum_J \int_{\frac d 2}^{\frac d 2 + i \oo} \fft{d\Delta}{2\pi i} I(\Delta,J)\Psi^{\Delta_i}_{\Delta,J}(x_i)\,.
\ee
Here, $\Psi^{\Delta_i}_{\Delta,J}(x_i)$ is the conformal partial wave. It is essentially a linear combination of the conformal block and its shadow block, ensuring that it is single-valued in Euclidean signature:
\be
\Psi^{\Delta_i}_{\Delta,J}(x_i)=S_{\tl{\Delta},J}^{\Delta_3,\Delta_4}G_{\Delta,J}^{\Delta_i}(x_i)+S_{\Delta,J}^{\Delta_1,\Delta_2}G_{\tl{\Delta},J}^{\Delta_i}(x_i)\,,
\ee
where $S_{\Delta,J}^{\Delta_1,\Delta_2}$ is the shadow coefficient \cite{Simmons-Duffin:2012juh} (for its spinning generalization, see \cite{Karateev:2018oml})
\be
S_{\Delta,J}^{\Delta_1,\Delta_2}=\frac{\pi ^{d/2} \Gamma \left(\Delta -\frac{d}{2}\right) \Gamma (J+\Delta -1) \Gamma \left(\frac{1}{2} \left(d+J-\Delta +\Delta _1-\Delta _2\right)\right) \Gamma \left(\frac{1}{2} \left(d+J-\Delta -\Delta _1+\Delta _2\right)\right)}{\Gamma (\Delta -1) \Gamma \left(\frac{1}{2} \left(J+\Delta +\Delta _1-\Delta _2\right)\right) \Gamma \left(\frac{1}{2} \left(J+\Delta -\Delta _1+\Delta _2\right)\right) \Gamma (d+J-\Delta )}\,.\label{eq: shadow coefficients}
\ee
The OPE data $I(\Delta,J)$ encode the information of the physical spectrum and OPE coefficients appearing in the conformal block decomposition of $\cG$. More precisely, its physical poles encode the primary operators that are exchanged, while the residues give OPE coefficients
\be
c(\Delta,J)=I(\Delta,J)S_{\tl{\Delta},J}^{\Delta_3,\Delta_4}\sim \fft{c_{\Delta_{\rm phys},J}}{\Delta-\Delta_{\rm phys}}\,.
\ee

Reference \cite{Caron-Huot:2017vep} provides an elegant way to systematically compute $c(\Delta,J)$ from the double discontinuity of four-point correlators. This formula is known as the Lorentzian inversion formula:
\be
 c^t(\Delta,J)=\fft{\kappa_{\Delta+J}}{4}\int_0^1 dzd\bar{z}\mu^{ab}(z,\bar{z})G_{J+d-1,\Delta-d+1}^{ab}(z,\bar{z}){\rm dDisc}[\mathcal{G}(z,\bar{z})]\,,
\ee
where $(z,\bar{z})$ are the cross-ratios. This formula reconstructs the OPE data 
\be
c(\Delta,J)=c^t(\Delta,J)+(-1)^J c^u(\Delta,J)\,,
\ee
where the $u$-channel contribution is the same but with the integration over the $u$-channel configuration. The ingredients of the Lorentzian inversion formula are given below
\be
\mu^{ab}(z,\bar{z})=\left|\fft{z-\bar{z}}{z\bar{z}}\right|^{d-2}\fft{\big((1-z)(1-\bar{z})\big)^{a+b}}{(z\bar{z})^2}\,,\quad \kappa_{\beta} = \frac{\Gamma(\frac{\beta}{2} - a) \Gamma(\frac{\beta}{2} + a) \Gamma(\frac{\beta}{2} - b) \Gamma(\frac{\beta}{2} + b)}{2\pi^2 \Gamma(\beta - 1) \Gamma(\beta)}\,,
\ee
where $a=\frac 1 2 (\Delta_2-\Delta_1)$ and $b=\frac 1 2 (\Delta_3 - \Delta_4)$.
The double discontinuity is essentially the double commutator \cite{Caron-Huot:2017vep,Simmons-Duffin:2017nub}, and it is computed by
\be
\text{dDisc}[\mathcal{G}(z,\bar{z})] = \cos(\pi (a + b)) \mathcal{G}(z,\bar{z}) - \frac{e^{-i(a+b)}}{2} \mathcal{G}^{\circlearrowleft}(z,\bar{z}) - \frac{e^{i(a+b)}}{2} \mathcal{G}^{\circlearrowright}(z,\bar{z})\,.
\ee

\subsection{Diagonalizing the kernel $K_\alpha$}
\label{app: diagonalize K}

We now use the Lorentzian inversion formula to diagonalize the kernel $K_\alpha$ for horizontal detectors in \eqref{eq: horizontal K} and then obtain \eqref{eq: dia1}.

We should think of $K_\alpha$ as a conformal four-point function in $(d-2)$-dimensions, where the celestial sphere realizes its embedding space. Writing it explicitly, we find
\be
K_\alpha(z_i)&=\langle\varphi_{d-3-\alpha+2\epsilon}(z_1)\varphi_{d-3+\alpha+2\epsilon}(z_2)\varphi_1(z_3)\varphi_1(z_4)\rangle \nn\\
&=\frac{z_{14}^{\alpha } z_{24}^{-\alpha } z_{12}^{-d-2 \epsilon +3}}{z_{34}}\frac{\pi  u \csc (\pi  \alpha ) 2^{d+2 \epsilon -4} v^{-\alpha } \left(v^{\alpha }-1\right)}{v-1}\,,
\ee
where $u=z_{12}z_{34}/(z_{13}z_{24}), v=z_{14}z_{23}/(z_{13}z_{24})$ are cross-ratios. We then need to figure out its OPE data
\be
K_{\alpha}(z_i)=\sum_J \int_{\frac{d-2}{2}}^{\frac {d-2}{2}+i\oo} \fft{d\delta}{2\pi i}I_\alpha(\delta,j)\Psi^{d-3-\alpha+2\epsilon,d-3+\alpha+2\epsilon,1,1}_{\Delta,J}(z_i)\,.
\ee

One might now conclude that the Lorentzian inversion formula doesn't help because $K_\alpha$ naively has vanishing dDisc. However, it is not correct that the dDisc is vanishing --- it has distributional terms localized at the $t$-channel OPE limit. We can use a simple trick to extract their contribution: we deform $a=\alpha\rightarrow \alpha-\gamma$, which makes the dDisc nonzero for generic cross ratios, and then later take $\gamma\to 0$. The dDisc of the deformed correlator is
\be
\hat{{\rm dDisc}}[\mathcal{G}_K]=\frac{2 \pi  z \bar{z} \sin (\pi  \gamma) (1-z)^{-\alpha } (1- \bar{z} )^{-\alpha }}{z (- \bar{z} )+z+ \bar{z} }\,.
\ee
To perform the integral, we expand the integrand of the Lorentzian inversion formula in $z \rightarrow 0$ to arbitrarily high orders. The integral over $z$ at each order contributes to different twists $n$ in $\delta = 2 + 2n + j$ by automatically picking up the corresponding residues. For all $n \in \mathbb{Z}$, we find that the OPE coefficients exhibit the general structure
 \be
\sin(\gamma\pi)\Big(\fft{1}{j}+\text{regular in $j$}\Big)\,.
\ee
The singularity at $j=0$ is a $(d-2)$-dimensional Regge pole, indicating nontrivial physics at $j=0$. We therefore regulate $j=0$ by setting $j=\gamma$ and then taking the limit $\gamma\rightarrow 0$, which gives rise to the celestial OPE coefficients as follows
\be
c_{2+2n,0}\sim \frac{\sqrt{\pi } 2^{1-2 n} \Gamma \left(\frac{d}{2}-2 n+1\right) \Gamma (n-\alpha ) \Gamma (n+\alpha )}{\Gamma \left(n+\frac{1}{2}\right) \Gamma \left(\frac{d}{2}-n\right)}\,.
\ee
The Watson-Sommerfeld resummation allows us to reconstruct the full analytic OPE data by $I_\alpha(\delta,0) S^{\delta_3,\delta_4}_{\tl{\delta}}=c(\delta,0)$, where
\be
c(\delta,0)=\fft{\pi ^2 \alpha  2^{3-2 \delta } \csc (\pi  \alpha ) \csc \left(\frac{\pi  \delta }{2}\right) \left(2-\frac{d}{2}\right)_{\frac{\delta }{2}-1} (1-\alpha )_{\frac{\delta }{2}-1} (\alpha +1)_{\frac{\delta }{2}-1}}{\left(\frac{3}{2}\right)_{\frac{\delta }{2}-1} \left(1-\frac{d}{4}\right)_{\frac{\delta }{2}-1} \left(\frac{3}{2}-\frac{d}{4}\right)_{\frac{\delta }{2}-1}}\,.
\ee
We have verified that this OPE data is shadow symmetric. 

To obtain $\kappa_\alpha$, we again use the conformal bubble integral, but now in a $(d-2)$-dimensional celestial space \cite{Karateev:2018oml}
\be
& \int D^{d-2}z_1 D^{d-2}z_2\langle \varphi_{\delta}(z_5)\varphi_{d-3+2\epsilon-\alpha}(z_1)\varphi_{d-3+2\epsilon+\alpha}(z_4)\rangle \langle \varphi_{\delta'}(z_6)\varphi_{1+\alpha-2\epsilon}(z_1)\varphi_{1-\alpha-2\epsilon}(z_4) \rangle=\nn\\
&  -\frac{4 \pi ^{\frac{3 d}{2}+1} \csc \left(\frac{1}{2} \pi  (d-2 \delta )\right)}{(d-2 \delta ) \Gamma \left(\frac{d}{2}\right) \Gamma (\delta ) \Gamma (d-\delta )} 2\pi \delta(s-s') \delta(z_5-z_6)\,.
\ee
We also note that around the singularity $2d+J_{L1}+J_{L2}+4\epsilon-6=0$ we have
\be
1+\alpha-2\epsilon=d-2+J_{L1}\,,\quad 1-\alpha-2\epsilon=d-2+J_{L2}\,.
\ee
We thus find \eqref{eq: dia1} with \eqref{eq: dia2}. We verified that this answer reproduces the $d=4$ result derived in \cite{Caron-Huot:2022eqs}. Additionally, it can be easily checked by explicit conformal block resummation in $d=2,4$, where the conformal block is known explicitly.
 
\subsection{Diagonalizing the kernels in Bethe-Salpeter equations}
\label{app: sole BS}

We now use the Lorentzian inversion formula to compute the OPE data \eqref{eq: OPE IK} from \eqref{eq: Kernal BFKL}
\be
K^{\rm BFKL}=\fft{1}{|y_{12}|^{2(d-2)}|y_{34}|^{2(d-2)}} \fft{\Gamma(\fft{d}{2}-1)^4}{256 \pi^{2d}}u^{d-2}v^{1-\fft{d}{2}}\,.
\ee

Our strategy is again to expand $K^{\rm BFKL}$ as a Taylor series around $z\rightarrow 0$. This trivializes the integral over $z$ and allows us to solve for the OPE coefficients twist by twist (where $\Delta=2(d-2)+J+2n$). We find that all twists contain the same remaining integral over $\bar{z}$
\be
\mathcal{G}^{\rm BFKL}(z,\bar{z})=\sum_{i=0} A_i z^{d-2+i} (1-\bar{z})^{\fft{2-d}{2}}\bar{z}^{d-2}\,.
\ee
We have the following generating functional
\be
&Z_\beta=\int_0^1 \fft{d\bar{z}}{\bar{z}^2}\, k_\beta^{00}(\bar{z})\kappa_{\beta}^{00}\,{\rm dDisc}[(1-\bar{z})^{\fft{2-d}{2}}\bar{z}^{d-2}]\nn\\
&=\frac{\pi  \Gamma \left(\frac{\beta }{4}\right) 2^{-\frac{\beta }{2}+d-2} \sin \left(\frac{\pi  d}{2}\right) \Gamma \left(2-\frac{d}{2}\right)}{\Gamma \left(1-\frac{\beta }{4}\right) \Gamma \left(\frac{\beta }{2}-\frac{1}{2}\right) \Gamma \left(\frac{d}{2}-1\right) \left(\sin \left(\frac{\pi  d}{2}\right)-\sin \left(\frac{1}{2} \pi  (\beta +d)\right)\right) \Gamma \left(-\frac{d}{2}+\frac{5}{2}-\frac{\beta }{4}\right) \Gamma \left(-\frac{d}{2}+\frac{\beta }{4}+2\right)}\,,
\ee
where $k_\beta^{ab}(z)$ is the SL$(2,{\rm R})$ block
\be
k_\beta^{ab}(z)=z^{\beta /2} \, _2F_1\left(a+\frac{\beta }{2},b+\frac{\beta }{2};\beta ;z\right)\,.
\ee
For each twist $n$, we then need to evaluate the following sum
\be
\sum_{m=-n}^n B_{nm}^{00}Z_{\beta+2m}A_n\,,
\ee
where $B_{nm}^{ab}$ are coefficients that can be computed recursively using the quadratic Casimir equation \cite{Caron-Huot:2017vep}. We then compute this quantity to obtain the OPE coefficients for all integer $n$. Next, we apply Watson-Sommerfeld resummation to reconstruct the shadow-symmetric OPE data $I_K(\Delta,J)S^{d-2,d-2}_{d-\Delta,J}=c_K(\Delta,J)$, leading to \eqref{eq: OPE IK}. We have also verified that this result reproduces the one obtained in \cite{Caron-Huot:2022eqs} for $d=4$.

\section{More complete contributions to horizontal detector operators}
\label{app: more horizontal}


In this appendix, we compute more complete contributions to horizontal detector operators that provide additive renormalization. This calculation is only performed in the detector frame; understanding its analogue in the distribution frame is a problem for the future.

In addition to the BFKL-type diagram, there are also disconnected diagrams contributing to horizontal detector operators. More generally, we consider the singlet sector, which gives rise to the diagrams shown in Fig \ref{fig: detector operators horizontal disc}. All necessary ingredients are provided in section \ref{sec: detector operator}; we collect them here and obtain
\be
& \langle \mathcal{H}_{J_{L1},J_{L2},J_L}\rangle_{\rm bare}= \big(X_{J_{L1}}+X_{J_{L2}}\big)\langle \mathcal{H}_{J_{L1},J_{L2},J_L}\rangle
+\hat{S}(-J_L,d-2+J_{L2},d-2+J_{L1}) Y_{J_{L1}}\langle \mathcal{H}_{2-d-J_{L1},J_{L2},J_L}\rangle\nn\\
&+\hat{S}(-J_L,d-2+J_{L1},d-2+J_{L2}) Y_{J_{L2}}\langle \mathcal{H}_{J_{L1},2-d-J_{L2},J_L}\rangle\nn\\
& -\fft{\mu^{4\epsilon}}{N}\fft{C_\sigma^2 \kappa_{\alpha}}{J_{L1}+J_{L2}+2d+4\epsilon-6}\langle H_{3-d,3-d,J_L}\rangle+\cdots\,,\label{eq: bare H all}
\ee
where $X,Y$ are read off from the separate single detector operator computations that encode $1/\epsilon$ and $1/(J_{L_i}-J_{L_i,{\rm div}})$ poles. The coefficients $\hat{S}(a,b,c)$ are essentially $(d-2)$-dimensional shadow coefficient divided by the normalization of the shadow transform \eqref{eq: norm shadow}.
\be
\hat{S}(a,b,c)=\frac{\Gamma (c) \Gamma \left(\frac{a}{2}+\frac{d}{2}-\frac{b}{2}-\frac{c}{2}-1\right) \Gamma \left(-\frac{a}{2}+\frac{b}{2}+\frac{d}{2}-\frac{c}{2}-1\right)}{\Gamma (-c+d-2) \Gamma \left(\frac{a}{2}+\frac{c}{2}-\frac{b}{2}\right) \Gamma \left(-\frac{a}{2}+\frac{b}{2}+\frac{c}{2}\right)}\,.
\ee

We find that it is not always necessary to include all disconnected pieces, as some of them may not mix with others under renormalization, according to the divergent structure of \eqref{eq: bare H all}. An interesting and simple subspace is
\be
\mathcal{H}^{(1)}=\left\{\mathcal{H}_{3-d,3-d,J_L},\mathcal{H}_{3-d,-1,J_L},\mathcal{H}_{-1,-1,J_L}\right\}\,,\label{eq: H subset 1}
\ee
where the first operator is of BFKL-type. The BFKL-type operator always enters because all horizontal detector operators contain divergences controlled by the BFKL-type detector, as shown in \eqref{eq: bare H all}. All detectors in \eqref{eq: H subset 1} mix with each other under renormalization. From \eqref{eq: bare H all}, we can extract the dilatation operator
\be
& \mathbf{\Delta}_L^{(1)}=\left(
\begin{array}{ccc}
 -2 & 0 & 0 \\
 0 & 2-d & 0 \\
 0 & 0 & 6-2 d \\
\end{array}
\right)+ \frac{2^{2d-5} (1-\cos (\pi  d)) \Gamma \left(\frac{d-1}{2}\right)^2}{\pi ^3 N} \left(
\begin{array}{ccc}
 \kappa (0) & 0 & 0 \\
 \kappa \left(\frac{d-4}{2}\right) & 0 & 0 \\
 \kappa (0) & 0 & 0 \\
\end{array}
\right)\nn\\
&+\frac{2^{d+1} (d-2) \sin \left(\frac{\pi  d}{2}\right) \Gamma \left(\frac{d+1}{2}\right)}{\pi ^{3/2} N \Gamma \left(\frac{d}{2}+1\right)} \left(
\begin{array}{ccc}
 1 & 0 & 0 \\
0 & a_{22} & 0\\
 0 & 0 & a_{33} \\
\end{array}
\right)\,,
\ee
where
\be
& a_{22}=\frac{d ((d-9) d+24)+\pi  (d-4) (d-3) (d-2) d \csc (\pi  d)-24}{2 (d-6) (d-2) (d-1)}\,,\nn\\
& a_{33}=\frac{(d-3) (\pi  (d-4) (d-2) d \csc (\pi  d)+4)}{(d-6) (d-2) (d-1)}\,.
\ee
We now also observe that this dilatation operator develops a singularity at $d\rightarrow 5$, which requires further treatment in the renormalization. We leave this problem for future studies.

A more complicated subspace of horizontal detectors is
\be
& \mathcal{H}^{(2)}=\nn\\
&\left\{\mathcal{H}_{3-d,3-d},\mathcal{H}_{J_{L1},J_{L1}},\fft{\mathcal{H}_{2-d-J_{L1},J_{L1}}-\mathcal{H}_{J_{L1},J_{L1}}}{J_{L1}-\fft{2-d}{2}},\fft{\mathcal{H}_{2-d-J_{L1},2-d-J_{L1}}-2\mathcal{H}_{2-d-J_{L1},J_{L1}}+\mathcal{H}_{J_{L1},J_{L1}}}{(J_{L1}-\fft{2-d}{2})^2}\right\}\,,
\label{eq: H subset 2}
\ee
aking $J_{L1}=J_{L2}\rightarrow (2-d)/2$. This particular linear combination cancels spurious singularities and degeneracy, following the spirit of \cite{Caron-Huot:2022eqs}. The general dimensional result of the dilatation operator for \eqref{eq: H subset 2} is lengthy; for simplicity, we only provide its $d=5$ result.
\be
&\mathbf{\Delta}_L^{(2)}=\left(
\begin{array}{cccc}
 -2 & 0 & 0 & 0 \\
 0 & -3 & 0 & 0 \\
 0 & 2 & -3 & 0 \\
 0 & 0 & 4 & -3 \\
\end{array}
\right) +\fft{64}{\pi^3 N} \left(
\begin{array}{cccc}
 \kappa (0) & 0 & 0 & 0 \\
 \kappa (0) & 0 & 0 & 0 \\
 0 & 0 & 0 & 0 \\
 -\kappa ''(0) & 0 & 0 & 0 \\
\end{array}
\right) + \fft{1024}{5\pi^2 N}\left(
\begin{array}{cccc}
 1 & 0 & 0 & 0 \\
 0 & a_{22} & a_{23} & 0 \\
 0 & a_{32} & a_{33} & a_{34} \\
 0 & a_{42} & a_{43} & a_{44} \\
\end{array}
\right)\,,
\ee
where
\be
& a_{22}=\frac{1}{96} (135 \hat{S}^{(0,0,1)}(\delta _1,\delta _2,\delta _3)+135 \hat{S}^{(0,1,0)}(\delta _1,\delta _2,\delta _3)+105 \hat{S}(\delta _1,\delta _2,\delta _3)+270 \log (2) \hat{S}(\delta _1,\delta _2,\delta _3)-379)\,,\nn\\
& a_{23}=\frac{45}{32} \hat{S}(\delta _1,\delta _2,\delta _3)\,,\quad a_{33}=\frac{1}{96} (-270 \hat{S}^{(0,1,0)}(\delta _1,\delta _2,\delta _3)-379)\,,\quad a_{34}=\frac{45}{64} \hat{S}(\delta _1,\delta _2,\delta _3)\,, \nn\\
& a_{32}=-\frac{5}{64}  ((14 \hat{S}^{(0,0,1)}(\delta _1,\delta _2,\delta _3)+9 \hat{S}^{(0,0,2)}(\delta _1,\delta _2,\delta _3)+14 \hat{S}^{(0,1,0)}(\delta _1,\delta _2,\delta _3)+18 \hat{S}^{(0,1,1)}(\delta _1,\delta _2,\delta _3)\nn\\
& +9 \hat{S}^{(0,2,0)}(\delta _1,\delta _2,\delta _3)+36 \log (2) \hat{S}^{(0,0,1)}(\delta _1,\delta _2,\delta _3)+36 \log (2) \hat{S}^{(0,1,0)}(\delta _1,\delta _2,\delta _3)+3 \pi ^2 \hat{S}(\delta _1,\delta _2,\delta _3)\nn\\
& +14 \hat{S}(\delta _1,\delta _2,\delta _3)+36 \log ^2(2) \hat{S}(\delta _1,\delta _2,\delta _3)+28 \log (2) \hat{S}(\delta _1,\delta _2,\delta _3)-114)\,,\nn\\
& a_{42}=\frac{5}{16} (3 \pi ^2 \hat{S}^{(0,1,0)}(\delta _1,\delta _2,\delta _3)+14 \hat{S}^{(0,1,0)}(\delta _1,\delta _2,\delta _3)+14 \hat{S}^{(0,1,1)}(\delta _1,\delta _2,\delta _3)+)9 \hat{S}^{(0,1,2)}(\delta _1,\delta _2,\delta _3)\nn\\
& +3 \hat{S}^{(0,3,0)}(\delta _1,\delta _2,\delta _3)+36 \log ^2(2) \hat{S}^{(0,1,0)}(\delta _1,\delta _2,\delta _3)+28 \log (2) \hat{S}^{(0,1,0)}(\delta _1,\delta _2,\delta _3)+36 \log (2) \hat{S}^{(0,1,1)}(\delta _1,\delta _2,\delta _3))\,,\nn\\
& a_{43}=-\frac{5}{32} (14 \hat{S}^{(0,0,1)}(\delta _1,\delta _2,\delta _3)+9 \hat{S}^{(0,0,2)}(\delta _1,\delta _2,\delta _3)-14 \hat{S}^{(0,1,0)}(\delta _1,\delta _2,\delta _3)-18 \hat{S}^{(0,1,1)}(\delta _1,\delta _2,\delta _3)\nn\\
& +9 \hat{S}^{(0,2,0)}(\delta _1,\delta _2,\delta _3)+36 \log (2) \hat{S}^{(0,0,1)}(\delta _1,\delta _2,\delta _3)-36 \log (2) \hat{S}^{(0,1,0)}(\delta _1,\delta _2,\delta _3)+3 \pi ^2 \hat{S}(\delta _1,\delta _2,\delta _3)\nn\\
& +14 \hat{S}(\delta _1,\delta _2,\delta _3)+36 \log ^2(2) \hat{S}(\delta _1,\delta _2,\delta _3)+28 \log (2) \hat{S}(\delta _1,\delta _2,\delta _3)-114)\,,\nn\\
& a_{44}=\frac{1}{96} (-135 \hat{S}^{(0,0,1)}(\delta _1,\delta _2,\delta _3)+135 \hat{S}^{(0,1,0)}(\delta _1,\delta _2,\delta _3)-105 \hat{S}(\delta _1,\delta _2,\delta _3)-270 \log (2) \hat{S}(\delta _1,\delta _2,\delta _3)-379)\,,
\ee
with $\delta_1=-J_L,\delta_2=\delta_3=3/2$.

\section{Revisiting rapidity renormalization}
\label{app: rapidity}

In this appendix, we revisit rapidity renormalization, closely following the spirit of \cite{Chiu:2011qc,Chiu:2012ir} to help readers understand section \ref{sec: RG distribution operator} in the main text.

\subsection{Rapidity divergences and the rapidity renormalization}

\begin{figure}[t]
\centering
    \includegraphics[width=0.5\textwidth]{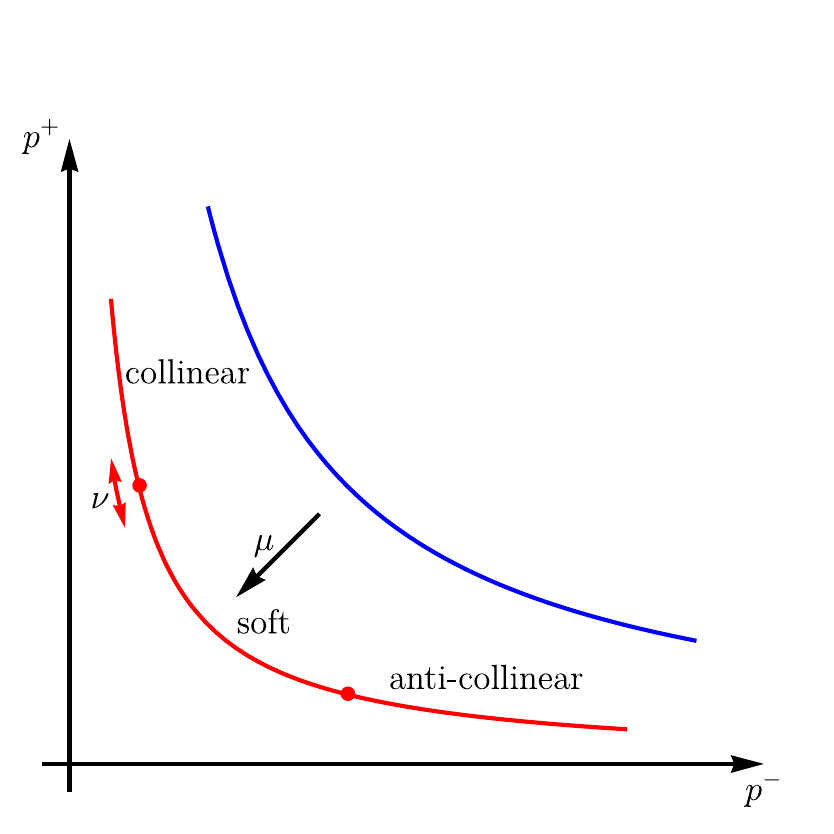} 
    \caption{The standard RG describes a flow from a high-energy hyperbola (blue) to a low-energy hyperbola (red) along $\mu$. The rapidity RG arises from the ambiguity in separating the low-energy hyperbola (red) into collinear, soft, and anti-collinear regions; this separation is controlled by $\nu$, which flows only along the red hyperbola.}
    \label{fig: rapidity flow}
\end{figure}

In momentum space, rapidity divergences arise from a momentum region with fixed $p^2$ but a divergence in the rapidity $p^+/p^-$. This is neither an IR nor UV divergence; rather, it arises from the factorization between soft and collinear (as well as anti-collinear) modes. A simple integral can conceptually illustrate the rapidity divergence.
\be
I=\int_{m}^Q \fft{dp^+}{p^+}\,,
\ee
where $m$ is the low-energy scale, and $Q$ is the hard scale, such that $m \ll Q$. This integral is not a complete integral; instead, it is factorized out of a finite integral (or CFT correlator) in the UV-complete picture by decoupling the hard and soft regions with cutoffs $Q$ and $m$. These hard and soft regions can be regulated using dimensional regularization in general or scaling dimensional regularization in the critical O$(N)$ model. Now, the integral $I$ receives contributions from both the collinear region, where $p^+ \sim Q$ (i.e., it is strongly directed along the $+$ direction), and the soft region, where $p^+ \sim m$ (i.e., its momentum is as small as the soft scale). If we now try to separate these regions and develop an effective description, we will find
\be
I=\int_{m}^\Lambda \fft{dp^+}{p^+} + \int_\Lambda^Q \fft{dp^+}{p^+}\rightarrow \int_{m}^{\infty} \fft{dp^+}{p^+} + \int_{0}^Q \fft{dp^+}{p^+}\,,
\ee
where the second approximation appears because EFTs do not know the ambiguous scale $m\ll\Lambda\ll Q$ unless we regulate them. This is precisely aligned with the spirit of UV divergence, where the UV divergence arises from the ambiguity of separating the low-energy and high-energy modes. Nevertheless, this divergence, like $\int_0^Q dp^+/p^+$, is not regulated by dimensional regularization (which has already played its role in regulating UV and IR divergences). A new regulator is needed to handle the rapidity divergence. The simplest approach is to reintroduce $\Lambda$ to obtain $\int_\Lambda^Q dp^+/p^+$, which follows the spirit of a hard cutoff regulator for UV divergences. A more canonical approach, following the spirit of dimensional regularization, is to introduce the rapidity regulator by smearing this integral, i.e.,
\be
\int_0^Q \fft{dp^+}{p^+}  \left|\fft{p^+}{\nu}\right|^{-\eta}\,,
\ee
with the rapidity scale $\nu$. Then, the ambiguity of $\nu$ induces the rapidity RG flow to run only in the rapidity direction and resum the large rapidity logarithms. These intuitions are conceptually illustrated in Fig.~\ref{fig: rapidity flow}

One may wonder why rapidity divergences can appear in CFTs, given that CFTs do not have any intrinsic scales. This follows the same spirit as UV divergences appearing in perturbative calculations of CFTs in the OPE limit: the small separation of two closely spaced operators sets the UV scale for expanding operators that are widely separated
\be
g \log\left(\fft{u}{1-v}\right)\Big|_{u\rightarrow 0, v\, \text{fixed}} \sim g \log\left(-\fft{x_{12}^2}{x_{23}^2}\right)\Big|_{|x_{12}|\ll |x_{23}|}+\text{finite}\,,\label{eq: log UV CFT}
\ee
where we take $x_{12}$ to be spacelike and $x_{23}$ to be timelike. This large logarithm still needs to be resummed by ``RG'', which is nothing but a scaling operation that builds this conformal frame. The rapidity divergences particularly appear in the Regge limit of CFTs \cite{Costa:2012cb}
\be
g\log u\Big|_{u\rightarrow 0,v\rightarrow 1} \sim g \log\left(\fft{-x_{12}^+ x_{12}^-}{x_{13}^2}\right)\Big|_{|x_{12}^+ x_{12}^-|\ll x_{13}^2}+\text{finite}\,,
\ee
where the large rapidity arises from the fact that $x_{12}$ approaches a null line, while $x_{13}$ is spacelike. Therefore, a rapidity RG is required to resum this type of large logarithm.

\subsection{The rapidity regulator by tilting or smearing}
\label{app: rapidity regulator}

We now provide a brief discussion on regulating the rapidity divergence.

Rapidity divergences can arise from the infinite rapidity of a light-ray operator. A natural way to regulate such divergences is to tilt the light-ray to have a constant rapidity $Y$. For example, in the case of a null integrated operator, we can instead consider
\be
\int dx^- \mathcal{O}(x^- Y,x^-,x^\perp)\,.
\ee
This is similar to introducing a small but finite mass as an IR regulator when dealing with IR divergences. Nevertheless, we find that this regulator is not practically easy to implement \cite{Balitsky:2007feb,Balitsky:2009xg}.

Another approach is to smear the light-ray operator over a small window (with an appropriate normalization)
\be
-\int \fft{d\delta x^+}{\nu} \mathcal{O}(\delta x^+,x^-,x^\perp)\left(\fft{\delta x^+}{\nu}\right)^{-1-\eta}\eta\, \theta(\delta x^+) w(\nu) f(x^\perp)\,.
\ee
Taking $\eta\rightarrow 0$ returns to the original definition via $(\delta x^+)^{-1-\eta}\theta(\delta x^+) \sim -\delta(\delta x^+)/\eta$. In momentum space, this smearing gives $(\ell^-)^{\eta}$. The on-shell cuts, e.g., in \eqref{eq: do Cs}, then translate $(\ell^-)^\eta$ to $f(\ell_i^\perp)^{\eta}(p^+\xi)^{-\eta}$, with some nontrivial but unimportant function of transverse loop momenta $f(\ell_i^\perp)$. Therefore, this smearing ensures the insertion of the regulator \eqref{eq: rapidity regulator} in the loop integrand, with an additional function $f(\ell_i^\perp)$ that can be dropped because it does not contribute to the divergence).

There are more ways of regulating the rapidity divergences, see .e.g, \cite{Collins:1981uk,Collins:1989bt,Chiu:2009yx,Li:2016axz,Balitsky:2007feb,Balitsky:2009xg}.

\bibliographystyle{JHEP}
\bibliography{refs}

\end{document}